\documentclass[a4paper,12pt]{article}

\usepackage{amsmath}
\usepackage{amssymb}
\usepackage{color}

\makeatletter
\@addtoreset{equation}{section}
\renewcommand{\theequation}{\thesection.\@arabic\c@equation}
\makeatother

\newcommand{\tr}{\mathrm{tr}}

\newcommand{\aw}{\alpha}
\newcommand{\awb}{\bar{\alpha}}

\newcommand{\awu}{\hat{\alpha}}
\newcommand{\AX}{a}
\newcommand{\MX}{m}
\newcommand{\MXo}{m}
\newcommand{\mw}{M}
\newcommand{\ms}{M_S}
\newcommand{\PX}{\Pi}
\newcommand{\pw}{\pi}
\newcommand{\cc}{\lambda}
\newcommand{\rc}{\tilde{\lambda}}
\newcommand{\OX}{\mathcal O}
\newcommand{\OS}{\mathcal S}
\newcommand{\sw}{\hat{S}_{w}}
\newcommand{\swb}{\hat{S}_{\bar w}}

\allowdisplaybreaks[1]

\begin{document}

\begin{titlepage}

\vspace*{-15mm}   
\baselineskip 10pt   
\begin{flushright}   
\begin{tabular}{r}    
{\tt OU-HET 999}\\   
February 2019
\end{tabular}   
\end{flushright}   
\baselineskip 24pt   
\vglue 10mm   

\begin{center}
{\Large\bf
Nuclear states and spectra in holographic QCD
}

\vspace{8mm}   

\baselineskip 18pt   

\renewcommand{\thefootnote}{\fnsymbol{footnote}}

%Author~Name\footnote[2]{e-mail@address}, 
Koji~Hashimoto%
${}^{a}$\footnote[2]{e-mail address: koji@phys.sci.osaka-u.ac.jp},
Yoshinori~Matsuo%
${}^{a,b}$\footnote[3]{e-mail address: matsuo@het.phys.sci.osaka-u.ac.jp}
and 
Takeshi~Morita%
${}^{c,d}$\footnote[4]{e-mail address: morita.takeshi@shizuoka.ac.jp}

\renewcommand{\thefootnote}{\arabic{footnote}}
 
\vspace{5mm}   

{\it  
% Affiliation
${}^{a}$Department of Physics, Osaka University, Toyonaka, Osaka 560-0043, Japan\\
${}^{b}$Department of Physics and Center for Theoretical Physics, \\
National Taiwan University, Taipei 106, Taiwan,
R.O.C.\\
${}^{c}$Department of Physics, Shizuoka University\\
  	836 Ohya, Suruga-ku, Shizuoka 422-8529, Japan \\
${}^{d}$Graduate School of Science and Technology, Shizuoka University\\
836 Ohya, Suruga-ku, Shizuoka 422-8529, Japan
}

\vspace{10mm}   

\end{center}

\begin{abstract}
A new method to study nuclear physics via holographic QCD is proposed. 
Multiple baryons in the Sakai-Sugimoto background are described by a matrix model
which is a low energy effective theory of D-branes of the baryon vertices. 
We study the quantum mechanics of the matrix model 
and calculate the eigenstates of the Hamiltonian. 
The obtained states are found to coincide with known nuclear and baryonic states, 
and have appropriate statistics and charges. 
Calculated spectra of the baryon/nucleus for small baryon numbers
show good agreement with experimental data.
For hyperons, the Gell-Mann--Okubo formula is approximately derived.
Baryon resonances up to spin $5/2$ and isospin $5/2$ and dibaryon spectra are obtained
and compared with experimental data. 
The model partially explains even the magic numbers of light nuclei, $N=2,8$ and 20.

\end{abstract}

\end{titlepage}

\newpage

\baselineskip 18pt

\noindent\rule{\textwidth}{1pt}

\setcounter{tocdepth}{2}
\tableofcontents

\vskip 12pt

\noindent\rule{\textwidth}{1pt}

\vskip 12pt

%\clearpage

%%%%%%%%%%%%%%%%%%%%%%%%%%%%%%%%%%%%%%%%%%%%%%%%%%%%%%%%%%%%%%%%%%%%%%%%%%%%%%%%
%%%%%%%%%%%%%%%%%%%%%%%%%%%%%%%%%%%%%%%%%%%%%%%%%%%%%%%%%%%%%%%%%%%%%%%%%%%%%%%%
%%%%%%%%%%%%%%%%%%%%%%%%%%%%%%%%%%%%%%%%%%%%%%%%%%%%%%%%%%%%%%%%%%%%%%%%%%%%%%%%

\section{Introduction}

One of the long-standing problems in QCD is to reproduce profound nuclear physics.
The strong coupling nature of QCD prevents us from solving it analytically, and even numerical
simulations have a limitation such as the volume of the atomic nucleus versus the lattice size.
It is quite important 
to bridge the particle physics and the nuclear physics, by solving QCD to derive typical fundamental
notions of the nuclear physics, such as the magic numbers, the nuclear binding energy and the
nuclear shell model.

Holographic QCD is an analytic method to approach these problems, in the strong coupling limit
and at a large $N_c$. The nuclear matrix model \cite{Hashimoto:2010je} is a
many-body quantum effective mechanics for multiple baryons, derived by the AdS/CFT
correspondence \cite{Maldacena:1997re,Gubser:1998bc,Witten:1998qj}
applied to QCD. The derivation does not assume any empirical feature of nuclear systems, since
it is just a combination of D-branes and the general principles of the AdS/CFT.

In this paper, we solve the spectra of the nuclear matrix model
for small baryon numbers, and deduce important features of nuclear physics, including
(i) baryon resonance spectrum, (ii) hyperon spectrum and Gell-Mann--Okubo relation, 
(iii) dibaryon spectrum
and (iv) the magic number $N=2,8,20$. We find that the model appears to possess a natural
picture of a quark model, and relates to the nuclear shell model.

In the large $N_c$ QCD \cite{'tHooft:1973jz}, mesons are light while baryons
are with a large mass of ${\cal O}(N_c)$ \cite{Witten:1979kh}, and so the
baryons can be thought of as solitons, like Skyrmions \cite{Skyrme:1962vh}.
The Sakai-Sugimoto model of holographic QCD \cite{Sakai:2004cn,Sakai:2005yt} 
allows a natural analogue of that, as instantons. Therefore, so far, 
study toward nuclear physics has utilized this solitonic picture, and 
a successful comparison with experiments
have been made for the case of a single baryon or the inter-baryon forces:
baryon static properties \cite{Hong:2007kx,Hata:2007mb,Hata:2008xc},
interactions with mesons \cite{Hong:2007kx,Hong:2007ay}, 
baryon electromagnetic
form factors \cite{Hong:2007dq,Hashimoto:2008zw,Kim:2008pw},
nucleon-nucleon potential
\cite{Hashimoto:2009ys,Kim:2009sr,Kim:2008iy,Kaplunovsky:2010eh,Cherman:2011ve}, and 
higher isospin baryons \cite{Park:2008sp,Grigoryan:2009pp}, for example.\footnote{Simple holographic models
allow baryons in a similar manner, see \cite{Pomarol:2008aa} and subsequent developments.}
Holographic deuteron was also studied
\cite{Kim:2009sr,Pahlavani:2010zzb,Pahlavani:2014dma} 
(see also 
\cite{Baldino:2017mqq, Bolognesi:2013jba}).
The soliton picture has been extended to the study of infinite systems, with a
finite baryon density \cite{Kim:2007zm,Rozali:2007rx,Kim:2007vd,Rho:2009ym,Kaplunovsky:2012gb,Kaplunovsky:2015zsa}.

However, nuclear physics is intrinsically for a finite number of baryons,
and the soliton picture, such as the Skyrme model, has difficulties in nuclear modeling. 
On the other hand, the nuclear matrix model \cite{Hashimoto:2010je} is a multi-particle 
quantum mechanics for a finite number of baryons, and was derived such that
it could serve as a basis for possibly deducing nuclear physics from holographic QCD.
Using the nuclear matrix model, the followings were studied: the
coincidence with the solitonic baryon spectrum \cite{Hashimoto:2010je},
the universal repulsive core \cite{Hashimoto:2010je}, 
the three-body nuclear forces \cite{Hashimoto:2010ue}, 
the nucleon statistics \cite{Hashimoto:2010rb},
the hyperon repulsive core \cite{Aoki:2012th}, and the $N$-body forces  
\cite{Hashimoto:2009as}.\footnote{
A large baryon number limit was studied
to describe the formation of an atomic nucleus \cite{Hashimoto:2011nm}.
Previously, those gravity duals were considered \cite{Hashimoto:2008jq,Hashimoto:2009pe}.}
However, in all of these analyses, the solitonic picture was implicitly in mind ---
in fact, the nuclear matrix model was treated such that it coincides with the
soliton picture. So the full merit of the model --- a quantum mechanical model
of nuclear physics, as is the case for the nuclear shell model --- has not been 
enjoyed.

In the present paper, 
we perform a detailed analysis of the nuclear matrix model, based on a new
quantization scheme, and find that 
the model possesses naturally a similarity to the quark model.
While the previous work of the nuclear matrix model solved the ADHM equation \cite{Atiyah:1978ri}
which is obtained as the condition of the minimum of the potential
at the large 't Hooft coupling limit, here instead, we quantize the model simply with the potential 
as an interaction, and 
consider the fluctuations and excitations in the potential. 
The potential is a consequence of the D-brane interaction
\cite{Witten:1995gx,Douglas:1995bn}, therefore it is intriguing that the D-brane dynamics
naturally provides a part of the nuclear physics, just based on the fact that
baryons are D-branes \cite{Witten:1998xy,Gross:1998gk} in the AdS/CFT correspondence. 

The nuclear matrix model has another constraint, 
which comes from the equation of motion of the non-dynamical gauge field. 
Instead of just integrating out the gauge field, 
we impose the constraint on the Hamiltonian eigenstates.%
\footnote{%
A similar procedure is utilized in studies of quantum Hall states, matrix models 
and their relation to solitons \cite{Tong:2015xaa, Dorey:2016mxm, Barns-Graham:2017zpv}. 
} 
This constraint extracts states which have particular charges and symmetries, 
and then, the resultant states turn out to have appropriate properties of baryons/nuclei. 
Then we investigate the spectra of the baryon/nucleus by calculating the eigenvalues of the Hamiltonian. 
This alternative procedure remarkably provides extensive applications 
in nuclear physics for small baryon numbers: 
the baryon spectra which cover states beyond the soliton limit, 
the dibaryon spectra and even a part of the magic numbers.

The organization of this paper is as follows. In Sec.~\ref{sec:Model}, we review the Lagrangian
of the nuclear matrix model, and explain our new quantization method and obtain
ground state wave functions. 
In Sec.~\ref{sec:Linear}, we consider the linear order perturbation and 
splitting of degenerated energy levels for allowed states. 
In Sec.~\ref{sec:Hyperon}, we study hyperon spectrum and derive Gell-Mann--Okubo(-like) relation.
In Sec.~\ref{sec:Dibaryon}, we investigate the interaction terms in detail, and consider dibaryons.
In Sec.~\ref{sec:Resonance}, we calculate baryon resonance which cannot be obtained from the
Skyrme model.
In Sec.~\ref{sec:MagicNumber}, we explain magic numbers of the nuclear matrix model, and 
obtain $N=2,8,20$.
The last section is for a summary and discussions.
Appendix \ref{app:gen} studies general interaction terms and their effects on the dibaryon spectrum.
In Appendix \ref{app:Mass}, we discuss several proposals for determining the mass parameter of the baryon vertices.
Appendix \ref{app:more} is for detailed analyses of the dibaryon spectra with different ansatz.

%%%%%%%%%%%%%%%%%%%%%%%%%%%%%%%%%%%%%%%%%%%%%%%%%%%%%%%%%%%%%%%%%%%%%%%%%%%%%%%%
%%%%%%%%%%%%%%%%%%%%%%%%%%%%%%%%%%%%%%%%%%%%%%%%%%%%%%%%%%%%%%%%%%%%%%%%%%%%%%%%
%%%%%%%%%%%%%%%%%%%%%%%%%%%%%%%%%%%%%%%%%%%%%%%%%%%%%%%%%%%%%%%%%%%%%%%%%%%%%%%%

\section{Nuclear matrix model and its quantization}
\label{sec:Model}

The nuclear matrix model which is studied in this paper \cite{Hashimoto:2010je}  
comes from the baryon vertex in the Sakai-Sugimoto model of holographic QCD 
\cite{Sakai:2004cn,Sakai:2005yt}. 
In order to investigate $A$ baryons, we introduce $A$ D4-branes corresponding to the baryon vertices which are embedded in $N_f$ flavor D8-branes 
and are wrapped on the $S^4$-cycle in the color D4-brane background.
After the dimensional reduction for the compact directions of $S^4$, 
the effective theory is described by a $U(A)$ matrix quantum mechanics. 
After an appropriate redefinition of fields and coupling constants, 
the action is expressed as 
\begin{align}
 S &= S_0 + N_c \int dt\,\tr A_t \ , 
\label{action-o}
\\
 S_0 &= \int dt\,\tr
 \biggl[
  \frac{1}{2} \left(D_t X^I\right)^2 
  + \frac{1}{2} D_t \bar w^{\dot \alpha i} D_t w_{\dot \alpha i}
  - \frac{1}{2} \mw^2 \bar w^{\dot \alpha i} w_{\dot \alpha i}
\notag\\
&\qquad\qquad\qquad
  + \frac{1}{4\cc} (D^I)^2 + D^I\left(2 i\epsilon^{IJK} X^J X^K 
  + \bar w^{\dot \alpha i} (\tau^I)_{\dot \alpha}{}^{\dot\beta} w_{\dot \beta i} \right)
 \biggr] 
\ , 
\label{action}
\end{align}
where $D_t = \partial_t - i A_t$ is the covariant derivative and 
$A_t$ is the gauge field of $U(A)$ on the $A$ D4-branes. 
The indices $I, J, K$ label the three dimensional space in which 
the flavor D8-branes are extended but the baryon D4-branes are localized, 
and tangent to the color D4-branes.
The other indices $i,j,k$ and $\dot \alpha, \dot \beta, \dot \gamma$ 
stand for the (anti-)fundamental representations of flavor $SU(N_f)$ 
and spin $SU(2)$, respectively.  
The dynamical fields $X^I$, $w$ and $\bar w$ are scalar fields 
which come from the D4-D4 and D4-D8 open strings and 
behave as adjoint, fundamental and anti-fundamental in $U(A)$, respectively.
We neglected the fermionic fields, since they are heavy because of 
the supersymmetry breaking.%
\footnote{%
	\label{ftnt-large-L}
The interaction between baryon D4-branes can be described by the off-diagonal components 
of the baryon $U(A)$ only if the distances between the D4-branes are sufficiently small. 
When the distance $L$ is large such that $L \gg M_{\rm KK} \lambda^{-2/3} \sim (N_c/\lambda_{\rm QCD})^{2/3}/M_{\rm KK}$, effects of the fermionic fields are not negligible and modify the action \eqref{action}. 
}  
The massive Kaluza-Klein modes on the  $S^4$ and the scalar fields which describe fluctuations in the directions away from 
the tip of the cigar geometry will also be very heavy and were neglected.%
\footnote{
In the original model \cite{Hashimoto:2010je}, there 
exists another scalar field $X^z$ which specifies the location of the baryon D4-branes along the curved D8-branes in the curved space.
In addition, there exists a scalar field $X^y$ which measures the distance between
the baryon D4-brane and the flavor D8-branes. These two scalars 
are heavy and in this paper we shall not study this excitation 
for simplicity.}

The mass $M$ and the coupling constant $\lambda$ are 
related to $M_\text{KK}$ and $\lambda_\text{QCD}$ as 
\begin{align}
 M &= \frac{1}{\sqrt 6} M_\text{KK} \ , 
& 
 \lambda &= \frac{\lambda_\text{QCD} M_\text{KK}^3}{54 \pi N_c} \ ,  
 \label{coupling}
\end{align}
where $M_\text{KK}$ is the inverse of the radius of $S^1$ of the extra dimension in the color D4-branes, 
and $\lambda_\text{QCD}$ is the 't Hooft coupling of QCD, or equivalently, 
the 't Hooft coupling on the color D4-brane effective theory after the dimensional reduction of $S^1$. 
Since our model comes from the baryon vertex in the Sakai-Sugimoto model, one can choose these constants such that the Sakai-Sugimoto model reproduces, for example, the meson spectra. 
In this paper, however, we treat them as free parameters 
and determine by using experimental data such that 
our model becomes suitable to describe the nuclear spectra. 
It should be noted that this model is based on the Sakai-Sugimoto background, 
which is justified only in the large-$N_c$ limit with $\lambda_\mathrm{QCD}$ fixed, 
while we mostly interested in nuclear states at $N_c=3$. 
Hence we study the model for finite $N_c$, $\cc$, and so $\lambda_\mathrm{QCD}$. 
Assuming that we can still utilize the large-$N_c$ expansion for $N_c=3$, 
there would be the higher order corrections, which we will simply neglect in this paper.

We first integrate out the auxiliary field $D^I$ to obtain 
\begin{align}
 S_0 
&= 
 \int dt\,\tr
 \biggl[
  \frac{1}{2} \left(D_t X^I\right)^2 
  + \frac{1}{2} D_t \bar w^{\dot \alpha i} D_t w_{\dot \alpha i}
  - \frac{1}{2} \mw^2 \bar w^{\dot \alpha i} w_{\dot \alpha i} 
\notag\\
&\quad\qquad\qquad
  + 2 \cc \left[X^I , X^J \right]^2
  + 4 i \cc \epsilon^{IJK} X_A^J X_B^K f^{AB}{}_C\, 
    \bar w^{\dot \alpha i}_a (\tau^I)_{\dot \alpha}{}^{\dot\beta} (t^C)^a{}_b w_{\dot \beta i}^b 
\notag\\
&\quad\qquad\qquad
  - \cc \left(\bar w^{\dot \alpha i} (\tau^I)_{\dot \alpha}{}^{\dot\beta} w_{\dot \beta i}\right)^2 
 \biggr] 
\ , 
\end{align}
where the indices $a, b$ label the (anti-)fundamental representations of $U(A)$, and 
the $t^A$ and $f^{ABC}$ are the generator and structure constant. 

Next, we consider the equation of motion for the gauge field $A_t$. 
In the matrix quantum mechanics, 
the gauge field behaves as an auxiliary field and gives a constraint. 
The variation of the action with respect to the gauge field $A_t$ 
is nothing but the $U(A)$ charge of the fields $X^I$, $w$ and $\bar w$; 
\begin{equation}
 Q_{U(A)} = J^t_{U(A)} = - \frac{\delta S_0}{\delta A_t} 
 = i X^I D_t X^I + \frac{i}{2} \bar w^{\dot \alpha i} D_t w_{\dot \alpha i} 
 - \frac{i}{2} (D_t\bar w^{\dot \alpha i}) w_{\dot \alpha i} \ . 
\end{equation}
The equation of motion for $A_t$ gives a constraint on the $U(A)$ charge 
\begin{equation}
 Q_{U(A)} - N_c \mathbb I_A = 0 \ , 
\label{Const}
\end{equation}
where the charge $Q_{U(A)}$ above is expressed as an $A\times A$ matrix 
and $\mathbb I_A$ is $A\times A$ unit matrix. 
Here, we do not substitute \eqref{Const} back to the action. 
We will study the matrix quantum mechanics by using the Hamiltonian formalism 
and impose the constraint \eqref{Const} on the eigenstates of Hamiltonian. 
The condition \eqref{Const} can also be written as 
\begin{align}
 Q_{SU(A)} &= 0 \ , & 
 Q_{U(1)} &= N_c A \ , 
\end{align}
implying that physical states must be 
singlet of the $SU(A)$ symmetry and the total charge of the overall $U(1)$ in 
$U(A)$, $Q_{U(1)} = \tr Q_{U(A)}$ must be $N_c A$. 

Hamiltonian of the model is given by 
\begin{align}
 H_\text{full} &= H - \tr A_t \left(Q_{U(A)} - N_c \right) \ , 
\\
 H
&=
 \frac{1}{2} \tr (\PX^I)^2 + 2 \bar\pw_{\dot \alpha i}^a \pw^{\dot \alpha i}_a 
 + \frac{1}{2} \mw^2 \bar w^{\dot \alpha i}_a w_{\dot \alpha i}^a 
\notag\\
&\quad
  - 2 \cc \tr \left[X^I , X^J \right]^2
  - 4 i \cc \epsilon^{IJK} X_A^J X_B^K f^{AB}{}_C\, 
    \bar w^{\dot \alpha i}_a (\tau^I)_{\dot \alpha}{}^{\dot\beta} (t^C)^a{}_b w_{\dot \beta i}^b 
\notag\\
&\quad
  + \cc \left(\bar w^{\dot \alpha i} (\tau^I)_{\dot \alpha}{}^{\dot\beta} w_{\dot \beta i}\right)^2 \ , 
\end{align}
where $\PX$, $\pw$ and $\bar\pw$ are canonical conjugate momenta of $X$, $w$ and $\bar w$, respectively. 
The second term in $H_\text{full}$ gives the constraint \eqref{Const}, 
and $H$ is the Hamiltonian without the constraint. 
We will consider the eigenstates of $H$ and impose the constraint \eqref{Const}. 

Although $X$ has flat directions at classical level, 
it is known that the potential of the form $[X^I,X^J]^2$ provides 
a bound state of the eigenvalues of $X^I$ with $U(A)$ invariance for ground state
\cite{Luscher:1982ma}.%
\footnote{%
Although this interaction term provides attractive force between arbitrarily separated baryons, 
the action \eqref{action} itself is not appropriate for sufficiently separated baryons. 
It is expected that the interactions between baryons should be suppressed in such a case. 
\label{ftnt-long-distance}
}
It is also known that the potential can be approximated by a harmonic potential 
if the number of the spatial dimensions is large \cite{Mandal:2009vz,Hashimoto:2011nm}.%
\footnote{%
The approximation of the commutator squared potential by a harmonic potential at large number of the spatial dimensions is so called ``large-$D$ expansion'' \cite{Hotta:1998en,Mandal:2009vz, Morita:2010vi, Mandal:2011hb}.
As we will discuss in  App.~\ref{B:list}, there are numerical evidences which support this approximation in the case of the matrix models obtained through the dimensional reduction of higher dimensional pure Yang-Mills theories \cite{Hotta:1998en, Aharony:2004ig, Aharony:2005ew,  Kawahara:2007fn, Azuma:2014cfa}.
The large-$D$ expansion seems to work quantitatively even for 
small number of the spatial dimensions such as three \cite{Azuma:2014cfa}, 
though it is nontrivial whether the large-$D$ expansion can be applied 
for the nuclear matrix model since it is not just a dimensionally-reduced Yang-Mills theory. 
Moreover, there are other possible methods that approximate the models 
with the commutator squared potential by the harmonic potential as we show in App.~\ref{app:Mass}. 
In this paper, we assume that some of these approximations work in the nuclear matrix model.
%In App.~\ref{app:Mass}, we discuss possible ways to determine the value of the mass.
} 
Although the spatial dimensions is just three in our model, 
we assume that the approximation by using a harmonic potential 
is still good in this case.%
\footnote{%
According to the previous study such as \cite{Hashimoto:2009ys}, 
at the short distances the holographic baryons feel a universal repulsive force. And also at
long distances, the inter-baryon force should disappear. In our approximation
using the harmonic potential, these features seem not captured. However, note that
(i) these features are for given classical values of the diagonal components of the matrices $X$
which solve the classical equation of motion of the matrix model, since the components
provide the physical meaning of the distance between the baryons. In this paper, on the
other hand, we provide a quantum analysis of the matrix model at which we put no distinction between the diagonal and off-diagonal components, with the $U(A)$ symmetry, to look at bound state properties
of the multi-baryon system. 
%The harmonic potential ${\rm tr} [(X^I)^2]$ vanishes at $X^I=0$, but it does not immediately lead to no repulsive core.
Therefore, it should be understood 
that the typical hadron features described above cannot be seen in the matrix operators $X^I$ 
but will emerge in calculations of some physical quantities such as expectation values.
(ii) For the repulsive force at the short distance, 
the effect of $w$ fields and the constraint from the gauge field $A_t$ play important roles. 
The harmonic potential ${\rm tr} [(X^I)^2]$ vanishes at $X^I=0$, but it does not immediately lead to no repulsive core.
The repulsive force would be reproduced, if possible, once $w$ fields are integrated. 
(iii) We assume that nucleons form a bound state and the distance is not very large for the perturbation. 
Also, our model itself is not appropriate at long distances. 
See footnote \ref{ftnt-large-L} and \ref{ftnt-long-distance}.
} 
We will consider perturbation around this harmonic potential,%
\footnote{%
We simply assume that the perturbation works at this moment, 
but there are several references in which similar perturbative analyses 
were considered \cite{Kabat:1999hp,Kabat:2000zv,Kabat:2001ve,Iizuka:2001cw,Nishimura:2001sx,Kawai:2002jk,Nishimura:2002va}. 
We also check the validity after calculating the perturbative corrections. 
In \cite{Hashimoto:2010je} it was argued that the relevant terms are of order $\lambda$
and therefore in the limit $\lambda \to \infty$ they are treated as a classical potential.
Here in this paper, instead, we treat them as a perturbation, which is a working hypothesis.
We expect that in the approach of \cite{Hashimoto:2010je} if we solve the ADHM equation and
consider only the bottom of the ADHM potential, the energy at the bottom is not so large 
so that it can be accessible by the perturbation of our approach. 
See also a discussion provided in Sec.\ref{subsec-largeNc} at the large $N_c$ limit.
} 
and difference from the original interaction terms are treated as perturbation. 
Hence, the Hamiltonian is separated into the unperturbed pert $H_0$ and the perturbation $V$ as
\begin{align}
 H &= H_0 + V \ , 
\label{H}
\\
 H_0 
&= 
 \frac{1}{2} \tr (\PX^I)^2 + \frac{1}{2} \MX^2 \tr (X^I)^2 + 2 \bar\pw_{\dot \alpha i}^a \pw^{\dot \alpha i}_a 
 + \frac{1}{2} \mw^2 \bar w^{\dot \alpha i}_a w_{\dot \alpha i}^a \ , 
\label{H0}
\\
 V 
&= 
  - \frac{1}{2} \MX^2 \tr (X^I)^2 
  - 2 \cc \tr \left[X^I , X^J \right]^2
\notag\\
&\quad
  - 4 i \cc \epsilon^{IJK} X_A^J X_B^K f^{AB}{}_C \, 
    \bar w^{\dot \alpha i}_a (\tau^I)_{\dot \alpha}{}^{\dot\beta} (t^C)^a{}_b w_{\dot \beta i}^b 
 + \cc \left(\bar w^{\dot \alpha i} (\tau^I)_{\dot \alpha}{}^{\dot\beta} w_{\dot \beta i}\right)^2 \ . 
\label{V}
\end{align}
Since the harmonic potential (mass term) $ \frac{1}{2} \MX^2 (X^I)^2 $  comes from the interaction term $[X^I,X^J]^2$, 
the overall $U(1)$, or equivalently trace part of $X^I$ will not get mass from the interaction term. 
The trace part of $X^I$ represents the position of the center of mass of the nucleus 
and simply describes its free propagation. 
Thus the trace part can be separated from the others and 
we consider only the $SU(A)$ part of $X^I$. 
The effective mass $\MX$ would be generated dynamically and it generally depends on the physical state.
However the value depends on the scheme of the perturbation, 
although the results are expected to be independent of $\MX$ 
if the higher order corrections are taken into account. 
We discuss the determination of the effective mass $\MX$ in App.~\ref{app:Mass}.%
\footnote{%
Although the results of the perturbative expansion would not 
be very sensitive to the coefficient of the harmonic potential, 
or equivalently, the mass of $X^I$ in these approximation methods, 
it will be more sensitive if only lower order terms are considered. 
We need to choose a suitable value since the expansion is truncated at lower order, in this paper. 
}

It would be convenient to introduce the creation and annihilation operators to describe 
eigenstates for quantum mechanics with harmonic potential. 
For adjoint field $X^I$, they are given by 
\begin{align}
 \AX^I &= \sqrt{ \frac{\MX}{2}}X^I + i \frac{1}{\sqrt{2\MX}} \PX^I \ , 
\\
 \AX^{\dag I} &= \sqrt{ \frac{\MX}{2}} X^I - i \frac{1}{\sqrt{2\MX}} \PX^I \ , 
\end{align}
and those for $w$ and $\bar w$ are 
\begin{align}
 \aw &= \frac{\sqrt{\mw}}{2} \bar w + i \frac{1}{\sqrt{\mw}} \pw \ , 
\label{aw}
\\ 
 \aw^\dag &= \frac{\sqrt{\mw}}{2} w - i \frac{1}{\sqrt{\mw}} \bar\pw \ , 
\\ 
 \bar\aw &= \frac{\sqrt{\mw}}{2} w + i \frac{1}{\sqrt{\mw}} \bar\pw \ , 
\\ 
 \bar\aw^\dag &= \frac{\sqrt{\mw}}{2} \bar w - i \frac{1}{\sqrt{\mw}} \pw \ . 
\label{bawd}
\end{align}
In terms of the creation and annihilation operators, 
the Hamiltonian $H_0$ is expressed as 
\begin{equation}
 H_0 = \MX \AX^{\dag I} \AX^I 
 + \mw \aw^{\dag\,a}_{\dot \alpha i} \aw^{\dot \alpha i}_a 
 + \mw \bar \aw^{\dag\dot \alpha i}_{a} \bar \aw_{\dot \alpha i}^a \ , 
\label{H0a}
\end{equation}
up to a constant from the normal ordering. 

Now, we consider the ground state for the unperturbed Hamiltonian $H_0$. 
Without the perturbation $V$, the Hamiltonian $H_0$ 
is that for the harmonic oscillators of $X$, $w$ and $\bar w$. 
The ground state $|0\rangle$ for $H_0$ is defined by the conditions
\begin{align}
 \AX|0\rangle &= 0 \ , 
&
 \aw|0\rangle &= 0 \ , 
&
 \bar \aw|0\rangle &= 0 \ ,  
\end{align}
or equivalently, in terms of the wave function, 
\begin{equation}
 |0\rangle \sim \exp\left[-\tr(X^I)^2 - \bar w^{\dot \alpha i}_a  w_{\dot \alpha i}^a \right] \ . 
\end{equation}
However, the physical states must satisfy the constraint \eqref{Const}, 
and have the overall $U(1)$ charge of $N_c A$, 
while $|0\rangle$ does not carry any charge. 
The physical ground state is constructed from 
the unconstrained ground state $|0\rangle$ 
by introducing an appropriate number of the excitations of $X$, $w$ and $\bar w$. 
As $X$, $w$ and $\bar w$ has $U(1)$ charge of $0$, $1$ and $-1$, respectively, 
the excitations in the physical states must satisfy 
\begin{equation}
 N_w - N_{\bar w} = N_c A \ , 
\label{NPhys}
\end{equation}
where $N_w$ and $N_{\bar w}$ are the numbers of the excitations of $w$ and $\bar w$, respectively. 
The physical ground state is the lowest energy state 
which satisfies the condition \eqref{Const}, 
and hence, has the smallest number of the excitations under the condition \eqref{NPhys}. 
Therefore, the physical ground state satisfies 
\begin{align}
 N_w &= N_c A \ , 
& 
 N_{\bar w} &= 0 \ .  
\label{NGS}
\end{align}

The constraint \eqref{Const} also implies that 
the physical state must be singlet in the $SU(A)$ symmetry. 
Naively, a singlet state which satisfies \eqref{NGS} 
is given by
\begin{equation}
 \left[
  \epsilon_{a_1\cdots a_A} \aw_{\dot \alpha_1 i_1}^{\dag\ a_1} 
  \cdots \aw_{\dot \alpha_A i_A}^{\dag\ a_A} 
 \right]^{N_c} |0\rangle \ . 
 \label{naivesinglet}
\end{equation}
Note here that the total number of 
distinct $w^{\dot \alpha i}$ with $\dot \alpha = 1,2$ and $i = 1,\cdots,N_f$ 
is $2N_f$ for $N_f$ flavors. 
So, in the case of $A > 2 N_f$, it is impossible to form anti-symmetric combinations of $\aw$,
and resultantly, we have to introduce additional operators in the fundamental representation of $SU(A)$ 
by using the creation operators of $X^I$ as 
\begin{equation}
 \left(\OX_n(\AX^\dag)\right)^a{}_b\,\aw^{\dag\,b}_{\dot \alpha i} \ , 
\label{Of}
\end{equation}
where the subscript $n$ labels different combinations of $\AX^\dag$, which are, for example, 
\begin{equation}
 \left(\OX_n(\AX^\dag)\right)^a{}_b 
 = 
 \delta^a{}_b \ , 
 \quad (\AX^{\dag I})^a{}_b 
 \ , 
 \quad (\AX^{\dag I}\AX^{\dag J})^a{}_b  
 \ , 
 \quad \cdots \ , 
 \quad \text{etc.}
\label{OX}
\end{equation}
Then, singlet operators of $SU(A)$ are constructed 
from the creation operators $\AX^\dag$ and $\aw^\dag$ as 
\begin{equation}
 \OS_{\{n\},\{\dot\alpha\},\{i\}} = 
 \epsilon_{a_1\cdots a_A} \left(\OX_{n_1}(\AX^\dag)\aw^{\dag}_{ \dot \alpha_1 i_1}\right)^{a_1} 
 \cdots \left(\OX_{n_A}(\AX^\dag)\aw^{\dag}_{ \dot \alpha_A i_A}\right)^{a_A}  \ , 
\label{OS}
\end{equation}
which has the overall baryon $U(1)$ charge $Q_{U(1)} = A$. 
The physical ground state $|\psi_0\rangle$ can be obtained by 
acting $N_c$ singlet operators $\OS_{\{n\},\{\dot\alpha\},\{i\}}$ 
to the unconstrained ground state $|0\rangle$, 
\begin{align}
 |\psi_0\rangle 
 &= 
 \prod_{r=1}^{N_c} \OS_{\{n\}_r,\{\dot\alpha\}_r,\{i\}_r} 
 |0\rangle \ .  
\label{psi0}
\end{align}
Here, the singlet operators $\OS_{\{n\},\{\dot\alpha\},\{i\}}$ must have the smallest number of 
the creation operators $\AX^\dag$ for the physical ground state, 
and given by, for example, 
\begin{align}
 \OS_{\{n\},\{\dot\alpha\},\{i\}} &= \epsilon_{a_1\cdots a_A} (\aw^{\dag}_{\dot \alpha=1, i=1})^{a_1} (\aw^{\dag}_{\dot \alpha=2, i=1})^{a_2} 
  (\aw^{\dag}_{\dot \alpha=1, i=2})^{a_3} (\aw^{\dag}_{\dot \alpha=2, i=2})^{a_4} 
 \cdots (\aw^{\dag}_{\dot \alpha=2,i={N_f}})^{a_{2N_f}} 
\notag\\&\quad\times
 (\AX^{\dag I=1} \aw^{\dag}_{\dot \alpha=1, i=1})^{a_{2N_f+1}} 
 (\AX^{\dag I=1} \aw^{\dag}_{\dot \alpha=2, i=1})^{a_{2N_f+2}} 
 \cdots (\AX^{\dag I=1} \aw^{\dag}_{ \dot \alpha=2,i={N_f}})^{a_{4N_f}} 
\notag\\&\quad\times
  (\AX^{\dag I=2} \aw^{\dag}_{\dot \alpha=1, i=1})^{a_{4N_f+1}} 
  (\AX^{\dag I=2} \aw^{\dag}_{\dot \alpha=2, i=1})^{a_{4N_f+2}} 
 \cdots (\AX^{\dag I=2} \aw^{\dag}_{ \dot \alpha=2,i={N_f}})^{a_{6N_f}} 
\notag\\&\quad\times
 \cdots (\AX^{\dag I}\cdots \AX^{\dag J}\aw^{\dag}_{ \dot \alpha_A i_A})^{a_A}  \ .  
\end{align}
In general, $N_c$ singlet operators $\OS_{\{n\},\{\dot \alpha\},\{i\}}$ 
can be different from each other. 
The operator $\OX_n$ can also contain the combination of $w$ and $\bar w$ 
such as $\left(\bar w^{\dot \alpha i} w_{\dot \alpha i} \right)_a{}^b$, 
but then it will have higher energy than those only with $X^I$ if 
the effective mass of $X^I$ is much smaller than the mass of $w$ and $\bar w$.   

Although, we obtained the physical ground state for arbitrary $A$, 
by introducing the creation operators of $X^I$ for $A > 2 N_f$, 
in this paper, we will focus on the case of $A \leq 2 N_f$, in which 
the physical ground state contains only the excitations of $w$. 
In this case, the physical ground state, at the 0-th order of the perturbative expansion, 
\eqref{psi0} with \eqref{OX} and \eqref{OS}, is simply expressed as 
\begin{align}
 |\psi_0\rangle 
 &= 
 \prod_{r=1}^{N_c} \OS_{\{\dot\alpha\}_r,\{i\}_r} 
 |0\rangle \ , 
\label{psi0w}
\\
 \OS_{\{\dot \alpha\},\{i\}} &= 
  \epsilon_{a_1\cdots a_A} \aw_{\dot \alpha_1 i_1}^{\dag\ a_1} 
  \cdots \aw_{\dot \alpha_A i_A}^{\dag\ a_A} \ , 
\label{OSw}
\end{align}
and diagonalizes $H_0$ as 
\begin{align}
 H_0 |\psi_0\rangle &= E_0 |\psi_0 \rangle \ , 
\end{align}
where the energy at the leading order of the perturbative expansion $E_0$ 
is simply given in terms of the number of the excitation $N_w = N_c A$ and the mass of $w$ as 
\begin{align}
 E_0 &= N_w \mw = N_c A \mw \ . 
\label{E0w}
\end{align}

Here, we make comments on the physical interpretation of the excitations in the state. 
The excitations $w$ represent quarks as it is from 
the string connecting the baryon D4-branes and the flavor D8-branes. 
Charges carried by $w$ can be identified with those of quarks. 
Although $w$ does not have color, it is consistent to the fact 
that the color is hidden by confinement. 
The constraint $N_w = N_c A$ is also consistent to confinement, 
since there are $N_c$ excitations of $w$ for each D4-brane, 
as it has been already discussed for D4-D8 strings. 
Although the mass of quarks cannot directly be identified with that of $w$, 
the mass of the baryon will be identified with sum of 
the mass of $w$, mass of $X^I$ and the tension of the baryon D4-brane. 
Thus, the mass of $w$ is related to the constituent quark mass even though they are not identical. 
The excitations of $X^I$ in \eqref{Of} are interpreted as 
motion of the quark which corresponds to the excitation of $w$ in the same \eqref{Of}, 
and the mass would be related to the energy of the motion. 
It would also be consistent with confinement that 
each excitation of $w$ does not have different operators of $X^I$, 
but $X^I$ comes from those of $A$ baryon D4-branes. 

Let us also point out that our physical ground state $|\psi_0\rangle$ 
is consistent with the fact that baryons are fermions.
The state is given as an antisymmetric combination
of fundamental representation of $U(A)$ to form a singlet of $SU(A)$. 
The indices of the fundamental representation of $U(A)$ of the $w$ field labels 
$A$ baryons, but the anti-symmetricity of the indices 
represents the anti-symmetric properties of the quarks. 
Since the state $|\psi_0\rangle$ has $N_c$ copies of these anti-symmetric combinations, 
it is symmetric under the exchange of two baryons if $N_c$ is even, 
and anti-symmetric if $N_c$ is odd. 
So we have a consistent statistics for the baryons under
their exchange.

In this section, we have considered the physical ground state 
for the Hamiltonian $H_0$. 
This contains only the effects of the mean field potential 
at the leading order of the perturbative expansion. 
We will discuss the effects of the perturbation $V$ in the next section.

%%%%%%%%%%%%%%%%%%%%%%%%%%%%%%%%%%%%%%%%%%%%%%%%%%%%%%%%%%%%%%%%%%%%%%%%%%%%%%%%
%%%%%%%%%%%%%%%%%%%%%%%%%%%%%%%%%%%%%%%%%%%%%%%%%%%%%%%%%%%%%%%%%%%%%%%%%%%%%%%%
%%%%%%%%%%%%%%%%%%%%%%%%%%%%%%%%%%%%%%%%%%%%%%%%%%%%%%%%%%%%%%%%%%%%%%%%%%%%%%%%

\section{Quantized states for small baryon numbers}
\label{sec:Linear}

In this section, we study the linear order corrections in the perturbative expansion. 
We focus on the case of $A\leq 2N_f$ with $N_f=2$, 
and the physical ground state $|\psi_0\rangle$, and 
its eigenvalue of Hamiltonian $E_0$ are given by 
\eqref{psi0w} with \eqref{OSw}, and \eqref{E0w}, respectively. 
The linear order correction for the energy of the ground state is simply given by 
taking the expectation value of the Hamiltonian for $|\psi_0\rangle$; 
\begin{align}
 E = \langle\psi_0|H|\psi_0\rangle = E_0 + \langle \psi_0|V|\psi_0\rangle  \ . 
\end{align}
Here, however, the ground state $|\psi_0\rangle$ is not unique 
but all states which take the form of \eqref{psi0w} and \eqref{OSw} 
are degenerated at the 0-th order of the perturbative expansion 
and have the same energy \eqref{E0w}. 
In order to find the lowest energy state at the linear order of the perturbative expansion, 
we have to diagonalize the perturbation $V$ in the space of the 0-th order ground states 
\eqref{psi0w} with \eqref{OSw}.

%%%%%%%%%%%%%%%%%%%%%%%%%%%%%%%%%%%%%%%%%%%%%%%%%%%%%%%%%%%%%%%%%%%%%%%%%%%%%%%%
%%%%%%%%%%%%%%%%%%%%%%%%%%%%%%%%%%%%%%%%%%%%%%%%%%%%%%%%%%%%%%%%%%%%%%%%%%%%%%%%

\subsection{4-point interaction of $w$}\label{ssec:4pt-w}

Now, we diagonalize $V$ with respect to the 0-th order ground states for $A\leq 2N_f$. 
Since the states have the excitations of $w$ only, 
terms which contain $X^I$ are irrelevant 
and we only need to consider the last term in \eqref{V}. 
In terms of the creation and annihilation operators of $w$, 
it can be expressed as% 
\footnote{%
	The interaction term \eqref{V} also contain $\awb$ and $\awb^\dag$, 
	whose contraction gives correction to the mass term of $w$, 
	which can be absorbed by the redefinition of $\mw$. 
} 
\begin{equation}
 V =  \frac{\cc}{\mw^2} 
 \left[\alpha^\dag{}^a_{\dot\alpha i} (\tau^I)^{\dot\alpha}{}_{\dot\beta} \alpha^{\dot\beta i}_b\right]
 \left[\alpha^\dag{}^b_{\dot\gamma j} (\tau^I)^{\dot\gamma}{}_{\dot\delta} \alpha^{\dot\delta j}_a\right]
 \ . 
\label{w-4pt}
\end{equation}
For $A=1$, since the baryon $U(A)$ gauge indices are trivial, 
this interaction term reduces to the quadratic Casimir operator of spin $SU(2)$.
For $A\geq 2$, we need to calculate \eqref{w-4pt}  further to obtain a simpler expression. 
In general, generators $t^A$ in the fundamental representation of $U(N)$ satisfy the following formula;  
\begin{equation}
 \sum_A (t^A)_a{}^b \, (t^A)_c{}^d = \frac{1}{2} \delta_a{}^d \,\delta_c{}^b \ , 
 \label{U(N)}
\end{equation}
or equivalently, 
\begin{equation}
 \sum_A (t^A)_a{}^b \, (t^A)_c{}^d 
 = 
 \frac{1}{2} \left(\delta_a{}^d \,\delta_c{}^b - \frac{1}{N} \delta_a{}^b \,\delta_c{}^d \right) \ ,
 \label{SU(N)}
\end{equation}
for $SU(N)$. 
We first apply the formula \eqref{SU(N)} for the $SU(2)$ spin generators to obtain %, which are the half of the Pauli matrices $\tau^I$, and obtain 
\begin{equation}
 (\aw^{\dag\,a}_{{\dot\alpha} i} (\tau^I)^{\dot\alpha}{}_{\dot\beta} \aw^{{\dot\beta} i}_b) 
 (\aw^{\dag\,b}_{\dot\gamma j} (\tau^I)^{\dot\gamma}{}_{\dot\delta} \aw^{\dot\delta j}_a)
 = 
 2 (\aw^{\dag\,a}_{{\dot\alpha} i} \aw^{{\dot\beta} i}_b) 
 (\aw^{\dag\,b}_{{\dot\beta} j} \aw^{{\dot\alpha} j}_a)
 - 
 (\aw^{\dag\,a}_{{\dot\alpha} i} \aw^{{\dot\alpha} i}_b) 
 (\aw^{\dag\,b}_{{\dot\beta} j} \aw^{{\dot\beta} j}_a) \ . 
\end{equation}
Next, we use \eqref{SU(N)} for the flavor $SU(N_f)$ symmetry to the first term and 
\eqref{U(N)} for the baryon $U(A)$ to the second term. 
Then, we have the following expression;  
\begin{equation}
 2 (\aw^{\dag\,a}_{{\dot\alpha} i} (t_{U(N_f)}^M)^i{}_j \aw^{{\dot\alpha} j}_a) 
 (\aw^{\dag\,b}_{{\dot\beta} k} (t_{U(N_f)}^M)^k{}_l \aw^{{\dot\beta} l}_b)
 - 
 (\aw^{\dag\,a}_{{\dot\alpha} i} (t^A)_a{}^b \aw^{{\dot\alpha} i}_b) 
 (\aw^{\dag\,c}_{{\dot\beta} j} (t^A)_c{}^d \aw^{{\dot\beta} j}_d)   \ , 
\label{w-4-U(N_f)}
\end{equation}
where $t_{U(N_f)}^M$ is the generator of $U(N_f)$ which is obtained 
by adding the overall $U(1)$ to the flavor $SU(N_f)$ symmetry. 
The second term is the quadratic Casimir of $U(A)$, 
which gives only the square of the overall $U(1)$ charge 
since the physical states must be singlet of $SU(A)$ due to the constraint.   
Then, the interaction term is rewritten in the following form; 
\begin{equation}
 V = \frac{4\cc}{\mw^2} (\alpha^\dag t_f^M \alpha)(\alpha^\dag t_f^M \alpha) 
 + \frac{\cc}{\mw^2} \frac{2A-N_f}{N_fA} (\alpha^\dag \alpha)^2 \ ,  
\label{VFCasimir}
\end{equation}
where $t_f^M$ is the generator of the flavor $SU(N_f)$. 
The operator $\alpha^\dag t_f^M \alpha$ plays the role of 
the generator of the flavor $SU(N_f)$ symmetry 
for the excitation of $w$. 
Since the physical ground state $|\psi_0\rangle$ has no excitation of $\bar w$, 
it can simply be treated as the $SU(N_f)$ generator for $|\psi_0\rangle$, 
and $V$ is nothing but the quadratic Casimir operator 
of the flavor $SU(N_f)$ for $|\psi_0\rangle$. 
For $N_f=2$, the expectation value of $V$ gives 
the square of the isospin of the state $|\psi_0\rangle$. 
This implies that the ground state with the linear order correction of 
the perturbative expansion must have the minimum isospin 
in the ground state at the 0-th order. 

Although we focus on the case of $A\leq 2N_f$, without excitations of $X^I$, here, 
the 4 point interaction term of $w$, \eqref{w-4pt} will give a similar effect for $A > 2N_f$. 
Then, nuclei with smaller isospin have smaller energy, and hence, 
those with the same numbers of protons and neutrons will be 
%most stable. 
the lowest energy states.
This is consistent with nuclei in real. 
The interaction term \eqref{w-4pt} gives a similar effect to 
the symmetry term in the Bethe-Weizs\"acker mass formula 
though the relation is not very clear.

%%%%%%%%%%%%%%%%%%%%%%%%%%%%%%%%%%%%%%%%%%%%%%%%%%%%%%%%%%%%%%%%%%%%%%%%%%%%%%%%
%%%%%%%%%%%%%%%%%%%%%%%%%%%%%%%%%%%%%%%%%%%%%%%%%%%%%%%%%%%%%%%%%%%%%%%%%%%%%%%%

\subsection{Allowed states}

In the previous section, 
we have shown that the lowest energy state 
at the linear order of the perturbative expansion 
has the minimum isospin of the ground state at the 0-th order. 
In this section, we consider 
a few examples for $N_f=2$, and 
see the spin and isospin of the lowest energy %most stable 
states. 
Although in general it is difficult to study the stability, lowest energy states are expected to be
stable, so we focus on the lowest energy states.

%%%%%%%%%%%%%%%%%%%%%%%%%%%%%%%%%%%%%%%%%%%%%%%%%%%%%%%%%%%%%%%%%%%%%%%%%%%%%%%%

\subsubsection{$A=1$ and $N_c=3$}\label{sssec:nucleon}

In this case, the baryon symmetry is $U(1)$ 
and there are no constraint for the baryon $SU(A)$ symmetry. 
The creation operators of $w$ do not need to form 
the singlet operator $\mathcal S$, or equivalently, 
the creation operator of $w$ can be treated as the singlet operator; 
\begin{equation}
 \mathcal S_{\dot \alpha i} = \aw^\dag_{\dot \alpha i} \ . 
\end{equation}
Here, the index of the baryon symmetry is omitted, since it is $U(1)$. 
The physical ground state is obtained by 
multiplying $N_c$ creation operators of $w$ to $|0\rangle$. 
For $N_c=3$, it is expressed as 
\begin{align}
 |\psi_0\rangle 
 = 
 \aw_{\dot \alpha_1 i_1}^{\dag} 
 \aw_{\dot \alpha_2 i_2}^{\dag}
 \aw_{\dot \alpha_3 i_3}^{\dag} |0\rangle \ . 
\end{align}
Here, the creation operators $\aw^\dag$ must be symmetric 
under the exchange of any pair of three $\aw^\dag$ above. 
The spin and isospin must have the same symmetry under the exchange, 
and hence, must be the same. 
The allowed states are 
\begin{align}
 (J,I) = \left(\tfrac{1}{2},\tfrac{1}{2}\right) \ , \quad  \left(\tfrac{3}{2},\tfrac{3}{2}\right) \ , 
\end{align}
which can be identified as nucleon and $\Delta$, respectively. 
The lowest energy state is nucleon, which has smaller isospin, 
\begin{align}
 (J,I) = \left(\tfrac{1}{2},\tfrac{1}{2}\right) \ . 
\end{align}

Here, we estimate the value of the first order perturbation $\langle V \rangle$ 
from the mass of nucleon, 939 MeV and that of $\Delta$, 1232 MeV. 
The difference comes from that of the first order perturbation 
\begin{equation}
 \langle V \rangle = \frac{4\cc}{\mw^2} I (I+1)
\end{equation}
for $I=\frac{1}{2}$ (nucleon) and $I=\frac{3}{2}$ ($\Delta$). 
From this condition, we obtain 
\begin{align}
 E_0 &= 866 \mathrm{[MeV]} \ , 
 &
 \langle V \rangle_{\mathrm N} &= 73 \mathrm{[MeV]} \ , 
 &
 \langle V \rangle_{\Delta} &= 366 \mathrm{[MeV]} \ . 
\end{align}
The first order perturbation $\langle V \rangle$ is smaller than the
0-th order mass $E_0$, and, hence, the perturbative expansion would be valid 
for these states. 

%%%%%%%%%%%%%%%%%%%%%%%%%%%%%%%%%%%%%%%%%%%%%%%%%%%%%%%%%%%%%%%%%%%%%%%%%%%%%%%%

\subsubsection{$A=2$ and $N_c=1$}

Since the physical states for $A=2$ are more complicated than $A=1$, 
it is useful to start with the case of $N_c=1$. 
In this case, the 0-th order ground state $|\psi_0\rangle$ has 
only one antisymmetric combination and contains the two excitations of $w$; 
\begin{align}
 |\psi_0\rangle 
 &= 
 \epsilon_{a_1 a_2} \aw_{\dot \alpha_1 i_1}^{\dag\ a_1} \aw_{\dot \alpha_2 i_2}^{\dag\ a_2}
 |0\rangle \ . 
\end{align}
In order to form an antisymmetric combination by the spin and flavor indices, 
one of them can be antisymmetric and the other must be symmetric. 
Therefore, the allowed states are those with 
\begin{align}
 (J,I) = (1,0) \ , \quad (0,1) \ . 
\end{align}
In these states, $I=0$ is the minimum of isospin, and the lowest energy state has 
\begin{align}
 (J,I) = (1,0) \ . 
\end{align}

%%%%%%%%%%%%%%%%%%%%%%%%%%%%%%%%%%%%%%%%%%%%%%%%%%%%%%%%%%%%%%%%%%%%%%%%%%%%%%%%

\subsubsection{$A=2$ and $N_c=3$}\label{sssec:A=2,Nc=3}

In this case, allowed states are given by 
symmetric combinations of those of $A=2$ and $N_c=1$. 
Then, the allowed states are 
\begin{align}
 (J,I) &= 
 (1,0) \ , \quad
 (3,0) \ , \quad
 (1,2) \ , \quad
 (0,1) \ , \quad
 (0,3) \ , \quad
 (2,1) \ . 
\end{align}
In these states, $I=0$ is the minimum of isospin. 
The %most stable 
lowest energy 
states are 
\begin{align}
 (J,I) &= 
 (1,0) \ , \quad
 (3,0) \ . 
\end{align}
The state with $(J,I) = (1,0)$ corresponds to the deuteron, and 
the other state is another dibaryon state. 
At the linear order of the perturbative expansion in this model, 
there is no difference of the energy in these two states. 
We will discuss more on the dibaryon states in Sec.~\ref{sec:Dibaryon}.

%%%%%%%%%%%%%%%%%%%%%%%%%%%%%%%%%%%%%%%%%%%%%%%%%%%%%%%%%%%%%%%%%%%%%%%%%%%%%%%%

\subsubsection{$A=3$ and $N_c=1$}  

In this case, only one singlet operator $\OS_{\{\dot \alpha\}.\{i\}}$ is acting on $|0\rangle$, 
which contains three creation operators of $w$ in the ground state at the 0-th order $|\psi_0\rangle$. 
It is expressed as 
\begin{align}
 |\psi_0\rangle 
 &= 
 \OS_{\dot\alpha_1\dot\alpha_2\dot\alpha_3,i_1 i_2 i_3}  
 |0\rangle \ , 
\\
 \OS_{\dot\alpha_1\dot\alpha_2\dot\alpha_3,i_1 i_2 i_3}  
 &= 
 \epsilon_{a_1 a_2 a_3} \aw_{\dot \alpha_1 i_1}^{\dag\ a_1} 
 \aw_{\dot \alpha_2 i_2}^{\dag\ a_2}
 \aw_{\dot \alpha_3 i_3}^{\dag\ a_3} \ . 
\end{align}
Here, the creation operator $\aw^\dag$ must be symmetric under 
the exchange of any pair of three $\aw^\dag$, since they are bosonic operators. 
As indices $a$ of $U(A)$ are totally antisymmetric, 
the other indices of spin $\dot \alpha$ and flavor $i$ must form another 
totally antisymmetric combination. 
However, neither spin and isospin can solely form an antisymmetric combination of three states, 
and hence, both spin and isospin must have an antisymmetric pair. 
The total spin $J$ and isospin $I$ of the ground state $|\psi_0\rangle$ 
must be 
\begin{align}
  (J,I) = \left(\tfrac{1}{2},\tfrac{1}{2}\right) \ . 
\end{align}

%%%%%%%%%%%%%%%%%%%%%%%%%%%%%%%%%%%%%%%%%%%%%%%%%%%%%%%%%%%%%%%%%%%%%%%%%%%%%%%%

\subsubsection{$A=3$ and $N_c=3$}  

In this case, the 0-th order ground state $|\psi_0\rangle$ has 
the same singlet operator $\OS_{\{\dot \alpha\}.\{i\}}$ 
to that in the case of $A=3$ and $N_c=1$ 
but three $\OS_{\{\dot \alpha\}.\{i\}}$ are acting on $|0\rangle$, 
which is written as 
\begin{align}
 |\psi_0\rangle 
 &= 
 \OS_{\dot\alpha_1\dot\alpha_2\dot\alpha_3,i_1 i_2 i_3}  
 \OS_{\dot\alpha_4\dot\alpha_5\dot\alpha_6,i_4 i_5 i_6}  
 \OS_{\dot\alpha_7\dot\alpha_8\dot\alpha_9,i_7 i_8 i_9}  
 |0\rangle \ , 
\\
 \OS_{\dot\alpha_1\dot\alpha_2\dot\alpha_3,i_1 i_2 i_3}  
 &= 
 \epsilon_{a_1 a_2 a_3} \aw_{\dot \alpha_1 i_1}^{\dag\ a_1} 
 \aw_{\dot \alpha_2 i_2}^{\dag\ a_2}
 \aw_{\dot \alpha_3 i_3}^{\dag\ a_3} \ , 
\end{align}
and similarly for the other two operators of $\OS_{\{\dot \alpha\}.\{i\}}$. 
Each of $\OS_{\{\dot \alpha\}.\{i\}}$ in the expression above 
gives the spin and isospin with the same norm as those in $A=3$ and $N_c=1$, 
but their direction is not necessary to be the same. 
As the operator $\OS_{\{\dot \alpha\}.\{i\}}$ is bosonic 
and three of them in the above are indistinguishable, 
they must form a symmetric combination. 
Then, allowed states can be obtained by 
the symmetric combinations of the three physical state for $A=3$ and $N_c=1$, 
or equivalently, $I=\frac12$ and $J=\frac12$. 
Then, the allowed states have the spin and isopsin 
\begin{align}
 (J,I) = \left(\tfrac{1}{2},\tfrac{1}{2}\right) \ , \quad  \left(\tfrac{3}{2},\tfrac{3}{2}\right) \ . 
\end{align}
In these states, that with $I=\frac{1}{2}$ has a smaller isospin. 
Therefore, the %most stable 
lowest energy 
state has 
\begin{align}
 (J,I) = \left(\tfrac{1}{2},\tfrac{1}{2}\right) \ . 
\end{align}
This is  ${}^3$H and ${}^3$He. 

%%%%%%%%%%%%%%%%%%%%%%%%%%%%%%%%%%%%%%%%%%%%%%%%%%%%%%%%%%%%%%%%%%%%%%%%%%%%%%%%

\subsubsection{$A=4$}  

In this case, the singlet operator has the maximum number 
of the creation operators of $w$, to form singlet without excitations of $X^I$. 
The singlet operator must contain the creation operators of $w$ with 
all varieties of the spin and flavor indices. 
The spin and flavor indices must form a totally antisymmetric combination 
and $\mathcal S$ is the singlet of both the rotation and flavor symmetries. 
The physical ground state is given by 
\begin{align}
 |\psi_0\rangle 
 &= 
 \OS^{N_c} 
 |0\rangle \ , 
\\
 \OS 
 &= 
 \epsilon_{a_1 a_2 a_3 a_4} 
 \epsilon^{\dot \alpha_1\dot \alpha_2} \epsilon^{\dot \alpha_3\dot \alpha_4}
 \epsilon^{i_1 i_3} \epsilon^{i_2 i_4} 
 \aw_{\dot \alpha_1 i_1}^{\dag\ a_1} 
 \aw_{\dot \alpha_2 i_2}^{\dag\ a_2}
 \aw_{\dot \alpha_3 i_3}^{\dag\ a_3} 
 \aw_{\dot \alpha_4 i_4}^{\dag\ a_4} \ , 
\end{align}
and has the spin and isospin 
\begin{align}
 (J,I) = \left(0,0\right) \ . 
\end{align}
This is nothing but ${}^4$He.

%%%%%%%%%%%%%%%%%%%%%%%%%%%%%%%%%%%%%%%%%%%%%%%%%%%%%%%%%%%%%%%%%%%%%%%%%%%%%%%%

\subsubsection{ Relation to \cite{Hashimoto:2010je} and large-$N_c$}  
\label{subsec-largeNc}

Here, we consider the case of large-$N_c$ and discuss the relation to 
results in \cite{Hashimoto:2010je}.
We first review the analyses in \cite{Hashimoto:2010je}. 
In \cite{Hashimoto:2010je}, static configurations 
in the cases of single baryon are studied. 
For the static configurations for $A=1$, 
the conjugate momentum of $w$ is given by 
\begin{align}
 \pw^{\dot \alpha i}
 &= 
 \frac{1}{2} \left(\partial_t \bar w^{\dot \alpha i} + i A_t \bar w^{\dot \alpha i}\right) 
 = \frac{i}{2} A_t \bar w^{\dot \alpha i} \ , 
\\
 \bar \pw_{\dot \alpha i} 
 &= 
 \frac{1}{2} \left(\partial_t w_{\dot \alpha i} - i A_t w_{\dot \alpha i}\right) 
 = - \frac{i}{2} A_t w_{\dot \alpha i} \ , 
\end{align}
and the constraint \eqref{Const} is expressed as 
\begin{equation}
 A_t \left(\bar w^{\dot \alpha i} w_{\dot \alpha i}\right) + N_c = 0  \ . 
\end{equation}
Then, the Hamiltonian is expressed as 
\begin{align}
 H 
 = 
 \frac{1}{2} \frac{N_c^2}{\left(\bar w^{\dot \alpha i} w_{\dot \alpha i}\right)} 
 + \frac{1}{2} \mw^2 \left(\bar w^{\dot \alpha i} w_{\dot \alpha i}\right) 
 + \cc \left(\bar w^{\dot \alpha i} (\tau^I)_{\dot \alpha}{}^{\dot\beta} w_{\dot \beta i}\right)^2 \ . 
\label{staticH}
\end{align}
By using the perturbative expansion with respect to $\lambda$, 
the lowest energy configuration is given by 
\begin{equation}
 \left(\bar w^{\dot \alpha i} w_{\dot \alpha i}\right) = \frac{N_c}{\mw} \ , 
\label{staticSol}
\end{equation}
and the energy at the leading order of the perturbative expansion is given by 
\begin{align}
 H_0 
 &= 
 \frac{1}{2} \frac{N_c^2}{\left(\bar w^{\dot \alpha i} w_{\dot \alpha i}\right)} 
 + \frac{1}{2} \mw^2 \left(\bar w^{\dot \alpha i} w_{\dot \alpha i}\right) 
 = 
 N_c M \ . 
\label{staticE}
\end{align}
This result agrees with our result \eqref{E0w}. 

For $N_f = 1$, however, the Hamiltonian \eqref{staticH} can be rewritten as 
\begin{align}
 H 
 = 
 \frac{1}{2} \frac{N_c^2}{\left(\bar w^{\dot \alpha} w_{\dot \alpha}\right)} 
 + \frac{1}{2} \mw^2 \left(\bar w^{\dot \alpha} w_{\dot \alpha}\right) 
 + \cc \left(\bar w^{\dot \alpha} w_{\dot \alpha}\right)^2 \ , 
\label{staticH1f}
\end{align}
and then, the interaction term becomes 
%larger than the 0-th order Hamiltonian 
%for large-$N_c$; 
\begin{equation}
 V = \cc \left(\bar w^{\dot \alpha} w_{\dot \alpha}\right)^2 
 = \frac{\cc}{\mw^2} N_c^2 \propto N_c M \lambda_\mathrm{QCD} \ . 
\end{equation}
Recall that the Sakai-Sugimoto model is justified for the large-$N_c$ limit with $\lambda_\mathrm{QCD}$ kept fixed as a large value. 
Thus the interaction term is larger than the leading term \eqref{staticE} by the factor of $\lambda_\mathrm{QCD}$, 
and the perturbative expansion breaks down. 
In this case, as is discussed in \cite{Hashimoto:2010je}, 
the first term and third term in \eqref{staticH1f} must balance 
at the leading order approximation, 
and give 
\begin{align}
 & \left(\bar w^{\dot \alpha} w_{\dot \alpha}\right) = 2^{2/3}N_c^{2/3} \cc^{-1/3} \ ,  
\\
 H &= 3 \cdot 2^{-4/3} N_c^{4/3} \cc^{1/3} \ . 
\end{align}

For $N_f=2$, the interaction term vanishes, for example, 
for the configuration 
\begin{align}
 w_{\dot \alpha i} 
 &= \rho\,\delta_{\dot\alpha i} \ , 
& 
 \bar w^{\dot \alpha i} 
 &= \rho\,\delta^{\dot\alpha i} \ , 
\end{align}
and then, the perturbative expansion is still valid. 
The solution and the energy are given by 
\eqref{staticSol} and \eqref{staticE}, respectively. 

Now, we consider the case of large-$N_c$ by using our procedure, 
and compare with the results of \cite{Hashimoto:2010je} above. 
For $N_f=1$, the physical ground state is given by 
\begin{align}
 |\psi_0\rangle 
 = \prod_{n=1}^{N_c} \aw^\dag_{\dot \alpha_n} |0\rangle \ . 
\end{align}
Since there is no $SU(N_f)$ flavor symmetry, 
the interaction term \eqref{VFCasimir} becomes%
\footnote{%
We ignore the contraction terms in the normal ordering, 
which give contribution of $\mathcal O(N_c)$. 
} 
\begin{equation}
 V
 = 
 \frac{\cc}{\mw^2} (\aw^\dag\aw) = \frac{\cc}{\mw^2} N_c^2  \ . 
\end{equation}
and hence, the perturbative expansion breaks down. 
This is consistent with the result in \cite{Hashimoto:2010je}. 
We do not pursue this case, here. 

For $N_f=2$, the physical ground state has both spin and flavor indices;  
\begin{align}
 |\psi_0\rangle 
 = \prod_{n=1}^{N_c} \aw^\dag_{\dot \alpha_n i_n} |0\rangle \ . 
\end{align}
Because of the statistics of the creation operators $\aw^\dag$, 
the spin and isospin of the state $|\psi_0\rangle$ must be the same, 
but can take the minimum of 
\begin{align}
 (J,I) 
 = 
\begin{cases}
 \left(\tfrac{1}{2},\tfrac{1}{2}\right) & N_c\text{ : odd}
\\
 \left(0,0\right) & N_c\text{ : even} 
 \label{State-Large-Nc}
\end{cases}
\ . 
\end{align}
As the second term in \eqref{VFCasimir} vanishes for $A=1$ and $N_f=2$, 
the interaction term does not become large even for large-$N_c$; 
\begin{equation}
 V = 
\begin{cases}
 \frac{3\cc}{\mw^2} & N_c\text{ : odd}
\\
 0 & N_c\text{ : even} 
 \label{V_Large-Nc}
\end{cases}
\ . 
\end{equation}
This is much smaller than the 0-th order energy \eqref{E0w}, which is of $\mathcal O(N_c)$. 
In the 't Hooft limit, $N_c\to\infty$ with $\lambda_\mathrm{QCD} \propto N_c \lambda$ fixed, 
the perturbation above is suppressed as $1/N_c$. 
The perturbative expansion is valid and 
our result is consistent with that in \cite{Hashimoto:2010je}, again. 
The expectation value of $w$ can also be calculated as 
\begin{equation}
 \left\langle\bar w^{\dot \alpha i} w_{\dot \alpha i}\right\rangle = \frac{N_c}{\mw} \ , 
\end{equation}
at the 0-th order of the perturbative expansion. 
This also agrees with \eqref{staticSol}. 
Note that with a proper normalization of the operator $w$, physically 
this quantity would correspond to the size of the nucleon \cite{Hashimoto:2010je}. 

For $N_f > 2$, the result is similar to that for $N_f=2$. 
For $A=1$, the perturbation $V$, \eqref{w-4pt} is exactly 
the quadratic Casimir operator of the spin $SU(2)$ symmetry. 
For $N_f\geq 2$, the state $|\psi_0\rangle$ can take the minimum spin of \textcolor{red}{$J=0$ or $\frac{1}{2}$}, 
%\eqref{State-Large-Nc}, 
and then, the expectation value of $V$ is given by the quadratic Casimir of the spin, 
and becomes the same to the case of $N_f=2$ as \eqref{V_Large-Nc}.%
\footnote{%
For $A=1$, the interaction term \eqref{w-4pt} becomes $V= \frac{4\lambda}{M^2} J(J+1) $ for spin $J$ states.
%as the quadratic Casimir of spin $SU(2)$, 
%we obtain a similar expression to \eqref{VFCasimir} for spin $SU(2)$ 
%without the second term, for arbitrary $N_f$. 
For $N_f \geq 2$, the ground state has spin $J=\frac{1}{2}$ for odd $N_c$ and $J=0$ for even $N_c$, 
and then, the perturbation $\langle V \rangle$ becomes the same to \eqref{V_Large-Nc}. 
In \eqref{VFCasimir}, this suppression can be understood 
as cancellation between the first and second term. 
For $N_f>2$, the states with spin $J=0$ or $\frac{1}{2}$ 
are in non-trivial representations of the flavor $SU(N_f)$, 
and the quadratic Casimir of flavor, or equivalently the first term in \eqref{VFCasimir}, 
will be very large and positive but canceled with the second term that is very large and negative. 
(The second term is always $O(\lambda_{QCD} M N_c)$.)
Such cancellation can be generalized to the $A \ge 2$ case, if  $N_f > 2A$ 
where the second term \eqref{VFCasimir} is negative.
Note that \eqref{w-4pt} comes from $(D^I)^2$ in \eqref{action} and is non-negative.
Thus, although the second term \eqref{VFCasimir} takes a large negative value at large-$N_c$ in the $N_f > 2A$ case, the first term must take a larger value. % so that \eqref{w-4pt} becomes non-negative.
Particularly, the state can take various representations of the flavor $SU(N_f)$ and the first term \eqref{VFCasimir} could be various values.
Thus $\langle V \rangle$ would be tuned to be small for the low energy states, and  the perturbation in large-$N_c$ would work.
Therefore the necessary condition for the validity of the perturbation in large-$N_c$ is $N_f \ge 2A $.
Note that this condition is for large-$N_c$, and it might be reasonable that the number of flavor must be larger than real.
%Note that this conditions are necessary only for validity of the perturbation in  
%
%
%In general, the state can take various representations of flavor $SU(N_f)$, 
%and the first and second term \eqref{VFCasimir} can be canceled if the second term is negative. 
%This would happen for the lowest energy state 
%because \eqref{w-4pt} comes from $(D^I)^2$ and is non-negative. 
%For $A \ge 2$, $N_f$ should also be larger in large-$N_c$ for the cancellation in large-$N_c$. 
%though $N_f$ does not need to be so large as $N_c$ but $N_f \geq 2 A$ would be sufficient. 
%Note that this conditions are necessary only for validity of the perturbation in large-$N_c$, 
%in which it might be reasonable that the number of flavor must be larger than real. 
%In our main analysis for $A \ge 2$, we take $N_c=3$ and simply ignore the $1/N_c$ corrections through the quantum gravity effects.
}
Thus, 
the perturbation is suppressed as $1/N_c$ in the 't Hooft limit and 
the perturbative expansion is valid for $N_f\geq 2$. 
The case of single flavor $N_f=1$ is the exception since 
the spin must be in the totally symmetric representation, 
and hence the expectation value of $V$ becomes 
larger than 0-th order terms in the 't Hooft limit with large $\lambda_\mathrm{QCD}$. 
Our perturbative analysis cannot be used for $N_f=1$, but 
is valid for realistic cases of $N_f \geq 2$. 
%of $\mathcal O(N_c^2)$, only for $N_f=1$. 

%%%%%%%%%%%%%%%%%%%%%%%%%%%%%%%%%%%%%%%%%%%%%%%%%%%%%%%%%%%%%%%%%%%%%%%%%%%%%%%%
%%%%%%%%%%%%%%%%%%%%%%%%%%%%%%%%%%%%%%%%%%%%%%%%%%%%%%%%%%%%%%%%%%%%%%%%%%%%%%%%
%%%%%%%%%%%%%%%%%%%%%%%%%%%%%%%%%%%%%%%%%%%%%%%%%%%%%%%%%%%%%%%%%%%%%%%%%%%%%%%%

\section{Hyperons}
\label{sec:Hyperon}

%%%%%%%%%%%%%%%%%%%%%%%%%%%%%%%%%%%%%%%%%%%%%%%%%%%%%%%%%%%%%%%%%%%%%%%%%%%%%%%%
%%%%%%%%%%%%%%%%%%%%%%%%%%%%%%%%%%%%%%%%%%%%%%%%%%%%%%%%%%%%%%%%%%%%%%%%%%%%%%%%

\subsection{Strange quark mass}

In the previous section, 
we have considered only two flavors of $u$ and $d$. 
The model can be generalized to arbitrary number of flavors 
by considering $SU(N_f)$ flavor symmetry. 
In this section, we introduce strangeness 
and consider $SU(3)$ flavor symmetry. 
The hypernuclei, which have strangeness, 
have higher mass compared with those without strangeness 
(in the same irreducible representation of $SU(3)$)
because of the mass difference of the strange quark to the other two flavors. 
In order to reproduce the mass spectrum of the hypernuclei, 
we need to take this effect into account.

As we have discussed in Sec.~\ref{sec:Model}, 
$w$ and $\bar w$ are related to the quarks and anti-quarks. 
The constituent mass of quarks may be related to the mass of $w$ and $\bar w$. 
The mass difference between the strange quark and the other two flavors 
would appear in our model as a similar mass difference for $w$ and $\bar w$. 
The current quark mass in the Sakai-Sugimoto model was given in
the open Wilson-loop approach \cite{Aharony:2008an,Hashimoto:2008sr}
and in the tachyon condensation approach \cite{Casero:2007ae,Bergman:2007pm,Dhar:2007bz}), 
and the three-flavor mass dependence of holographic 
baryons based on the Wilson-loop approach \cite{Hashimoto:2009hj} 
was studied in \cite{Hashimoto:2009st} (see also \cite{Bigazzi:2018cpg,Druks:2018hif} for related works). However, it is difficult to derive the mass difference in  
$w$ and $\bar w$ directly from those approaches.
Here, we assume that $w$ and $\bar w$ with strangeness 
will get a different mass as a consequence of these effects, 
and introduce the mass difference by modifying the model as 
\begin{equation}
 \sum_{i=1}^3 \mw^2 \bar w_a^{\dot \alpha i} w^a_{\dot \alpha i} 
 \to 
 \sum_{i=1}^2 \mw^2 \bar w_a^{\dot \alpha i} w^a_{\dot \alpha i} 
 + \ms^2 (\bar w_{(s)})_a^{\dot \alpha} (w_{(s)})^a_{\dot \alpha} \ , 
\end{equation}
where the strangeness is identified to the third component of 
the fundamental representation of the flavor $SU(3)$, $w_{(s)} = w_{i=3}$. 
The Hamiltonian at the leading order simply gives the expectation value 
\begin{equation}
 E_0 
 = 
 N_X \MX + N_w \mw + \left(\ms - \mw\right) N_S \ , 
 \label{MNs}
\end{equation}
where $N_S$ is the number of the excitations of $w$ with strangeness, $w_{(s)}$, 
and we have assumed that there are no excitations of $\bar w$.

%%%%%%%%%%%%%%%%%%%%%%%%%%%%%%%%%%%%%%%%%%%%%%%%%%%%%%%%%%%%%%%%%%%%%%%%%%%%%%%%
%%%%%%%%%%%%%%%%%%%%%%%%%%%%%%%%%%%%%%%%%%%%%%%%%%%%%%%%%%%%%%%%%%%%%%%%%%%%%%%%

\subsection{Evaluating 4-point interaction of $w$ for hyperon}
\label{ssec:hyperon-1st}

Now, we focus on nuclei with small numbers of baryons 
which satisfy $A\leq 2N_f = 6$, and evaluate the linear order perturbation.
Although we introduced the mass term 
which breaks the flavor $SU(3)$ symmetry, 
we assume that the interaction terms in \eqref{V} remain
the same and have 
no additional flavor symmetry breaking term. 
Although the interaction terms in \eqref{V} 
obeys the flavor symmetry, 
the effect of the flavor symmetry breaking mass term 
appears in the linear order corrections in the perturbative expansion. 

As we are considering the perturbation around 
the physical state without the excitations of $X$, 
only the 4-point interaction term of $w$ in \eqref{V} is relevant. 
As the physical state at the leading order approximation is 
expressed in terms of the harmonic oscillators of $w$, 
it is convenient to express the interaction term 
in terms of the creation and annihilation operators of $w$ as in the previous sections. 
However, the creation and annihilation operators \eqref{aw}-\eqref{bawd} depend on the mass, 
which should be replaced by $\ms$ for the creation and annihilation operators with strangeness. 
The same argument as in Sec.~\ref{ssec:4pt-w} can be applied for 
this case, but additional factors appear 
through the creation and annihilation operators with strangeness. 

It would be more convenient to repeat the same procedure as in Sec.~\ref{ssec:4pt-w} 
in terms of $w$ and $\bar w$. 
We obtain a similar expression to \eqref{VFCasimir} in terms of $w$ and $\bar w$; 
\begin{align}
 V 
 &= {4\cc} V_F + {\cc} \frac{2A-N_f}{N_fA} V_N \ , 
\label{VFwCasi}
\\
 V_F 
 &= 
 (w t_f^M \bar w)(w t_f^M \bar w)  \ , 
\label{V_F}
\\
 V_N 
 &= 
 (w \bar w)^2 \ . 
\label{V_N}
\end{align}
We first consider $V_F$ \eqref{V_F}, which gives the quadratic Casimir of $SU(3)$ 
in the absence of the flavor symmetry breaking. 
This term is separated into 
\begin{equation}
 (w t_f^M \bar w)(w t_f^M \bar w) 
 = \sum_{M=1}^3 (w t_f^M \bar w)(w t_f^M \bar w) 
 + \sum_{M=4}^7 (w t_f^M \bar w)(w t_f^M \bar w) 
 + (w t_f^8 \bar w)(w t_f^8 \bar w)  \ , 
\label{Vsepa}
\end{equation}
where we have chosen the generators of the flavor $SU(3)$ symmetry as 
\begin{equation}
 t_f^M = \frac{1}{2} \lambda^M \ , 
\end{equation}
and $\lambda^M$ is the Gell-Mann matrix, and the strangeness is identified 
to the third components (row and column) of the matrix. 
The Gell-Mann matrix $\lambda^M$ for $M=1,2,3$ are identified to 
the generators of the isospin $SU(2)$ subalgebra, 
they can be simply expressed in terms of the creation and annihilation operators as  
\begin{equation}
 \sum_{M=1}^3 (w t_f^M \bar w)(w t_f^M \bar w) 
 \sim \sum_{M=1}^3 \frac{1}{\mw^2} (\aw^\dag t_f^M \alpha)(\alpha^\dag t_f^M \aw) 
 \sim \frac{1}{\mw^2} I(I+1) \ , 
\end{equation}
where $I$ denotes the isospin of the state. 
Here the equivalence ``$\sim$'' is in the first order perturbation, 
or equivalently, for the expectation values of $|\psi_0\rangle$ 
in the irreducible representations of the symmetries. 
The Gell-Mann matrix $(\lambda^M)^i{}_j$ with $M=4,\cdots,7$ 
have non-zero components only in the off-diagonal components 
in the third row or column, namely, in $i=3$ and $j\neq 3$, or $i\neq3$ and $j=3$. 
Then, the second term in \eqref{Vsepa} can be expressed 
in terms of the creation and annihilation operators as 
\begin{equation}
 \sum_{M=4}^7 (w t_f^M \bar w)(w t_f^M \bar w) 
 \sim \sum_{M=4}^7 \frac{1}{\mw\ms} (\aw^\dag t_f^M \alpha)(\alpha^\dag t_f^M \aw) \ . 
\end{equation}
It can be further rewritten in terms of the isospin $I$ and 
the hypercharge $Y = \frac{2}{\sqrt 3} t_f^8$ as 
\begin{align}
 \sum_{M=4}^7 \frac{1}{\mw\ms} (\aw^\dag t_f^M \alpha)(\alpha^\dag t_f^M \aw) 
 &\sim 
 \frac{1}{\mw\ms} \left( C_f - I(I+1) - \frac{3}{4}Y^2 \right) \ , 
\end{align}
where $C_f$ is the quadratic Casimir of the flavor $SU(3)$. 
For the term with the generator $t_f^8$, 
\begin{align}
 (w t_f^8 \bar w)
 &\sim 
 \frac{1}{2\sqrt{3}} \aw^\dag
 \left(
 \begin{matrix}
  \frac{1}{\mw} & 0 & 0 \\
  0 & \frac{1}{\mw} & 0 \\
  0 & 0 & - \frac{2}{\ms} 
 \end{matrix}
 \right) \aw
\notag\\
 &= 
 \frac{1}{3} \left(\frac{1}{\mw} + \frac{2}{\ms}\right) (\aw^\dag t_f^8 \alpha)
 + \frac{1}{3\sqrt{3}} \left(\frac{1}{\mw} - \frac{1}{\ms}\right) (\aw^\dag \alpha) .
\end{align}
Thus, the third term in \eqref{Vsepa} gives 
\begin{equation}
 (w t_f^8 \bar w) (w t_f^8 \bar w)
 \sim 
 \frac{1}{3} \left[\frac{1}{2} \left(\frac{1}{\mw} + \frac{2}{\ms}\right) Y + 
 \frac{1}{3} \left(\frac{1}{\mw} - \frac{1}{\ms}\right) N_w \right]^2 \ . 
\end{equation}
To summarize, $V_F$ \eqref{V_F} is calculated as 
\begin{align}
 V_F &= 
 (w t_f^M \bar w)(w t_f^M \bar w) 
\notag\\
 &\sim \frac{1}{\mw\ms}C_f + \left[\frac{1}{\mw^2} - \frac{1}{\mw\ms}\right] I(I+1) 
 - \frac{3}{4\mw\ms} Y^2
\notag\\&\quad 
 + \frac{1}{3} \left[\frac{1}{2} \left(\frac{1}{\mw} + \frac{2}{\ms}\right) Y 
 + \frac{1}{3} \left(\frac{1}{\mw} - \frac{1}{\ms}\right) N_w \right]^2 \ . 
 \label{exwtfwwtfw}
\end{align}
In a similar fashion, $V_N$ \eqref{V_N} 
becomes 
\begin{equation}
 V_N = (w \bar w) (w \bar w)
 \sim  
 \left[\left(\frac{1}{\mw} - \frac{1}{\ms}\right) Y 
 + \frac{1}{3} \left(\frac{2}{\mw} + \frac{1}{\ms}\right) N_w \right]^2 \ . 
 \label{exwwww}
\end{equation}

%%%%%%%%%%%%%%%%%%%%%%%%%%%%%%%%%%%%%%%%%%%%%%%%%%%%%%%%%%%%%%%%%%%%%%%%%%%%%%%%
%%%%%%%%%%%%%%%%%%%%%%%%%%%%%%%%%%%%%%%%%%%%%%%%%%%%%%%%%%%%%%%%%%%%%%%%%%%%%%%%

\subsection{Hyperon spectrum formula and Gell-Mann--Okubo relation}

Let us consider the case $A=1$, the baryon spectrum. Using the results \eqref{MNs} \eqref{VFwCasi}
with the expressions \eqref{exwtfwwtfw} and \eqref{exwwww}, we find that the baryon spectrum is
written as
\begin{align}
M_{\rm hyperon} = & M_{\rm D4} +  3 \mw + \left(\ms - \mw\right) N_S
\nonumber \\
&
+4\lambda \left\{
\frac{1}{\mw\ms}C_f + \left[\frac{1}{\mw^2} - \frac{1}{\mw\ms}\right] I(I+1) 
 - \frac{3}{4\mw\ms} Y^2
 \right.
\notag\\&\quad \quad\qquad
 \left.
 + \frac{1}{3} \left[\frac{1}{2} \left(\frac{1}{\mw} + \frac{2}{\ms}\right) Y 
 + \left(\frac{1}{\mw} - \frac{1}{\ms}\right)  \right]^2
 \right\}
 \notag \\
 &
 -\frac13 \lambda 
 \left[\left(\frac{1}{\mw} - \frac{1}{\ms}\right) Y 
 + \left(\frac{2}{\mw} + \frac{1}{\ms}\right)  \right]^2 \ .
 \label{hyperonm}
\end{align}
Here we have put $A=1$, $N_f=3$, $N_w=3$ and $N_X=0$ for low-lying modes of a hyperon.
We have added $M_{\rm D4}$ which is the baryon vertex D4-brane mass and has not been included in the action \eqref{action}.
Although this mass can be written explicitly by the parameters of the model at the classical level, 
 it will be corrected quantum mechanically through the zero-point energies of the fields of the system and would become a complicated function of $\lambda$, $M$ and $M_S$.
Here we simply treat $M_{\rm D4}$ as a free parameter.%
\footnote{%
Indeed it is not easy to compute $M_{\rm D4}$, since it will depend on the fields which have been neglected in the action \eqref{action} and the method by which we introduced the strangness mass.
}

Now we apply the formula \eqref{hyperonm} to the baryon octet and the decuplet.
 The number of the strange quark, $N_S$, is related to the hypercharge $Y=(1/3)\mbox{diag}(1, 1, -2)$ as $Y=1-N_S$ for the states, so in total, the relation above
includes four unknown constants: $M_{\rm D4}, \mw, \ms$ and $\lambda$. 
Let us introduce a different parametrization of these constants, as
\begin{align}
&
\tilde{M}_{\rm D4} \equiv M_{\rm D4} + (2\mw+\ms) + {\cc}
\left(
\frac{1}{\ms^2}-\frac{4}{\mw\ms}
\right) \ ,
\\
&
\tilde{\lambda}\equiv \frac{\lambda}{\mw^2} \ , 
\\
&
\delta \equiv 1- \frac{\mw}{\ms} \ .
\end{align}
The last $\delta$ roughly measures the difference between the constituent u- and d- mass and the
strange quark mass.
Then the hyperon mass formula \eqref{hyperonm} is rewritten as
\begin{align}
M_{\rm hyperon} = 
&
\tilde{M}_{\rm D4} + 4\tilde{\lambda} (1-\delta)C_f
\notag \\
&
-\left( \ms \, \delta - 2 \tilde{\lambda}\,\delta\left(1 - \delta \right)\right) Y
+4\tilde{\lambda}\, \delta
\left(I(I+1) -\frac14 Y^2 \right) 
+ \tilde{\lambda}\, \delta^2 \, Y^2
\ .
\label{GMOo}
\end{align}
This is our hyperon mass formula. 
In this expression the free constants are $\tilde{M}_{\rm D4}, \ms, \tilde{\lambda}$ and $\delta$.

We compare this with the original
Gell-Mann--Okubo (GMO) mass formula \cite{GellMann:1961ky, Okubo:1961jc, Okubo:1962zzc},
\begin{align}
M_{\rm hyperon} = a_0 + a_1 Y + a_2 \left(I(I+1) -\frac14 Y^2 \right) \ . 
\label{GMO}
\end{align}
Here $a_0$, $a_1$ and $a_2$ are free constants. Note that $a_0$ needs to be
chosen different values for the octet and for the decuplet, so in effect there
are four free parameters. Comparing the GMO mass formula \eqref{GMO}
and our hyperon mass formula \eqref{GMOo}, we find that our formula
reproduces the original \eqref{GMO} except for the last term in \eqref{GMOo}
proportional to $Y^2$. Since this term has a coefficient proportional to $\delta^2$
which is at a higher order in the mass difference among the u- and d-quarks and the strange
quark, it is natural that it was ignored in the original GMO mass formula \eqref{GMO}.

As it is known that the GMO mass formula fits well the hyperon mass spectrum, let us
check that our hyperon mass formula \eqref{GMOo} can do as well.
The hyperon spectrum is summarized in Table \ref{table:hyp}. We fit the spectrum
numerically by our hyperon mass formula \eqref{GMOo}.
\begin{table}[tb]
\begin{center}
  \begin{tabular}{|ccc|}
  \hline
  \multicolumn{3}{|c|}{Octet}
   \\ \hline 
   mass & $I$ & $Y$ 
   \\ \hline\hline
    N(939) &  $\frac12$ & $1$ \\ \hline
    $\Lambda$(1116) &  $0$ & $0$ \\ \hline
    $\Sigma$(1193) &  $1$ & $0$ \\ \hline
    $\Xi$(1318) &  $\frac12$ & $-1$ \\ \hline
  \end{tabular}
  \hspace{8pt}
  \begin{tabular}{|ccc|}
  \hline
   \multicolumn{3}{|c|}{Decuplet} 
   \\ \hline 
   mass & $I$ & $Y$ 
   \\ \hline\hline
    $\Delta$(1232) & $\frac32$ & $1$ \\ \hline
    $\Sigma^*$(1385) & $1$ & $0$ \\ \hline
    $\Xi^*$(1533) & $\frac12$ & $-1$ \\ \hline
    $\Omega$(1672) & $0$ & $-2$ \\ \hline
  \end{tabular}
  \caption{The list of hyperons and their charges.}
  \label{table:hyp}
  \end{center}
\end{table}
Using 
\begin{align}
C_f = \left\{ 
\begin{array}{ll}
3 & \mbox{for octet}\\
6 & \mbox{for decuplet}
\end{array}
\right. \ , 
\end{align}
we find that our formula can fit the hyperon spectrum nicely. See Table \ref{table:rhyp}
for the numerical result.
The obtained constants are
\begin{align}
\tilde{M}_{\rm D4} = 933\, \mbox{[MeV]}, \quad
\ms = 603\, \mbox{[MeV]}, \quad
\tilde{\lambda}= 24.9 \, \mbox{[MeV]}, \quad
\delta = 0.339 \, .
\label{parameters}
\end{align}
In Table \ref{table:rhyp}, we also list the global fit of the hyperon spectrum by the GMO mass formula
\eqref{GMO}. The obtained constants are
\begin{align}
a_0^{\rm octet} = 1117 \, \mbox{[MeV]}, \quad
a_0^{\rm decuplet} = 1317\, \mbox{[MeV]}, \quad
a_1=-196\, \mbox{[MeV]}, \quad
a_2 = 33 \, \mbox{[MeV]} \, .
\end{align}
We find that our hyperon mass formula \eqref{GMOo} can fit the hyperon spectrum as nicely 
as the original GMO mass formula.

Here, we comment on the validity of the perturbation. 
The energy at the 0-th order of the perturbative expansion \eqref{MNs}
comes from the mass $\mw$ and $\ms$, 
while the first order corrections \eqref{VFwCasi} appear with the coupling constant $\rc$. 
The obtained parameters \eqref{parameters} imply that the first order corrections
are much smaller than the 0-th order terms, since 
$\mw \sim \ms \sim \mathcal O(100\text{MeV})$ while $\rc \sim \mathcal O(10\text{MeV})$.%
\footnote{% 
In the total mass of hyperon $M_\text{hyperon}$, the 0-th order energy comes from three excitations of 
$\mw \simeq \mathrm{400MeV}$ or $\ms \simeq \mathrm{600MeV}$, which is estimated as 
$E_0 \simeq \text{1200--1800MeV}$, while the first order perturbation is evaluated as 
$\langle V \rangle \simeq \text{143--473MeV}$, 
%(minimum of 143MeV for $\Lambda$ and maximum of 473MeV for $\Delta$), 
which is sufficiently smaller than $E_0$. 
Note that there is also a negative constant part of $M_\mathrm{D4}$.
}
Therefore, the perturbation gives a good approximation in this analysis. 
Although our result in Table~\ref{table:rhyp} appears to have only the errors less than 10MeV, 
those  from the perturbation may be larger but would still be of $\mathcal O(\text{10MeV})$.

\begin{table}[tb]
\begin{center}
  \begin{tabular}{|c|cccc|}
  \hline 
  Octet &  N(939) & $\Lambda$(1116)  & $\Sigma$(1193)  & $\Xi$(1318)   \\ \hline \hline
     GMO \eqref{GMO} &939&1117&1183&1328 \\ \hline
     Our \eqref{GMOo} & 941 & 1115 & 1182 & 1327 \\ \hline
  \end{tabular}
  
  \vspace{5mm}  

 \begin{tabular}{|c|cccc|}
  \hline 
  Decuplet    &
     $\Delta$(1232) & 
     $\Sigma^*$(1385) & 
     $\Xi^*$(1533) & 
     $\Omega$(1672)  \\ \hline \hline
     GMO \eqref{GMO} &1238&1383&1528&1673 \\ \hline
     Our \eqref{GMOo} & 1240 & 1380 & 1525 & 1676 \\ \hline
  \end{tabular}
  \caption{A numerical fit of the hyperon spectrum by the original GMO mass formula \eqref{GMO} 
  and our improved GMO mass formula \eqref{GMOo}.}
  \label{table:rhyp}
  \end{center}
\end{table}

%%%%%%%%%%%%%%%%%%%%%%%%%%%%%%%%%%%%%%%%%%%%%%%%%%%%%%%%%%%%%%%%%%%%%%%%%%%%%%%%
%%%%%%%%%%%%%%%%%%%%%%%%%%%%%%%%%%%%%%%%%%%%%%%%%%%%%%%%%%%%%%%%%%%%%%%%%%%%%%%%
%%%%%%%%%%%%%%%%%%%%%%%%%%%%%%%%%%%%%%%%%%%%%%%%%%%%%%%%%%%%%%%%%%%%%%%%%%%%%%%%

\section{Dibaryons}
\label{sec:Dibaryon}

In this section, we consider the dibaryons, 
which have baryon number $A=2$ (for a review, see \cite{Clement:2016vnl}). 
We have studied allowed dibaryon states in our model 
for $N_f=2$ and $N_c=3$ in Sec.~\ref{sssec:A=2,Nc=3}, 
and found that $(J,I) = (1,0)$ and $(3,0)$ are %most stable.
the lowest energy states. 
Although it is difficult to estimate the binding energy directly from our model,%
\footnote{%
As our model is based on the assumption that the baryons are sufficiently close to each other, 
unbounded state of two nuclei cannot be described in the same setup to the bound state. 
}
it is natural to expect that these two states would correspond to bound states. 
The state with spin 1, namely, $(J,I) = (1,0)$, 
is nothing but the deuteron, which has the mass of 1876~MeV. 

Now, a question would be what is the other state, $(J,I) = (3,0)$. 
The dibaryon state with $(J,I) = (3,0)$, which is referred to as $D_{03}$, 
is in fact observed recently at 2370~MeV \cite{Bashkanov:2008ih, Adlarson:2011bh}.%
\footnote{% 
There is another candidate of a dibaryon state in $(J,I) = (2,1)$. 
The observed resonance structure is very close to the $\Delta N$ threshold, 
and it has been argued that the observed structure 
is not a dibaryon state but a threshold phenomenon. 
Therefore, genuine dibaryon states are only for 
$(J,I) = (1,0)$ and $(3,0)$, at least as an opinion. 
} 
Since both dibaryon states with $(J,I) = (1,0)$ and $(3,0)$ are found experimentally, 
this appears consistent with our result that 
the %most stable 
lowest energy 
states are $(J,I) = (1,0)$ and $(3,0)$. 

However, our result at the linear order of the perturbative expansion 
implies that the energy of the deuteron and 
the other dibaryon $D_{03}$ are equal to each other, and do not agree with the observations. 
This is because the first order correction \eqref{VFCasimir} 
depends only on the flavor but independent of the spin. 
In order to solve this problem, 
we partially take into account the second order correction of the perturbative expansion. 
The second order perturbation in fact 
provides the spin dependence to the mass. 
We first consider the second order perturbation, and then, 
study the dibaryon spectra by using the resultant term.

We also study dibaryon states with strangeness. 
The most standard dibaryon with strangeness, H-dibaryon would 
consist of a couple of up quarks, a couple of down quarks and a couple of strange quarks, 
namely, a bound state of $uuddss$. 
There is another candidate of dibaryon, called di-Omega, 
which consists of 6 strange quarks. 
We estimate the masses of these dibaryons in our model.

%%%%%%%%%%%%%%%%%%%%%%%%%%%%%%%%%%%%%%%%%%%%%%%%%%%%%%%%%%%%%%%%%%%%%%%%%%%%%%%%
%%%%%%%%%%%%%%%%%%%%%%%%%%%%%%%%%%%%%%%%%%%%%%%%%%%%%%%%%%%%%%%%%%%%%%%%%%%%%%%%

\subsection{Second order perturbation}

In this section, we consider the second order corrections 
of the perturbative expansion of the nuclear matrix model given by the Hamiltonian \eqref{H}. 
We focus on the cases with $SU(N_f)$ invariance, $\ms = \mw$. 
Generalization to the case of $\ms \neq \mw$ is straightforward. 
Assuming that the unperturbed state $|\psi_0\rangle$ is diagonalized 
with respect to the perturbation $V$, 
the perturbative expansion of the energy is expressed as 
\begin{equation}
 E = E_0 + \langle V \rangle 
 - \sum_n \langle\psi_0|V|\psi_n\rangle \frac{1}{E_n - E_0}\langle\psi_n|V|\psi_0\rangle \ , 
\end{equation}
where $|\psi_n\rangle$ is the excited states 
which are orthogonal to the unperturbed ground state $|\psi_0\rangle$, 
and $E_n$ is the energy of $|\psi_n\rangle$. 
Here, we focus on the second order perturbation by the third term, 
\begin{equation}
% - 4 i \cc \epsilon^{IJK} X^J X^K 
%   \bar w^{\dot \alpha i} (\tau^I)_{\dot \alpha}{}^{\dot\beta} w_{\dot \beta i} \ . 
  - 4 i \cc \epsilon^{IJK} X_A^J X_B^K f^{AB}{}_C\, 
    \bar w^{\dot \alpha i}_a (\tau^I)_{\dot \alpha}{}^{\dot\beta} (t^C)^a{}_b w_{\dot \beta i}^b  \ . 
\label{XXww}
\end{equation}
The first and second terms in the expression \eqref{V} of the perturbation $V$ 
give the corrections which is independent of 
the flavor and spin of the physical state $|\psi_0\rangle$. 
It is also straightforward to see that the second order perturbation 
from the fourth term $(\bar w \tau^I w)^2$ gives only the same combination given in \eqref{w-4pt}, 
and does not provide spin dependence to the mass spectra.

Here, we consider the case of $A\leq 2 N_f$. 
As the state $|\psi_0\rangle$ does not have the excitations of $X^I$ in this case, 
two $X$'s in \eqref{XXww} must act as the excitation operator 
when it acts on $|\psi_0\rangle$, and 
the excited state $|\psi_n\rangle \sim V |\psi_0\rangle$ is obtained 
by acting the perturbation $V$ given by \eqref{V} to $|\psi_0\rangle$, 
it has two excitations of $X^I$. 
As we consider only the states without additional excitations of $w$ and $\bar w$ 
to those in $|\psi_0\rangle$, $w$ and $\bar w$ behave 
as the creation and annihilation operators of $w$. 
The excited state $|\psi_n\rangle$ has the same energy for $H_0$, 
\begin{equation}
 E_n = E_0 + 2 \MXo \ . 
\end{equation}
Here, the effective mass of $X^I$, $\MX$ should be determined as argued in App.~\ref{app:Mass}. 

The matrix element of the perturbation $V$ is expressed 
in terms of the creation and annihilation operators as 
\begin{align}
 \langle\psi_n|V|\psi_0\rangle 
 = 
 - \frac{2 \cc}{\MXo \mw} 
 \langle\psi_n| 
  \epsilon^{IJK} \AX_A^{\dag\,J} \AX_B^{\dag\,K} f^{AB}{}_C\, 
  \aw^{\dag\,a}_{\dot \alpha i} (\tau^I)^{\dot \alpha}{}_{\dot\beta} (t^C)_a{}^b \aw^{\dot \beta i}_b 
 |\psi_0\rangle \ . 
\end{align}
Then the second order correction of the perturbative expansion
of the energy is expressed as 
\begin{align}
 - \frac{2 \lambda^2}{\MXo^3 \mw^2} 
 \langle\psi_0|&
  \left[
   \epsilon^{IJK} \AX_A^J \AX_B^K f^{AB}{}_C\,
   \aw^{\dag\,a}_{\dot \alpha i} (\tau^I)^{\dot \alpha}{}_{\dot\beta} (t^C)_a{}^b \aw^{\dot \beta i}_b 
  \right] 
\notag\\
&\quad\times
  \left[
   \epsilon^{LMN} \AX_D^{\dag\,M} \AX_E^{\dag\,N} f^{DE}{}_F\, 
   \aw^{\dag\,c}_{\dot \gamma j} (\tau^L)^{\dot \gamma}{}_{\dot \delta} (t^F)_c{}^d \aw^{\dot \delta j}_d 
  \right]
 |\psi_0\rangle \ . 
\label{XXww2}
\end{align}
Taking the contractions of the creation and annihilation operators of $X^I$, 
the expression above can be written only in terms of 
the creation and annihilation operators of $w$ as 
\begin{align}
 - \frac{4 A\lambda^2}{\MXo^3 \mw^2} 
 \langle\psi_0|
  \biggl[&
   \aw^{\dag\,a}_{\dot \alpha i} (\tau^I)^{\dot \alpha}{}_{\dot\beta} \aw^{\dot \beta i}_b 
   \aw^{\dag\,b}_{\dot \gamma j} (\tau^I)^{\dot \alpha}{}_{\dot\beta} \aw^{\dot \delta j}_a 
%\notag\\&
   - \frac{1}{A} 
   \aw^{\dag\,a}_{\dot \alpha i} (\tau^I)^{\dot \alpha}{}_{\dot\beta} \aw^{\dot \beta i}_a 
   \aw^{\dag\,b}_{\dot \gamma j} (\tau^I)^{\dot \alpha}{}_{\dot\beta} \aw^{\dot \delta j}_b 
  \biggr]
 |\psi_0\rangle \ . 
\end{align}
As we discussed in Sec.~\ref{ssec:4pt-w}, 
the first term can be rewritten as the quadratic Casimir of flavor $SU(N_f)$, 
and the second term is nothing but the quadratic Casimir of spin $SU(2)$. 

In summary, 
the second order terms in the energy is given by 
the quadratic Casimirs of flavor $SU(N_f)$ and spin $SU(2)$ as 
\begin{align}
 \frac{4 \lambda^2}{\MXo^3 \mw^2} 
 \langle\psi_0|
  \left[
   - 4 A (\alpha^\dag t^M_f \alpha)(\alpha^\dag t^M_f \alpha) 
   - \frac{2A - N_f}{N_f} (\alpha^\dag \alpha)(\alpha^\dag \alpha)
   + (\alpha^\dag \tau^I \alpha)(\alpha^\dag \tau^I \alpha)
  \right]
 |\psi_0\rangle \ , 
\label{V-2nd}
\end{align}
up to the corrections independent of flavor or spin.

%%%%%%%%%%%%%%%%%%%%%%%%%%%%%%%%%%%%%%%%%%%%%%%%%%%%%%%%%%%%%%%%%%%%%%%%%%%%%%%%
%%%%%%%%%%%%%%%%%%%%%%%%%%%%%%%%%%%%%%%%%%%%%%%%%%%%%%%%%%%%%%%%%%%%%%%%%%%%%%%%

\subsection{Dibaryon spectrum from D-branes}\label{ssec:Dibaryon-Result}

Now, we consider the dibaryon spectrum by using the result from the second order 
in the perturbative expansion. 
The result of the second order perturbation \eqref{V-2nd} 
implies that the additional term of $V_S$ appears 
in the combination of 
\begin{align}
 V_S - A V_F - \frac{2A - 3}{12} V_N \ ,  \label{VS-2nd}
\end{align}
for $N_f = 3$, where 
\begin{equation}
 V_S = \frac{1}{4} \left(w \tau^I \bar w\right)^2 \ . 
\label{V_S}
\end{equation}
The interaction term $V_S$ simply gives the Casimir operator for the spin $SU(2)$ 
if the mass of $w$ is the same for all flavors. 
However, because of the different mass for $w_{(s)} = w_{i=3}$, 
$V_S$ deviates from that for the $SU(3)$ flavor invariant mass, 
in a similar fashion to $V_F$ and $V_S$ in Sec.~\ref{ssec:hyperon-1st}. 
Then, $V_S$ is calculated as 
\begin{align}
 V_S %\left(w \tau^I \bar w\right)^2 
 &\sim  
 \frac{1}{4} \left[ 
 \frac{1}{\mw^2} (\hat S_{(I)}^I)^2
  + \frac{1}{\mw \ms} \left((\hat S^I)^2 - (\hat S_{(I)}^I)^2 - (\hat S_{(S)}^I)^2\right)
 + \frac{1}{\ms^2} (\hat S_{(S)}^I)^2
 \right]
\notag\\
 &= 
 \frac{1}{\mw^2}
 \bigl[ (1- \delta) J(J+1) 
  + \delta s_I(s_I+1) - \delta (1-\delta) s_S(s_S + 1) 
 \bigr] \ , 
\label{VS}
\end{align}
where $\hat S_{(I)}^I$, $\hat S_{(S)}^I$ and $\hat S^I$ are 
the spin operator for the excitations of $w$ with the flavor index $i=1,2$, and that for $i=3$. 
and the total spin operator, respectively. 
We refer to (the norm of) the total spin of the excitations with $i=1,2$, 
that of $i=3$, and that of the whole state as $s_I$, $s_S$ and $J$, respectively. 

In a similar fashion, the second term in \eqref{w-4-U(N_f)} 
is modified due to the effect of $\ms\neq \mw$, 
and the $SU(A)$ part no longer vanish identically, for $A\geq 2$. 
For $A=2$, it becomes 
\begin{align}
 V_B 
 &= 
 \left(w t^A_{SU(2)} \bar w\right)^2 
\notag\\
 &= 
 \left[ 
 \frac{1}{\mw^2} (\hat B_{(I)}^A)^2
 + \frac{1}{\mw \ms} \left((\hat B^A)^2 - (\hat B_{(I)}^A)^2 - (\hat B_{(S)}^A)^2\right)
 + \frac{1}{\ms^2} (\hat B_{(S)}^A)^2
 \right]
\notag\\
 &= 
 \frac{\delta}{\mw^2} \left[ 
  b_I(b_I+1) - (1-\delta) b_S(b_S + 1) 
 \right] \ , 
\label{VB}
\end{align}
where $(\hat B_{(I)}^A)^2$, $(\hat B_{(S)}^A)^2$ and $(\hat B^A)^2$ 
are the quadratic Casimir of the baryon $SU(2)$ symmetry for the flavor $i=1,2$ part, 
strange part of $i=3$ and whole dibaryon states, respectively. 
The physical state must be singlet of the baryon $SU(2)$ due to the constraint \eqref{Const}, 
and the expectation value of $(\hat B^A)^2$ is always zero, 
while each of $i=1,2$ part and $i=3$ part can be in a non-trivial representation. 
The eigenvalues of $(\hat B_{(I)}^A)^2$ and $(\hat B_{(S)}^A)^2$ are 
given in terms of ``spin'' of baryon $SU(2)$, $b_I$ and $b_S$ as 
$b_I(b_I+1)$ and $b_S(b_S + 1)$, respectively. 
In order to form a singlet state from ``spin'' $b_I$ and $b_S$, they must satisfy $b_I = b_S$. 
If there are no excitation with strangeness, 
we have $b_I = b_S = 0$, and hence, $V_B$ has no contribution.

The mass formula is now given by 
\begin{align}
 M_\text{nucl} 
 &= 
 {M}_{\rm D4} A - M_S \left[(-3 + 2 \delta) A + \delta Y\right] 
\notag \\
&\quad
 + 4\tilde{\lambda} \mw^2 \tilde V_F 
+  4 c(A) \tilde \lambda' \mw^2 \left(V_S - A \tilde V_F \right)
 \ , 
\label{M_nucl3}
\end{align}
where 
\begin{equation}
 \tilde V_F = V_F + \frac{2A - 3}{12A} V_N \ . 
\end{equation}
Here, 
the bare D4-brane mass $M_\text{D4}$ and the coefficient of the new term \eqref{VS-2nd}, $\rc'$
in principle depend on the baryon number $A$. 
The new term \eqref{VS-2nd} is absent for $A=1$ 
as it comes from the interaction term \eqref{XXww}. 
We introduced the factor $c(A)$, which is defined as 
$c(1)=0$ and $c(2)=1$, in order to reproduce the absence of this term for $A=1$. 
Then, $\rc'$ can be treated as an $A$-independent parameter 
since we are considering only $A=1$ and $A=2$ states here. 

The bare D4-brane mass $M_\text{D4}$ is assumed to be an $A$-independent constant. 
There are other contributions from zero-point motion of $X^I$, 
and other fields which we have neglected in our model, 
for example, fermions and Kaluza-Klein modes. 
Some of these additional contributions behave as $A^2$ or 
have more complicated $A$-dependence. 
We assume that these contributions are much less than 
the bare D-brane tension and ignored here.

The coupling constant $\rc'$, in principle, 
can be determined by calculating the second and higher order perturbations 
and fixing $\MX$ by using the criteria discussed in App.~\ref{app:Mass}. 
However, it is difficult to find the most appropriate $\MX$.% 
\footnote{%
Furthermore, some of the results in App.~\ref{app:Mass} implies that 
the second order correction \eqref{V-2nd} is comparable to 
the first order interaction terms \eqref{w-4pt}. 
Hence, it is difficult to calculate a very precise value of the coupling constant $\rc'$, 
while it is clear that the second (and possibly higher) order corrections yield 
the interaction term \eqref{VS-2nd}. 
}
See App.~\ref{app:more} for calculation of the dibaryon spectra 
by using a few example of $\MX$ from the criteria in App.~\ref{app:Mass}.
Here we treat $\rc'$ as a free parameter, 
and fix it by fitting \eqref{M_nucl3} to experimental data. 
%The interaction term \eqref{VS-2nd} in general 
%appears with an $A$-dependent factor, $c(A)$. 
%Since \eqref{VS-2nd} comes from the interaction term \eqref{XXww}, 
%which is absent for $A=1$, the factor $c(A)$ is effectively obeys $c(1)=0$. 
%Since we are considering only $A=1$ and $A=2$ states here, 
%we can take $c(2)=1$ without loss of generality. 

Although our model is derived from the D-brane setup, 
the two points above, the $A$-dependence of constant term $M_\text{D4}A$, 
and the independence of the constant $\rc'$ might be considered as 
artificial modifications of the model. 

We numerically fit our mass formula \eqref{M_nucl3} to 
8 hyperons and 2 dibaryons (deuteron $D$ and $(J,I) = (3,0)$ dibaryon $D_{03}$). 
Although the dibaryon $D_{03}$ may not be well-established, 
we use this as an input for the fitting since 
the information of the hyperons and deuteron are not sufficient 
to determine the coupling constant of $V_S$ \eqref{VS}. 
For $A=1$, the spin and flavor are related to each other 
due to the constraint from the statistics, 
or equivalently, $V_F$ and $V_S$ are related to each other by the formula \eqref{SU(N)}. 
Thus, hyperons do not give information to fix the coefficient of $V_S$. 
Information of deuteron is used to determine the $A$ dependence of the mass, 
and hence, is not sufficient to fix another parameter. 
Therefore, information of $D_{03}$ is necessary. 
In fact, the additional term $V_S$ is introduced to 
explain the mass difference between deuteron and $D_{03}$, 
and hence, we should take it into account.

\begin{table}[tb]
\begin{center}
  \begin{tabular}{|r@{}l|ccccc|}
  \hline
   & mass & $I$ & $Y$ & $J$ & $C_f$ & $A$ \\ \hline\hline
    $D$ & (1876) & 0 & 2 & 1 & 6 & 2 \\ \hline 
    $D_{12}$ & (2160?) & 1 & 2 & 2 & 8 & 2 \\ \hline 
    $D_{03}$ & (2370) & 0 & 2 & 3 & 6 & 2 \\ \hline 
    \multicolumn{2}{|c|}{$H$-dibaryon} & 0 & 0 & 0 & 0 & 2 \\ \hline 
  \end{tabular}  
  \hspace{8pt}
  \begin{tabular}{|l|ccccc|}
  \hline
   & $I$ & $Y$ & $J$ & $C_f$ & $A$ \\ \hline\hline
    $D_{10}$ & 1 & 2 & 0 & 8 & 2 \\ \hline 
    $D_{21}$ & 2 & 2 & 1 & 12 & 2 \\ \hline 
    $D_{30}$ & 3 & 2 & 0 & 18 & 2 \\ \hline 
    $\Omega \Omega$ & 0 & $-4$ & 0 & 18 & 2 \\ \hline 
  \end{tabular}
  \caption{The list of hyperons, dibaryons and their charges.}
  \label{table:hyp-di}
  \end{center}
\end{table}

We also consider the H-dibaryon which is singlet 
of spin $SU(2)$, flavor $SU(3)$ and baryon $SU(2)$ symmetries. 
The state of H-dibaryon is given by 
\begin{align}
 |\psi\rangle 
 &= 
 \epsilon^{i_1 i_3 i_5} 
 \epsilon^{i_2 i_4 i_6} 
 \left(
 \epsilon_{a_1 a_2} 
 \epsilon^{\dot \alpha_1 \dot \alpha_2} 
 \aw_{\dot \alpha_1 i_1}^{\dag\ a_1} 
 \aw_{\dot \alpha_2 i_2}^{\dag\ a_2}
 \right)
 \left(
 \epsilon_{a_3 a_4} 
 \epsilon^{\dot \alpha_3 \dot \alpha_4} 
 \aw_{\dot \alpha_3 i_3}^{\dag\ a_3} 
 \aw_{\dot \alpha_4 i_4}^{\dag\ a_4} 
 \right)
\notag\\&\qquad\qquad\qquad\times
 \left(
 \epsilon_{a_5 a_6} 
 \epsilon^{\dot \alpha_5 \dot \alpha_6} 
 \aw_{\dot \alpha_5 i_5}^{\dag\ a_5}
 \aw_{\dot \alpha_6 i_6}^{\dag\ a_6} 
 \right) |0\rangle \ .  
\label{H-dib}
\end{align}
It should be noted that this state diagonalize 
the interaction term \eqref{VS} and \eqref{VB} only up to $\mathcal O(\delta^2)$, 
and decomposed into a state with $s_I = s_S = b_I = b_S = 0$ 
and that with $s_I = s_S = b_I = b_S = 1$. 
However, $\mathcal O(\delta^2)$ terms give only contributions of $\mathcal O(1\text{MeV})$ 
since 
\begin{align}
 \delta^2 &\simeq 0.1 \ , 
& 
 \rc \sim \rc' \sim \mathcal O(10\text{MeV}) \ , 
\end{align}
as we will see below, we will ignore $\mathcal O(\delta^2)$ terms for H-dibaryon. 
Di-Omega consists of 6 strange quarks and the corresponding state in our model is given by 
\begin{align}
 |\psi\rangle 
 &= 
 \left(
 \epsilon_{a_1 a_2} 
 \epsilon^{\dot \alpha_1 \dot \alpha_2} 
 \aw_{\dot \alpha_1 i=3}^{\dag\ a_1} 
 \aw_{\dot \alpha_2 i=3}^{\dag\ a_2}
 \right)
 \left(
 \epsilon_{a_3 a_4} 
 \epsilon^{\dot \alpha_3 \dot \alpha_4} 
 \aw_{\dot \alpha_3 i=3}^{\dag\ a_3} 
 \aw_{\dot \alpha_4 i=3}^{\dag\ a_4} 
 \right)
\notag\\&\quad\times
 \left(
 \epsilon_{a_5 a_6} 
 \epsilon^{\dot \alpha_5 \dot \alpha_6} 
 \aw_{\dot \alpha_5 i=3}^{\dag\ a_5}
 \aw_{\dot \alpha_6 i=3}^{\dag\ a_6} 
 \right) |0\rangle \ .  
\label{di-Om}
\end{align}
The hyperons, dibaryons and their charges are summerized in Table~\ref{table:hyp-di}. 

We first fit our mass formula \eqref{M_nucl3} to 8 hyperons, deuteron and $D_{03}$, 
and then, give a prediction for the masses of H-dibaryon \eqref{H-dib} and di-Omega \eqref{di-Om}, 
by using the parameters obtained from the fitting. 
The result of the numerical fit is shown in Table~\ref{table:rhyp-di5}.  
We fit our formula to 8 hyperons, deuteron and $D_{03}$. 
The data of $D_{03}$ is necessary to determine $\rc'$. 
The parameters are determined as 
\begin{align}
&{M}_{\rm D4} = 198\, \mbox{[MeV]}, \quad
M_S = 403\, \mbox{[MeV]}, \quad
\notag\\
&\tilde{\lambda}= 28.0 \, \mbox{[MeV]}, \quad
\tilde{\lambda}' = 10.2 \, \mbox{[MeV]}, \quad
\delta = 0.496 \, .
\label{para-diB}
\end{align}

\begin{table}[tb]
\begin{center}
  \begin{tabular}{|c|cccc|}
  \hline 
  Octet &  N(939) & $\Lambda$(1116)  & $\Sigma$(1193)  & $\Xi$(1318)   \\ \hline \hline
     GMO \eqref{GMO} &939&1117&1183&1328 \\ \hline
     Our \eqref{M_nucl3} & 975 & 1126 & 1237 & 1347 \\ \hline
  \end{tabular}
  
  \vspace{5mm}  

 \begin{tabular}{|c|cccc|}
  \hline 
  Decuplet    &
     $\Delta$(1232) & 
     $\Sigma^*$(1385) & 
     $\Xi^*$(1533) & 
     $\Omega$(1672)  \\ \hline \hline
     GMO \eqref{GMO} &1238&1383&1528&1673 \\ \hline
     Our \eqref{M_nucl3} & 1311 & 1407 & 1516 & 1639 \\ \hline
  \end{tabular}
  \vspace{5mm}  

 \begin{tabular}{|c|cccc|}
  \hline 
  Dibaryon    &
     $D$(1876) & 
     $D_{03}$(2370) & 
     $H$ & 
     $\Omega \Omega$ \\ \hline \hline
     Our \eqref{M_nucl3} & 1876 & 2285 & 2084 & 3007 \\ \hline
  \end{tabular}

  \vspace{5mm}  

 \begin{tabular}{|c|cccc|}
  \hline 
  Dibaryon    &
     $D_{12}$(2160?) & 
     $D_{21}$ & 
     $D_{10}$ & 
     $D_{30}$ 
     \\ \hline \hline
     Our \eqref{M_nucl3} & 
     2100
& 
     2058 
& 
     1855
& 
     2157
\\ \hline
  \end{tabular}

  \end{center}
   \caption{A numerical fit of the hyperon and dibaryon spectrum 
   by our formula \eqref{M_nucl3}.}
  \label{table:rhyp-di5}
\end{table}

The result agrees with experiments up to the error of 100 MeV, 
for those whose signals are found in experiments. 
The dibaryons with isopsin larger than spin 
have not found in experiments, and hence, 
expected to have large mass. 
However, our result implies that such dibaryons 
also have similar and even slightly smaller than those with $I<J$. 
This would be the problem of the fitting | 
we do not have experimental data on the isospin dependence 
of the dibaryon mass to be fitted by the formula. 
Therefore, the results for $I>J$, for which the isospin dependence is important, 
are not very reliable.%
\footnote{%
See App.~\ref{app:more}, for results which are good even for $I>J$, from other fitting. 
} 
Comparison to the threshold is also shown in Table~\ref{table:bind1}. 
We show two different threshold | one is from the experimental data of hyperons 
and the other is from our result of hyperons in the same fitting. 
The dibaryon $D_{03}$ is about 100--300 MeV below the threshold. 
The dibaryons with strangeness | H-dibaryon and di-Omega 
are also about 150 MeV and 300 MeV below the threshold. 
Even taking the error of 100 MeV into account, 
out result implies that they would compose bound states. 
Since the dibaryons with $I>J$ have not found in experiments so far, 
the error for them may comes to 300 MeV. 
Since the H-dibaryon and the di-Omega have $I=0$, and the same applies to the  
deuteron and $D_{03}$, our result for them would be better than those for $I\neq 0$, 
but could have a similar error possibly.

\begin{table}[tb]
\begin{center}

 \begin{tabular}{|c|cccc|}
  \hline 
  Dibaryon    &
     $D$(1876) & 
     $D_{03}$(2370) & 
     $H$ & 
     $\Omega \Omega$ \\ \hline
     Our \eqref{M_nucl3} & 1876 & 2285 & 2084 & 3007 \\ \hline\hline
     Threshold & N$+$N & $\Delta$$ + \Delta$ & $\Lambda$$ +\Lambda$ & $\Omega$$ + \Omega$ \\\hline
     Experiment & 1878 & 2464 & 2232 & 3344 \\\hline 
     Our \eqref{M_nucl3} & 1950 & 2622 & 2252 & 3278 \\\hline
  \end{tabular}

  \vspace{5mm}  

 \begin{tabular}{|c|cccc|}
  \hline 
  Dibaryon    &
     $D_{12}$(2160?) & 
     $D_{21}$ & 
     $D_{10}$ & 
     $D_{30}$ 
     \\ \hline 
     Our \eqref{M_nucl3} & 
     2100
& 
     2058 
& 
     1855
& 
     2157
\\ \hline\hline
     Threshold & N$+ \Delta$ & N$+ \Delta$ & N$+$N & $\Delta$$ + \Delta$ \\\hline
     Experiment & 2171 & 2171 & 1878 & 2464 \\\hline 
     Our \eqref{M_nucl3} & 2286 & 2286 & 1950 & 2622 \\\hline
  \end{tabular}

  \end{center}
   \caption{Comparison with the threshold for \eqref{M_nucl3}.}
  \label{table:bind1}
\end{table}

\subsubsection*{Without mass renormalization}

The interaction terms \eqref{V_F}, \eqref{V_N} and \eqref{V_S} 
also contains the creation and annihilation operators of $\bar w$. 
By taking the contraction between these operators, 
we obtain correction terms to the mass $\mw$ and $\ms$. 
They can be absorbed by redefinition of $\mw$ and $\ms$ 
and do not affect the result for the hyperon mass in Sec.~\ref{sec:Hyperon}. 
However, an ambiguity appears when we use 
the result in Sec.~\ref{sec:Hyperon} to calculate the dibaryon mass ---
which should be identified as the same parameters for $A=1$ and $A=2$, 
those before the redefinition or those after the redefinition. 
So far, we assumed that the renormalized mass, 
after the absorption of the correction terms by the redefinition of $\mw$ and $\ms$, 
are physical and will be the same for $A=1$ and $A=2$. 
Then, we have just neglected the contraction terms, 
as are already absorbed by the redefinition. 
Here, we consider the other assumption | 
the bare mass, without using the redefinition, 
are the same for $A=1$ and $A=2$. 
In this case, we need to calculate the all contraction terms. 

Now, we take all terms from the contraction between 
the creation and annihilation operators into account. 
Then, $\tilde V_F$ and $V_S$ are calculated as 
\begin{align}
 \mw^2 \tilde V_F
 &= 
 (1-\delta)C_f
 - \delta\left[1 + (2-\delta)\left(1 + \frac{1}{4A} - 3 A \right)\right] Y
 + \delta \left(I(I+1) -\frac14 Y^2 \right) 
\notag\\&\quad
 + \left(\frac{1}{2} - \frac{1}{4A}\right) \delta^2 Y^2
 -\frac{3}{4}\left[(1+3A-6A^2)(3-2 \delta) + (1+A-6A^2) \delta^2 \right]
 \ , 
\label{VFcont}
\\
 \mw^2 V_S 
 &= 
 (1- \delta) J(J+1) 
  + \delta s_I(s_I+1) - \delta (1- \delta) s_S(s_S + 1) 
\notag\\&\quad
  + \frac{3}{4}\left[3A \left(3 - 2 \delta + \delta^2 \right) + \left(2 - \delta \right) \delta Y\right]
 \ . 
\label{VScont}
\end{align}

We consider the global fit of our mass formula by treating 
the coupling constant of $V_S$ as a free parameter. 
The mass formula is now given by \eqref{M_nucl3} with \eqref{VFcont} and \eqref{VScont}, 
and we fit it to 8 hyperons, deuteron and $D_{03}$. 
The parameters are determined as 
\begin{align}
&{M}_{\rm D4} = -1067\, \mbox{[MeV]}, \quad
M_S = 701\, \mbox{[MeV]}, \quad
\notag\\
&\tilde{\lambda}= 28.1 \, \mbox{[MeV]}, \quad
\tilde{\lambda}' = 11.2 \, \mbox{[MeV]}, \quad
\delta = 0.402 \, .
\label{para-diB-cont}
\end{align}
The result is shown in Table~\ref{table:rhyp-di4}. 
The error to experiments for hyperons and dibaryons 
whose signals are already found in experiments 
are smaller than the previous result in Table~\ref{table:rhyp-di5}, 
and now is up to 50 MeV. 
The results for dibaryons for $I<J$ are still not very good, 
because of the same reason | 
absence of the information on the isospin dependence. 
Note that our result of deuteron mass is above the threshold 
but is inside the error up to 50 MeV. 
Results for $D_{03}$, H-dibaryon and di-Omega are all below the threshold.

\begin{table}[tb]
\begin{center}
  \begin{tabular}{|c|cccc|}
  \hline 
  Octet &  N(939) & $\Lambda$(1116)  & $\Sigma$(1193)  & $\Xi$(1318)   \\ \hline \hline
     GMO \eqref{GMO} &939&1117&1183&1328 \\ \hline
     Our \eqref{M_nucl3} & 925 & 1099 & 1189 & 1326 \\ \hline
  \end{tabular}
  
  \vspace{5mm}  

 \begin{tabular}{|c|cccc|}
  \hline 
  Decuplet    &
     $\Delta$(1232) & 
     $\Sigma^*$(1385) & 
     $\Xi^*$(1533) & 
     $\Omega$(1672)  \\ \hline \hline
     GMO \eqref{GMO} &1238&1383&1528&1673 \\ \hline
     Our \eqref{M_nucl3} & 1263 & 1391 & 1528 & 1675 \\ \hline
  \end{tabular}
  \vspace{5mm}  

 \begin{tabular}{|c|cccc|}
  \hline 
  Dibaryon    &
     $D$(1876) & 
     $D_{03}$(2370) & 
     $H$ & 
     $\Omega \Omega$ \\ \hline \hline
     Our \eqref{M_nucl3} & 1898 & 2345 & 2124 & 3147 \\ \hline
  \end{tabular}

  \vspace{5mm}  

 \begin{tabular}{|c|cccc|}
  \hline 
  Dibaryon    &
     $D_{12}$(2160?) & 
     $D_{21}$ & 
     $D_{10}$ & 
     $D_{30}$ 
     \\ \hline \hline
     Our \eqref{M_nucl3} & 
     2122
& 
     2036 
& 
     1854
& 
     2086
\\ \hline
  \end{tabular}

  \end{center}
   \caption{A numerical fit of the hyperon and dibaryon spectrum by our formula \eqref{M_nucl3} 
   with \eqref{VFcont} and \eqref{VScont}.}
  \label{table:rhyp-di4}
\end{table}

\begin{table}[tb]
\begin{center}

 \begin{tabular}{|c|cccc|}
  \hline 
  Dibaryon    &
     $D$(1876) & 
     $D_{03}$(2370) & 
     $H$ & 
     $\Omega \Omega$ \\ \hline
     Our \eqref{M_nucl3} & 1898 & 2345 & 2124 & 3147 \\ \hline\hline
     Threshold & N$+$N & $\Delta$$+\Delta$ & $\Lambda$$ +\Lambda$ & $\Omega$$ + \Omega$ \\\hline
     Experiment & 1878 & 2464 & 2232 & 3344 \\\hline 
     Our \eqref{M_nucl3} & 1850 & 2526 & 2198 & 3350 \\\hline
  \end{tabular}

  \vspace{5mm}  

 \begin{tabular}{|c|cccc|}
  \hline 
  Dibaryon    &
     $D_{12}$(2160?) & 
     $D_{21}$ & 
     $D_{10}$ & 
     $D_{30}$ 
     \\ \hline
     Our \eqref{M_nucl3} & 
     2122
& 
     2036 
& 
     1854
& 
     2086
\\ \hline\hline
     Threshold & N$+ \Delta$ & N$+ \Delta$ & N$+$N & $\Delta $$+ \Delta$ \\\hline
     Experiment & 2171 & 2171 & 1878 & 2464 \\\hline 
     Our \eqref{M_nucl3} & 2188 & 2188 &  1850 & 2526 \\\hline
  \end{tabular}

  \end{center}
   \caption{Comparison to the threshold for our formula \eqref{M_nucl3} 
   with \eqref{VFcont} and \eqref{VScont}.}
  \label{table:bind2}
\end{table}

%%%%%%%%%%%%%%%%%%%%%%%%%%%%%%%%%%%%%%%%%%%%%%%%%%%%%%%%%%%%%%%%%%%%%%%%%%%%%%%%
%%%%%%%%%%%%%%%%%%%%%%%%%%%%%%%%%%%%%%%%%%%%%%%%%%%%%%%%%%%%%%%%%%%%%%%%%%%%%%%%
%%%%%%%%%%%%%%%%%%%%%%%%%%%%%%%%%%%%%%%%%%%%%%%%%%%%%%%%%%%%%%%%%%%%%%%%%%%%%%%%

\section{Baryon resonance}
\label{sec:Resonance}

In the previous sections, 
the physical ground states, which have no excitation of $\bar w$, have been studied. 
The overall $U(1)$ part of the constraint \eqref{Const} 
gives the condition on the baryon $U(1)$ charge, 
and the excitations which satisfy \eqref{NPhys} are allowed. 
Thus, the same number of the excitations of $w$ and $\bar w$ 
can be introduced for excited states, in general. 
In our model, these excitations would be identified to 
the internal excitations in the baryon bound states. 
In this section, we focus on the case of $N_f=2$, $N_c=3$ and $A=1$  
and consider a pair of the excitations of $w$ and $\bar w$. 
Since the condition of $A=1$ gives single baryon states, 
the excited states would correspond to some of the baryon resonances. 

At the 0-th order of the perturbative expansion, 
the expectation value of the energy is simply given by 
\begin{equation}
 E_0 = \MX N_X + \mw \left(N_w + N_{\bar w}\right) \ . 
\end{equation}
Here, we consider the states without the excitations of $X^I$, 
and with a pair of the additional excitation of $w$ and $\bar w$
to the physical ground state \eqref{E0w}, namely, 
\begin{align}
 N_X &= 0 \ , 
& 
 N_w &= 4 \ , 
& 
 N_{\bar w} &= 1 \ . 
\end{align}
The 0-th order state $|\psi_0\rangle$ for this excited state 
is obtained by introducing 4 creation operators of $w$ and 
1 creation operator of $\bar w$ to the unconstrained ground state $|0\rangle$; 
\begin{align}
 |\psi_0\rangle 
 = 
 \aw_{\dot \alpha_1 i_1}^{\dag} 
 \aw_{\dot \alpha_2 i_2}^{\dag}
 \aw_{\dot \alpha_3 i_3}^{\dag} 
 \aw_{\dot \alpha_4 i_4}^{\dag} 
 \awb^{\dag\,\dot \alpha_4 i_4} |0\rangle 
\ . 
\label{EState}
\end{align}
Notice that $w$ and $\bar w$, or equivalently, $\alpha$ and $\bar \alpha$ 
do not have indices of the baryon $U(A)$ symmetry for $A=1$. 

Now, we consider the linear order corrections in the perturbative expansion. 
As for the physical ground state, 
excited states with a $w\bar w$ pair \eqref{EState} 
are not unique but all combinations of the spin and flavor indices are degenerated. 
Thus, we have to diagonalize the perturbation $V$ in 
the space of the excited state \eqref{EState}. 
Since \eqref{EState} has no excitations of $X^I$, 
only the last term in \eqref{V} is relevant, 
but now, we have to take $\awb$ and $\awb^\dag$ into account.  

In terms of the creation and annihilation operators 
the last term of \eqref{V} is expressed as 
\begin{align}
 \cc \left[w_{\dot\alpha i} \tau^I{}^{\dot\alpha}{}_{\dot\beta} \bar w^{\dot\beta i}\right]^2  
 = 
 \frac{\cc}{\mw^2}
 \left[\left(\aw^\dag + \awb\right)_{\dot\alpha i} 
  \tau^I{}^{\dot\alpha}{}_{\dot\beta} 
  \left(\aw + \awb^\dag\right)^{\dot\beta i}\right]^2 \ . 
\label{w-4pt-a1}
\end{align}
Since both spin and flavor (isospin) symmetries are $SU(2)$, which is pseudo-real, 
fundamental and anti-fundamental representations of these symmetries are converted to each other. 
We raise and lower the indices of $\awb$ and $\awb^\dag$ as 
\begin{align}
 \awb^{{\dot\alpha} i} 
 &= 
 \epsilon^{{\dot\alpha} {\dot\beta}} \epsilon^{ij} \awb_{{\dot\beta} j} \ , 
&
 \awb_{{\dot\alpha} i}^\dag
 &= 
 \epsilon_{{\dot\alpha} {\dot\beta}} \epsilon_{ij} \awb^{\dag\,{\dot\beta} j} \ . 
\end{align}
In order to diagonalize the interaction term \eqref{w-4pt-a1} for the excited state \eqref{EState}, 
it is sufficient to consider the terms which do not change the number of the excitations, 
or equivalently, those with the same numbers of the creation and annihilation operators 
for each of $w$ and $\bar w$, namely, 
\begin{align}
 &\aw^\dag \aw \, \aw^\dag \aw \ , 
&
 &\aw^\dag \aw \, \awb^\dag \awb \ , 
&
 &\awb^\dag \awb \, \awb^\dag \awb \ . 
\end{align}
The permutations of the creation and annihilation operators 
give terms which have only a pair of the creation and annihilation operators 
\begin{align}
 \aw^\dag_{{\dot\alpha} i} \aw^{{\dot\alpha} i} 
 + \awb^{\dag\,\dot\alpha i} \awb_{{\dot\alpha} i} \ , 
\end{align}
which can be absorbed by the renormalization of the masses of $w$ and $\bar w$, 
and a constant term which can be treated as the renormalization of 
the D4-brane tension. 
Here, we ignore these contraction terms assuming that 
they are already absorbed by the redefinition of these constants. 
After some algebra, the interaction terms of our concern in \eqref{w-4pt-a1} are summarized as 
\begin{align}
 &
 \frac{\cc}{2 \mw^2} 
 \left[\aw^\dag_{\dot\alpha i} (\tau^I)^{\dot\alpha}{}_{\dot\beta} \aw^{\dot\beta i} 
  - \awb^\dag_{\dot\alpha i} (\tau^I)^{\dot\alpha}{}_{\dot\beta} \awb^{\dot\beta i}\right]^2 
 + \frac{\cc}{2\mw^2} 
 \left[\aw^\dag_{\dot\alpha i} (\tau^M)^{i}{}_{j} \aw^{\dot \alpha j}
  - \awb^\dag_{\dot\alpha i} (\tau^M)^{i}{}_{j} \awb^{\dot \alpha j}\right]^2 
\notag\\&\quad
 + \frac{2\cc}{\mw^2} (\aw^\dag_{{\dot\alpha} i} \aw^{{\dot\alpha} i}) 
 (\awb^\dag_{{\dot\beta} j} \awb^{{\dot\beta} j}) 
 - \frac{2\cc}{\mw^2} 
 (\aw^\dag_{{\dot\alpha} i} \aw^{{\dot\beta} j}) 
 (\awb^\dag_{{\dot\beta} j} \awb^{{\dot\alpha} i})  \ . 
\label{w-4pt-a2}
\end{align}
Now, we unify $\aw$ and $\awb$ to $\awu$, a doublet of the creation and annihilation operators, as  
\begin{align}
 \awu^{\alpha i p} &= (\aw^{\alpha i} , \awb^{\alpha i}) \ , 
&
 \awu^{\alpha i, p=1} &= \aw^{\alpha i} \ , 
& 
 \awu^{\alpha i, p=2} &= \awb^{\alpha i} 
 = \epsilon^{{\dot\alpha} {\dot\beta}} \epsilon^{ij} \awb_{{\dot\beta} j} \ . 
\label{awu}
\end{align}
Then, \eqref{w-4pt-a2} is expressed as 
\begin{align}
 V 
 &\sim 
 V_1 + V_2 \ , 
\label{V12}
\\
 V_1 
 &= 
 8 \rc (\sw^I)^2 
 + 8 \rc (\swb^I)^2
 - 2 \rc (\hat I^M)^2 
 - 2 \rc (\hat J^I)^2 \ , 
\label{V1}
\\
 V_2 
 &= 
 - \rc (\awu^\dag_{\alpha i p} \awu^{\alpha i q})
 (\awu^\dag_{\beta j q} \awu^{\beta j p})
 + \rc (\awu^\dag_{\alpha i p} \awu^{\alpha i p})
 (\awu^\dag_{\beta j q} \awu^{\beta j q}) 
\notag\\
 &= 
 - 2 \rc (\hat K^P)^2 
 + \frac{\rc}{2} \left(N_w + N_{\bar w}\right)^2 \ , 
\label{V2}
\end{align}
where ``$\sim$'' implies that the equivalence is only for 
the matrix element for the excited states \eqref{EState}, 
and up to the contraction terms which can be absorbed by the renormalization. 
The coupling constant $\rc$ is defined by $\rc = \frac{\cc}{M^2}$. 
The operators $\sw$, $\swb$, $\hat I$, $\hat J$ and $\hat K^P$ (where $P=1,2,3$)  
are the spin operator for the excitations of $w$, 
spin operator for $\bar w$, 
isospin operator, total spin operator, 
and the generator of the new $SU(2)$ which rotate the doublet of \eqref{awu}, respectively, 
and are defined by 
\begin{align}
 \sw^I 
 &= 
 \frac{1}{4} \aw^\dag_{\dot\alpha i} (\tau^I)^{\dot\alpha}{}_{\dot\beta} \aw^{\dot\beta i} \ , 
\\
 \swb^I 
 &=  
 \frac{1}{4} \awb^\dag_{\dot\alpha i} (\tau^I)^{\dot\alpha}{}_{\dot\beta} \awb^{\dot\beta i} \ , 
\\
 \hat I^M 
 &=  
 \frac{1}{4} \awu^\dag_{\dot\alpha i p} (\tau^M)^{i}{}_{j} \awu^{\dot \alpha j p} \ , 
\\
 \hat J^I 
 &= 
 \frac{1}{4} \awu^\dag_{\dot\alpha i p} (\tau^I)^{\dot\alpha}{}_{\dot\beta} \awu^{\dot\beta i p} \ , 
\\
 \hat K^P 
 &= 
 \frac{1}{4} \awu^\dag_{\dot\alpha i p} (\tau^P)^{p}{}_{q} \awu^{\dot \alpha i q} \ . 
\end{align}
Hereafter, we will refer to these $SU(2)$ symmetries as 
$SU(2)_w$, $SU(2)_{\bar w}$, $SU(2)_I$, $SU(2)_J$ and $SU(2)_K$. 
Squares of the generators of each symmetry give the quadratic Casimir operator of the symmetry. 
Although all of $(\sw^I)^2$, $(\swb^I)^2$, $(\hat I^M)^2$, $(\hat J^I)^2$ and $(\hat K^P)^2$ 
cannot be diagonalized simultaneously in general, 
their eigenvalues for their irreducible representation 
are given 
\begin{align}
 & s(s+1) \ , 
&
 & \bar s (\bar s + 1) \ , 
&
 & I(I+1) \ , 
&
 & J(J+1) \ , 
&
 & K(K+1) \ , 
\end{align}
where each of $s$, $\bar s$, $I$, $J$ and $K$ is integer or half-integer. 
It should be noted that the spin and isospin of the excitations of 
either $w$ or $\bar w$ satisfy 
\begin{align}
 \left[\aw^\dag_{\dot\alpha i} (\tau^I)^{\dot\alpha}{}_{\dot\beta} \aw^{\dot\beta i}\right]^2
 &= 
 \left[\aw^\dag_{\dot\alpha i} (\tau^M)^{i}{}_{j} \aw^{\dot \alpha j}\right]^2 \ , 
\\
 \left[\awb^\dag_{\dot\alpha i} (\tau^I)^{\dot\alpha}{}_{\dot\beta} \awb^{\dot\beta i}\right]^2
 &= 
 \left[\awb^\dag_{\dot\alpha i} (\tau^M)^{i}{}_{j} \awb^{\dot \alpha j}\right]^2 \ , 
\end{align}
for $A=1$, and hence the states can be specified without using 
the isospin of either $w$ or $\bar w$. 
Both the square of spin and that of isospin of $w$ excitations are given by $s(s+1)$, 
and those of $\bar w$ are both given by $\bar s(\bar s +1)$.

\subsection{Excited states with $I\neq J$ or $I=J=5/2$}

Now, we diagonalize the interaction term \eqref{V12}-\eqref{V2}. 
Although the quadratic Casimir operators $(\hat I^M)^2$ and $(\hat J^I)^2$ 
can simultaneously diagonalized with any of other quadratic Casimir operators, 
$(\sw^I)^2$ and $(\hat K^P)^2$ cannot be diagonalized simultaneously in general. 

For $N_w = 4$ and $N_{\bar w} = 1$, 
they can be simultaneously diagonalized if $I\neq J$ or $I=J=5/2$. 
This can be understood as follows. 
The excited states with $N_w = 4$ and $N_{\bar w} = 1$ 
are the eigenstates of $\hat K^3$ with the same eigenvalue $3/2$, 
which can be decomposed into the irreducible representations of $SU(2)_K$ with $K=3/2$ and $K=5/2$. 
Here, the excited state is constructed by 5 creation operators $\awu^\dag$, 
where each $\awu^\dag$ is in the fundamental representations of $SU(2)_I$, $SU(2)_J$ and $SU(2)_K$. 
Then, $K=5/2$ is the totally symmetric representation of $SU(2)_K$. 

Since $\awu^\dag$ is the bosonic operator, 
the states must be totally symmetric under exchange of 2 $\awu^\dag$, 
or equivalently, under the exchange of all 3 indices in 2 $\awu^\dag$. 
If $K=5/2$, or equivalently, $SU(2)_K$ is in the totally symmetric representation, 
the combination of $SU(2)_I$ and $SU(2)_J$ must give a totally symmetric representation, 
which is possible only if the representations of $SU(2)_I$ and $SU(2)_J$ 
have the same symmetry under the exchange of indices. 
This implies that $K=5/2$ is possible only for $I=J$. 
Therefore, the irreducible representation of $SU(2)_I$ and $SU(2)_J$ with $I\neq J$ 
must have $K=3/2$ and is the eigenstate of quadratic Casimir of $SU(2)_K$. 

Now, we consider a direct product of states with fixed $s$ and fixed $\bar s$. 
This state diagonalizes $(\sw^I)^2$, and 
can be decomposed into the irreducible representations of $SU(2)_I$ and $SU(2)_J$. 
If a state have $I\neq J$ in these irreducible representations, 
it must have $K=3/2$ and also diagonalizes $(\hat K^P)^2$. 
Therefore, $(\sw^I)^2$ and $(\hat K^P)^2$ can be simultaneously diagonalized for $I\neq J$. 

For $I=J=5/2$, the state is totally symmetric under the exchange of 
any 2 indices of $SU(2)_I$ and $SU(2)_J$. 
Since the state must be totally symmetric under the exchange of $\awu^\dag$, 
$SU(2)_K$ must be in the totally symmetric representation, and have $K=5/2$. 
Since $I=J=5/2$ can be obtained only from $s=2$, 
the state with $I=J=K=5/2$ also diagonalizes $(\sw^I)^2$. 
Therefore, $(\sw^I)^2$ and $(\hat K^P)^2$ can be simultaneously diagonalized for $I=J=5/2$. 

Since all terms in the interaction term \eqref{V12}-\eqref{V2} can be simultaneously diagonalized 
for $I\neq J$ or $I=J=5/2$, 
the eigenvalues can simply be calculated as 
\begin{align}
 \langle V \rangle 
 &= 
 \rc\Bigl[ 8 s(s+1) + 8 \bar s (\bar s+1) - 2 I(I+1) - 2 J(J+1) 
\notag\\&\qquad\quad
 - 2 K(K+1) + \frac{1}{2} \left(N_w + N_{\bar w}\right)^2 \Bigr] \ . 
\end{align}
Since the states have only 1 excitation of $\bar w$, they always have $\bar s = 1/2$. 
The parameter $K$ is also determined by $I$ and $J$; 
\begin{align}
 K &= 
 \frac{3}{2}  & \text{for}\quad& I\neq J \ , 
\\
 K &= 
 \frac{5}{2}  & \text{for}\quad& I=J=\frac{5}{2} \ . 
\end{align}
Since $s$ and $\bar s$ give norms of both spin and isospin of $w$ and $\bar w$, respectively, 
the total isospin $I$ and total spin $J$ for $I\neq J$ must be 
\begin{align}
 I &= s \pm \frac{1}{2} \ , 
& 
 J &= s \mp \frac{1}{2} \ , 
\end{align}
where the relative sign for $\pm$ in $I$ and $\mp$ in $J$ is fixed by the condition $I\neq J$. 
Thus the eigenstates can be specified by $(I,J)$ in this case. 
The result of the eigenvalues of the linear perturbation \eqref{V12}-\eqref{V2} 
are shown in Table~\ref{table:IneqJ}. 

\begin{table}
\renewcommand{\arraystretch}{1.2}
\begin{center}
\begin{tabular}{|c|ccccc|}\hline
$(I,J)$ & $(1/2,3/2)$ & $(3/2,1/2)$ & $(3/2,5/2)$ & $(5/2,3/2)$ & $(5/2,5/2)$ \\\hline
$\langle V \rangle / \rc$ & 18 & 18 & 34 & 34 & 14 \\\hline
\end{tabular}
\caption{%
The eigenvalues of $V$ for the excited states with $N_X=0$, $N_w = 4$, $N_{\bar w} = 1$, and 
$I\neq J$ or $I=J=5/2$.}
\label{table:IneqJ}
\end{center}
\end{table}

%%%%%%%%%%%%%%%%%%%%%%%%%%%%%%%%%%%%%%%%%%%%%%%%%%%%%%%%%%%%%%%%%%%%%%%%%%%%%%%%
%%%%%%%%%%%%%%%%%%%%%%%%%%%%%%%%%%%%%%%%%%%%%%%%%%%%%%%%%%%%%%%%%%%%%%%%%%%%%%%%

\subsection{Excited states with $I=J=1/2$}

We consider the excited state with $I=J=1/2$. 
In this case, $(\sw^I)^2$ and $(\hat K^P)^2$ cannot be diagonalized simultaneously, 
and hence, we have to diagonalize the total interaction term \eqref{V12}. 
The eigenstate is different from those of $V_1$ or those of $V_2$, 
but given by linear combinations of those of $V_1$ or $V_2$. 
Here, we first consider the eigenstate of $V_1$ and 
calculate the matrix elements of $V_1$ and $V_2$ for the eigenstate. 
Then, we diagonalize \eqref{V12}. 

The eigenstate of $V_1$ \eqref{V1} can be obtained 
from the eigenstates of $(\sw^I)^2$. 
Direct products of an irreducible representation of $SU(2)_w$ and that of $SU(2)_{\bar w}$ 
can be decomposed into irreducible representations of $SU(2)_J$, 
which are nothing but the eigenstates of $V_1$. 
The states with $I=J=1/2$ can be obtained from those with $s=0$ and $s=1$. 
We first consider $s=0$ and $s=1$ states with only 4 excitations of $w$, 
which are given by 
\begin{align}
 |s=0\rangle 
 &= 
 [\aw^\dag\aw^\dag][\aw^\dag\aw^\dag] |0\rangle \ , 
\\
 |s=1\rangle 
 &= 
 [\aw^\dag\aw^\dag]\{\aw^\dag\aw^\dag\} | 0 \rangle \ , 
\end{align}
respectively, where the bracket $[\cdot]$ and the brace $\{\cdot\}$ 
stand for the symmetric and anti-symmetric combination for 
both spin and isospin indices, respectively, or more specifically, 
\begin{align}
 [\aw^\dag \aw^\dag] 
 &= 
 \left(\epsilon^{\alpha \beta} \epsilon^{ij} 
 \aw^\dag_{\dot\alpha i} \aw^\dag_{\dot\beta j}\right) \ , 
&
 \{\aw^\dag \aw^\dag\}
 &= 
 2 \left(\alpha^\dag_{\dot\alpha i} \alpha^\dag_{\dot\beta j} 
 + \alpha^\dag_{\dot\alpha j} \alpha^\dag_{\dot\beta i}\right) \ , 
\end{align}
and similarly for those with some of $\aw^\dag$ are replaced by $\awb^\dag$ or else. 
The excited states with $N_w=4$ and $N_{\bar w}=1$ are constructed by 
multiplying $\awb^\dag$ to these state. 
For $s=0$, the state with $I=J=1/2$ is obtained just by multiplying $\awb^\dag$ to $|s=0\rangle$. 
For $s=1$, one of spin indices and one of flavor indices in $|s=1\rangle$ 
should be contracted with those of $\awb^\dag$. 
Thus, we obtain $I=J=1/2$ states $|(I,J,s)\rangle$ as 
\begin{align}
 |(\tfrac12,\tfrac12,0)\rangle 
 &= 
 [\aw^\dag\aw^\dag][\aw^\dag\aw^\dag] \awb^{\dag\,\dot\alpha i} |0\rangle \ , 
\label{1s0}
\\
 |(\tfrac12,\tfrac12,1)\rangle 
 &= 
 2 [\aw^\dag\aw^\dag]
 \left(\aw^\dag_{\dot\alpha i} [\aw^\dag \awb^\dag] + 
 (\aw^\dag \awb^\dag\aw^\dag)_{\dot\alpha i}\right) |0\rangle \ , 
\label{1s1}
\end{align}
where $(\alpha^\dag \bar \alpha^\dag \alpha^\dag)_{\alpha i}$ stands for 
\begin{equation}
 (\aw^\dag \awb^\dag \aw^\dag)_{\dot\alpha i} 
 \equiv
 \aw^\dag_{\dot\alpha j} \awb^{\dag\,\dot\beta j} \aw^\dag_{\dot\beta i} 
 = \aw^\dag_{\dot\alpha i} [\aw^\dag \awb^\dag] 
 - \frac{1}{2} \awb^\dag_{\dot\alpha i} [\aw^\dag \aw^\dag] \ , 
\end{equation}
and 
\begin{align}
 [\aw^\dag \awb^\dag] 
 &= 
 \left(\epsilon^{\alpha \beta} \epsilon^{ij} 
 \aw^\dag_{\dot\alpha i} \awb^\dag_{\dot\beta j}\right) 
 = 
 \aw^\dag_{\dot\alpha i} \awb^{\dag\,\dot\alpha i} \ . 
\end{align}
Then, $ |(\tfrac12,\tfrac12,1)\rangle$ can also be expressed as 
\begin{equation}
 |(\tfrac12,\tfrac12,1)\rangle 
 = 
 \left(4 [\aw^\dag\aw^\dag][\aw^\dag\awb^\dag] \aw^\dag_{\dot\alpha i} 
 - [\aw^\dag\aw^\dag][\aw^\dag\aw^\dag] \awb^{\dag}_{\dot\alpha i} \right)|0\rangle \ . 
\end{equation}

Now, we consider the matrix elements of $V_1$ and $V_2$. 
Since \eqref{1s0} and \eqref{1s1} diagonalize $V_1$, 
$V_1$ does not have off-diagonal components for these states 
and the diagonal components are given by 
\begin{equation}
 \rc\left[ 8 s(s+1) + 8 \bar s (\bar s+1) - 2 I(I+1) - 2 J(J+1) \right] \ . 
\end{equation}
The other terms of the interaction terms, $V_2$ can be rewritten as 
\begin{align}
 V_2 
 &= 
 - \rc (\awu^\dag_{\dot\alpha i p} \awu^{\dot\alpha i q})
 (\awu^\dag_{\dot\beta j q} \awu^{\dot\beta j p})
 + \rc (\awu^\dag_{\dot\alpha i p} \awu^{\dot\alpha i p})
 (\awu^\dag_{\dot\beta j q} \awu^{\dot\beta j q}) 
\notag\\
 &= 
 - 2 \rc (\aw^\dag_{\dot\alpha i} \awb^\dag_{\dot\beta j} \awb^{\dot\alpha i} \aw^{\dot\beta j}) 
 - \rc (\aw^\dag_{\dot\alpha i} \aw^\dag_{\dot\beta j} \aw^{\dot\alpha i} \aw^{\dot\beta j}) 
 - \rc (\awb^\dag_{\dot\alpha i} \awb^\dag_{\dot\beta j} \awb^{\dot\alpha i} \awb^{\dot\beta j}) 
\notag\\&\quad
 - 2 \rc (\awu^\dag_{\dot\alpha i p} \awu^{\dot\alpha i p})
 + \rc (\awu^\dag_{\dot\alpha i p} \awu^{\dot\alpha i p})
 (\awu^\dag_{\dot\beta j q} \awu^{\dot\beta j q})  
 \ , 
\end{align}
where the first term exchanges a pair of $\aw^\dag$ and $\awb^\dag$, for example, 
\begin{equation}
 (\aw^\dag_{\dot\gamma k} \awb^\dag_{\dot\delta l} \awb^{\dot\gamma k} \aw^{\dot\delta l}) 
 \aw^\dag_{\dot\alpha i} \awb^\dag_{\dot\beta j} 
 =  \awb^\dag_{\dot\alpha i} \aw^\dag_{\dot\beta j} \ . 
\end{equation}

The second term exchanges a pair of $\aw^\dag$, but the state must be totally 
symmetric under the exchange of any pair of $\aw$, and hence, 
just counts the number of pairs of $\aw^\dag$. 
The third term is the exchange of 2 $\awb^\dag$, which vanishes for $N_{\bar w}=1$. 
It should be noted that the third term has the same form 
as the 0-th order Hamiltonian, but it must not be absorbed by 
the renormalization of mass as it has not be ignored 
in the calculation for $I\neq J$. 
Then, $V_2$ is now expressed as 
\begin{equation}
 V_2 =  - 2 \rc (\aw^\dag_{\dot\alpha i} \awb^\dag_{\dot\beta j} \awb^{\dot\alpha i} \aw^{\dot\beta j}) 
 + \rc \left(2 N_w N_{\bar w} - N_w - N_{\bar w}\right) \ . 
\end{equation}

Now, it is straightforward to calculate the matrix elements of $V$. 
The matrix elements of $V_1$ is calculated as 
\begin{align}
 V_1  |(\tfrac12,\tfrac12,0)\rangle 
 &= 3 \rc |(\tfrac12,\tfrac12,0)\rangle \ ,  
\\
 V_1  |(\tfrac12,\tfrac12,1)\rangle 
 &= 19 \rc |(\tfrac12,\tfrac12,1)\rangle  \ , 
\end{align}
and those of $V_2$ is obtained as 
\begin{align}
 V_2 |(\tfrac12,\tfrac12,0)\rangle 
 &= 
 \rc |(\tfrac12,\tfrac12,0)\rangle - 2 \rc |(\tfrac12,\tfrac12,1)\rangle \ , 
\\
 V_2 |(\tfrac12,\tfrac12,1)\rangle 
 &= 
 - \rc |(\tfrac12,\tfrac12,1)\rangle 
 - 12 \rc |(\tfrac12,\tfrac12,0)\rangle \ . 
\end{align}
Then, by diagonalizing the sum of the above, 
the eigenvalues of \eqref{V12} are calculated as 
\begin{equation}
 \left(11 \pm \sqrt{73}\right) \rc \ . 
\end{equation}

%%%%%%%%%%%%%%%%%%%%%%%%%%%%%%%%%%%%%%%%%%%%%%%%%%%%%%%%%%%%%%%%%%%%%%%%%%%%%%%%
%%%%%%%%%%%%%%%%%%%%%%%%%%%%%%%%%%%%%%%%%%%%%%%%%%%%%%%%%%%%%%%%%%%%%%%%%%%%%%%%

\subsection{Excited states with $I=J=3/2$} 

We consider the excited states with $I=J=3/2$. 
The interaction term $V$ of \eqref{V12} can be diagonalized 
in a similar fashion to the case of $I=J=1/2$. 
We first consider the eigenstates of $(\sw^I)^2$, 
calculate the matrix elements of $V_1$ and $V_2$, 
and diagonalize the matrix representation of $V$. 

The excited states with $I=J=3/2$ are obtained from 
the states with $s=1$ and $s=2$. 
The $s=1$ and $s=2$ states with $N_w=4$ and $N_{\bar w}=0$ 
are given by 
\begin{align}
 |s=1\rangle 
 &= 
 [\aw^\dag\aw^\dag]\{\aw^\dag\aw^\dag\} | 0 \rangle \ , 
\\
 |s=2\rangle 
 &= 
 \{\aw^\dag\aw^\dag \aw^\dag\aw^\dag\} | 0 \rangle \ , 
\end{align}
respectively. 
By multiplying $\awb^\dag$ and taking appropriate combination for the indices, 
$I=J=3/2$ states with $N_w=4$ and $N_{\bar w}=1$ are obtained as 
\begin{align}
 |(\tfrac32,\tfrac32,1)\rangle 
 &= 
 [\aw^\dag\aw^\dag]\{\aw^\dag\aw^\dag\awb^\dag\} |0\rangle \ , 
\label{3s1}
\\
 |(\tfrac32,\tfrac32,2)\rangle 
 &= 
 2
 \left( [\aw^\dag\awb^\dag]\{\aw^\dag\aw^\dag\aw^\dag\} 
 + 3\{(\aw^\dag \awb^\dag\aw^\dag)\aw^\dag\aw^\dag\right) |0\rangle 
\notag\\
 &= 
 \left( 8 [\aw^\dag\awb^\dag]\{\aw^\dag\aw^\dag\aw^\dag\}  
 - 3 [\aw^\dag\aw^\dag]\{\aw^\dag\aw^\dag\awb^\dag\}\right) |0\rangle \ , 
\label{3s2}
\end{align}
Now, it is straightforward to calculate the matrix elements of $V$. 
The matrix elements of $V_1$ is obtained as 
\begin{align}
 V_1 |(\tfrac32,\tfrac32,1)\rangle &= 7 \rc |(\tfrac32,\tfrac32,1)\rangle \ , 
\\
 V_1 |(\tfrac32,\tfrac32,2)\rangle &= 39 \rc |(\tfrac32,\tfrac32,2)\rangle  \ , 
\end{align}
and those of $V_2$ are 
\begin{align}
 V_2 |(\tfrac32,\tfrac32,1)\rangle  
 &= - \frac{5}{2} \rc |(\tfrac32,\tfrac32,1)\rangle 
 - \frac{1}{2} \rc |(\tfrac32,\tfrac32,2)\rangle \ , 
\\
 V_2 |(\tfrac32,\tfrac32,2)\rangle 
 &= 
 \frac{5}{2} \rc |(\tfrac32,\tfrac32,2)\rangle 
 - \frac{75}{2} \rc |(\tfrac32,\tfrac32,1)\rangle \ . 
\end{align}
Then, the eigenvalues of \eqref{V12} is calculated as 
\begin{equation}
 4 \rc \ , \qquad 42 \rc \ . 
\end{equation}

\begin{table}
\renewcommand{\arraystretch}{1.2}
\begin{center}
\begin{tabular}{|c|cccc|}\hline
$(I,J,N)$ & $(\tfrac12,\tfrac12,3)$ & $(\tfrac32,\tfrac32,3)$ 
& $(\tfrac12,\tfrac12,5)$ & $(\tfrac32,\tfrac32,5)$ \\\hline
$\langle V \rangle / \rc$ & 0 & 12 & $11 \pm \sqrt{73}$ & $4\ ,\ 42$ \\\hline
\end{tabular}
\caption{%
The eigenvalues of $V$ for the excited states with $N_X=0$, $N_w = 3$ and $N_{\bar w} = 0$, 
and those with $N_w = 4$ and $N_{\bar w} = 1$ in the cases of $I=J=1/2$ and $I=J=3/2$. 
Here, $N$ is the total number of the excitations, $N = N_w + N_{\bar w}$. }
\label{table:I=J}
\end{center}
\end{table}

%%%%%%%%%%%%%%%%%%%%%%%%%%%%%%%%%%%%%%%%%%%%%%%%%%%%%%%%%%%%%%%%%%%%%%%%%%%%%%%%
%%%%%%%%%%%%%%%%%%%%%%%%%%%%%%%%%%%%%%%%%%%%%%%%%%%%%%%%%%%%%%%%%%%%%%%%%%%%%%%%

\subsection{Results on Baryon Resonances}

Here, we summarize the result for the baryon resonances. 
The baryon spectrum for the excited states in our model 
is given by the following expression; 
\begin{equation}
 M_\text{baryon} = M_\text{D4} + \mw (N_w + N_{\bar w}) 
 + \langle V \rangle  \ , 
\end{equation}
where $\langle V \rangle$ is the eigenvalues of $V$, 
which are summarized in Tables~\ref{table:IneqJ}~and~\ref{table:I=J}. 
Since $w$ and $\bar w$ is invariant (even) under the parity transformation, 
all our excited states have the same parity as the nucleon and $\Delta$, 
namely, parity is even. 

Our mass formula has 3 parameters. 
We fit the parameters $\rc$ and $M_\text{D4} + 3 \mw$ with 
the experimental data of the lowest energy states of $N$ and $\Delta$, 
which are 938 [MeV] and 1232 [MeV], respectively. 
The other parameter $\mw$ is fixed by identifying 
the excited state of $N$ at 1440 [MeV] to 
the lowest excited state in our result, 
which has with $I=J=1/2$ and $\langle V \rangle = 11-\sqrt{73}$. 
Table~\ref{table:bres} shows our results and experimental data from Particle Data Group\cite{Tanabashi:2018oca}.%
\footnote{%
	In the experiments \cite{Tanabashi:2018oca}, many baryon resonances have been observed including the ones which we cannot find the corresponding states in our model.
	They may be described by the excitations of the modes which are neglected in our action \eqref{action}. 
	Stringy higher spin modes may need to be employed too.
 }
Here, the parameters are determined as 
\begin{align}
 M_\text{D4} &=  717\,\text{[MeV]} \ , 
& 
 \mw &= 221\,\text{[MeV]} \ , 
& 
 \rc &= 24.5\,\text{[MeV]} \ . 
\end{align}

\begin{table}
\begin{center}
\begin{tabular}{|c|r@{}l|c|}\hline
$N(I=1/2)$ & \multicolumn{2}{c|}{PDG} & Ours 
\\\hline\hline
$J=1/2$ & 938&(****) & (938) 
\\
 & 1440&(****) & (1440) 
\\
 & 1710&(***) & 
\\
 & 1880&(***) & 1859 
\\
 & 2100&(*) & 
\\\hline
$J = 3/2$ & 1720&(****) & 1821 
\\
 & 1900&(***) & 
\\
 & 2040&(*) & 
\\\hline
 $J = 5/2$ & 1680&(****) & 
\\
 & 1860&(**) & 
\\
 & 2000&(**) & 
\\\hline
\end{tabular}
\hspace{8pt}
\begin{tabular}{|c|r@{}l|c|}\hline
$\Delta(I=3/2)$ & \multicolumn{2}{c|}{PDG} & Ours 
\\\hline\hline
$J=1/2$ & 1750&(*) & 
\\
 & 1910&(****) & 1821 
\\
 & & &
\\
 & & &
\\
 & & &
\\\hline
$J = 3/2$ & 1232&(****) & (1232) 
\\
 & 1600&(***) & 1478  
\\
 & 1920&(***) & 2409  
\\\hline
 $J = 5/2$ & 1905&(****) & 
\\
 & 2000&(**) & 2213 
\\
 & & &
\\\hline
\end{tabular}

\vspace{5mm}

\begin{tabular}{|c|c|c|}\hline
$I = 5/2$ & PDG & Ours 
\\\hline\hline
$J = 3/2$ & -- & 2213 \\\hline
$J=5/2$ & -- & 1723
\\\hline
\end{tabular}

\caption{%
Our results for the baryon resonance. Numbers in the parenthesis $(\cdot)$ in our results 
are used to fit and trivially agrees with the experimental data \cite{Tanabashi:2018oca}.
The number of $*$ denotes the experimental observation status. 
} 
\label{table:bres}
\end{center}
\end{table}

%%%%%%%%%%%%%%%%%%%%%%%%%%%%%%%%%%%%%%%%%%%%%%%%%%%%%%%%%%%%%%%%%%%%%%%%%%%%%%%%
%%%%%%%%%%%%%%%%%%%%%%%%%%%%%%%%%%%%%%%%%%%%%%%%%%%%%%%%%%%%%%%%%%%%%%%%%%%%%%%%
%%%%%%%%%%%%%%%%%%%%%%%%%%%%%%%%%%%%%%%%%%%%%%%%%%%%%%%%%%%%%%%%%%%%%%%%%%%%%%%%

\section{Nuclear spectra: magic numbers and shell model}
\label{sec:MagicNumber}

Nuclear physics has a long history, and the nuclear shell model provides a fundamental basis for the spectral properties of atomic nuclei.
The magic numbers with which the atomic nuclei are stable are one of the most
important properties which the nuclear shell model explains.
In this section, we point out a similarity between the nuclear matrix model
and the nuclear shell model, and derive the magic numbers $2, 8$ and $20$ 
for doubly magic nuclei with $Z=N$
from the nuclear matrix model. Here ``doubly magic'' commonly means that both of the number of protons $Z$ and
the number of neutrons $N$ are magic numbers. 
The fact that
the holographic description of QCD derives 
a fundamental property of atomic nuclei illustrates a surprising connection 
between string theory and nuclear physics.

\subsubsection*{Derivation of the magic numbers}

To explain the derivation of the magic numbers, 
we start with the case of $N_c=1$, and consider $N_c=3$ later. 
This is because when $N_c=1$, 
the baryons are not composite particles but 
are identical to the quarks, and 
each baryon corresponds to each single $w$ operator.  
Thus, the magic number would simply be defined for each flavor of quarks, which are the flavor-classified excitations of $w$. 
(The magic number for real nuclei is defined for either protons or neutrons separately.)
For $N_c\neq 1$, the proton and neutron numbers 
are not identical to the up and down quark numbers, and their relation has to be taken into consideration, as we will see later. 
Besides, it is more convenient to consider the magic number for the baryon number (or equivalently, mass number) $A$, first, rather than the magic number for the flavors of quarks.

Let us look at the spectral property of the ground state of the nuclear matrix model. 
The ground state is given by \eqref{psi0},
which is provided with the $SU(A)$ single operator $\OS$.
For $N_c = 1$ and $A\leq 2N_f$, as explained in \eqref{naivesinglet}, the ground state is
\begin{align}
|\psi_0\rangle
= \OS |0\rangle \ , \quad \mbox{with} \quad
\OS \equiv  
  \epsilon_{a_1\cdots a_A} \aw_{\dot \alpha_1 i_1}^{\dag\ a_1} 
  \cdots \aw_{\dot \alpha_A i_A}^{\dag\ a_A} \, .
  \label{defgS}
\end{align}
Here, we emphasize that
$\OS$ is a singlet in $SU(A)$ symmetry inherent to the nuclear matrix model,
and the constraint that the state need to be a singlet forces the explicit form 
of the state to be contracted with the totally antisymmetric tensor $\epsilon_{a_1\cdots a_A}$, resultantly ensures a fermionic appearance of
the operator $\alpha^\dagger$ although it is a bosonic operator.
So, at this baryon state, $\alpha^\dagger$ behaves as if it is a quark,
although it has no color index --- it rather has the baryon index which
is in the fundamental representation of $SU(A)$.

As briefly explained at the end of Sec.~\ref{sec:Model},
when $A>2N_f$, 
the number of 
the species of the ``quark" operator $\alpha^\dagger$, $2N_f$, is not
enough to satisfy the singlet condition for $SU(A)$: 
the operator $\OS$ defined in \eqref{defgS} vanishes.
Therefore 
we have to introduce creation operators $a^\dagger$ of $X^I$ 
so that we have additional species $[a^\dagger \alpha^\dagger]$
in addition to the original $\alpha^\dagger$. Both of these are
in the fundamental representation of $SU(A)$, so
we can use 
all of these to form a singlet operator.
Then, we are allowed to have a state formed by
\begin{align}
\OS = 
  \epsilon_{a_1\cdots a_A} \aw^{\dag a_1} 
  \cdots \aw^{\dag a_{2N_f}}
  \, \beta^{\dag a_{2N_f+1}} 
  \cdots \beta^{\dag a_A} \, ,
\end{align}
where $\beta^\dagger \equiv ((a^I)_{a}^{\; b}\alpha_b)^\dagger$
which is the second option in \eqref{OX}, and we have omitted the spin and the flavor indices.
The existence of the creation operator $a^\dagger$ contributes to the
energy of the state additionally, so, this derives that $A=2N_f$ is
the magic number for the baryon number at $N_c=1$. 
In this case, the wave function is singlet for all symmetries of 
baryon, spin and isospin, which is the closed shell structure in our model.\footnote{
As we have seen in Sec.~\ref{ssec:4pt-w}, 
nuclei in the singlet representation of flavor symmetry are more stable for arbitrary $A$ due to the $w^4$-interactions. 
However, we ignore the contribution of this interaction in the current argument for simplicity.
}

Then what is the next magic number? When we further consider a larger $A$,
we consume all possible $\beta$ with different $a^I (I=1,2,3)$.
For $A>8N_f$, we have to introduce a new combination with two $a^\dagger$'s
acting on a single $\alpha^\dagger$,
\begin{align}
\gamma^\dagger \equiv
((a^I)_{a}^{\; b}(a^J)_{b}^{\; c}\alpha_c)^\dagger\, ,
\end{align} 
and form a ground state with
\begin{align}
\OS = \epsilon_{a_1\cdots a_A} \aw^{\dag a_1} \cdots \aw^{\dag a_{2N_f}}
\beta^{\dag a_{2N_f+1}} \cdots \beta^{\dag a_{8N_f}}
\gamma^{\dag 8_{N_f+1}} \cdots \gamma^{\dag A} \, .
\label{stbg}
\end{align} 
Therefore, the second magic number for the baryon number is $A = (1+3)\times 2 N_f = 8 N_f$. 

To find the next magic number, we need to count the number of possible
species of $\gamma$. We note that the operator $a^{I\,\, b}_{\,a}$
is a matrix with $I=1,2,3$, so possible operators quadratic in
$a^I$ are $[a^I,a^J]$ and $\{a^I,a^J\}$, thus sum up to $3\times 3=9$
species of $\gamma$.  
However, it is highly expected that $[a^I,a^J]$ acquires higher energy than $\{ a^I, a^J\}$ due to the interaction term $[X^I,X^J]^2$ in the Lagrangian.
Therefore $[a^I,a^J]$ cannot be used for constructing the ground state, leaving us with just $\{ a^I, a^J\}$.
So the allowed number of $\gamma$ is $6$. This means that the state of the form
\eqref{stbg} is allowed up to $A=(1+3+6)\times 2N_f = 20 N_f$, meaning that
the next magic number is $20N_f$.

We play this game and find the magic number formula for the baryon number at $N_c=1$, 
\begin{align}
A_{\rm magic}^{(m)} =2 N_f \sum_{j=0}^m {}_{j+2}C_2 = \frac{N_f}{3}(m+1)(m+2)(m+3) \, .
\label{magic-baryon}
\end{align} 
The sequence goes as $A_{\rm magic}=2N_f,8N_f,20N_f,40N_f,\cdots$.

The argument above can be applied for each flavor separately. 
For example, we focus on the ``quark'' with flavor $i=1$, and then, 
spin indices of $\aw^{\dag\,a}_{\dot \alpha, i=1}$ 
in \eqref{defgS} must be totally antisymmetric, 
and hence only two $\aw^{\dag\,a}_{\dot \alpha, i=1}$ can be put in \eqref{defgS}. 
In order to consider the state with more ``quark'' operators with $i=1$, 
the operator $\beta_{i=1}$ should be introduces as we discussed above. 
Thus, the magic numbers $2,8,20$ are derived for each flavor, separately. 
We find that this coincides with the known magic numbers in nuclear physics at $N_c=3$, 
for the states 
at which the spin-orbit interaction is unimportant, $2,8,20$ for each of flavors. 

We now notice that the magic number for the baryon number \eqref{magic-baryon},
 $A_{\rm magic}=2N_f,8N_f,20N_f,40N_f,\cdots$,
 is a special case of  the ``doubly magic'', in which the quark numbers for each flavor are given by the same magic number, $2,8,20,40,\cdots$.

So far, we have considered the case of $N_c=1$, in which 
the nuclei are identical to the quarks. 
For $N_c \neq 1$, however, we have to take the relation 
between nuclei and quarks into consideration. 
In general, nuclei are in a different representation of the flavor symmetry 
than that of quarks, the fundamental representation. 
However, for the real nuclei, $N_c=3$ and $N_f=2$, ignoring $\Delta$, 
the proton and neutron have the isospin $1/2$, in the same representation to the up and down quarks. 
The relation between the proton and neutron numbers to 
those for quarks is still non-trivial, but much simpler than other general cases. 
Here, we consider the case of realistic nuclei, $N_c=3$ and $N_f=2$.

As we will see below, there is always a closed shell configuration of (a flavor of) $w$ 
for a given magic number of proton or neutron. 
Thus, our model reproduces the magic numbers. 
However, neither number of proton nor that of neutron may be in 
the magic numbers even if the number of a flavor of $w$ 
in the operator $\OS$ is in the magic numbers.%
\footnote{%
Such cases would be unstable because of the linear or higher order corrections 
in the perturbative expansion, since they have non-trivial representations of 
spin and flavor symmetries, but we will not pursue it in this paper. 
A simple example is the $A=2$ state made from 6 $\alpha_u^\dagger$'s that forms a closed shell but is in $(J,I)=(0,3)$ and unstable.
}
The closed shell configurations in our model 
may include those which do not correspond to the magic number of proton or neutron, in general.  
On the other hand, if the both numbers of up and down quarks are in magic numbers, 
and if the configuration has a consistent charge as a bound state of proton and neutron, 
there is one-to-one correspondence between 
the magic numbers of proton and neutron to those of quarks, for small $A$. 

Now, we derive the magic numbers $2,8,20$,
for doubly magic nuclei in the case of $N_c=3$ and $N_f=2$. 
The term ``doubly magic'' means that both of
the proton number $Z$ and the neutron number $N$ take one of the magic numbers, $2,8,20, \cdots$, independently.

It is convenient to start with the magic number for the baryon number. 
For the baryon number, the argument is completely parallel to that for $N_c=1$, 
since it is irrelevant from the relation of proton and neutron numbers to quarks. 
The ground state wave function takes the same form to that for $N_c=1$ but now has 
$N_c$ singlet operators $\OS$; 
\begin{equation}
 |\psi_0\rangle = \OS^{N_c} |0\rangle \ , 
\end{equation}
where $N_c$ of $\OS$ acting on $|0\rangle$ can be different for general states, 
but must be same if the baryon number is the magic number since 
$\OS$ for the closed shell configuration for a magic number is unique. 
The magic number is derived as $A_{\rm magic}=2N_f,8N_f,20N_f,40N_f,\cdots$,
in which the quark numbers for each flavor are given by the same magic number, $2,8,20,40,\cdots$. 
As the state is singlet if the up and down quarks take the same magic number, 
it is identified to that with the same magic number of protons and neutrons. 
Thus we have derived the magic numbers $2,8,20,\cdots$ for $N_c=3$ and $N_f=2$ 
with the same number of protons and neutrons. 

It is easy to demonstrate this for explicit cases. Let us consider first the case $(Z,N)=(2,2)$ and thus $A=Z+N=4$, 
the helium nucleus.
We find that the ground state wave function \eqref{naivesinglet} holds as a consistent state which is singlet under $SU(A)$ and $SU(N_c)$,
\begin{align}
|\psi_0\rangle
= (\OS)^3 |0\rangle \ , \quad \mbox{with} \quad
\OS \equiv  
  \epsilon_{a_1\cdots a_4} 
  \aw_{ \uparrow u}^{\dag\ a_1} 
  \aw_{ \downarrow u}^{\dag\ a_2} 
  \aw_{ \uparrow d}^{\dag\ a_3} 
  \aw_{ \downarrow d}^{\dag\ a_4} \, .
  \label{defgS23}
\end{align}
That is, the state is given by $\OS$, and each $\OS$ is made of two ``up quarks'' and two ``down quarks.'' 
If one wants to add more baryons, then one needs $\beta^\dagger$ 
which costs more energy than $\alpha^\dagger$, as explained. Therefore,
the case $(Z,N)=(2,2)$ is magic, doubly magic.%
\footnote{% 
Note that three $\beta$'s should be introduced for the lowest energy state of the next state ($A=5$), 
which carry three spatial index $I$'s in our model. 
This is not expected in the conventional shell model 
where the nucleon in the next energy level has orbital angular momentum $\ell = 1$. 
We expect that the interaction terms $V$, which we have not taken into account, 
would make such a state stabler than others in our model. 
We leave this problem for future studies.}

This logic can be applied to all the cases with $Z=N=2,8,20,\cdots$.

Then what about general doubly magic case, $Z\neq N$? 
In the following, we provide a plausible argument that
the doubly magic number can be
reasonably understood in the nuclear matrix model, though the derivation is not logically complete.
Let us consider the case $(Z,N)=(2,8)$.\footnote{This combination is not
realistic, but this is the first nontrivial doubly magic case with $Z\neq N$. Other cases can be
discussed in the same manner.}
The ``quark'' for each flavor in $\OS$ take the closed shell configuration if 
the number of $\aw^\dag$ for the flavor is in the magic number $2,8,20,\cdots$. 
If all ``quarks'' in $\OS$ take the closed shell configuration, 
the baryon number must be sum of two magic numbers, $A=4,10,16,22,28\cdots$. 
As $A=4$ corresponds to the case of $N=Z=2$, the first example with $N\neq Z$ is $A=10$. 
For this case, the operator $\OS$ for closed shell configurations are 
\begin{align}
\OS_1 \equiv  
  \epsilon_{a_1\cdots a_{10}} 
  (\aw_{ \uparrow u}^{\dag\ a_1} 
  \aw_{ \downarrow u}^{\dag\ a_2}) 
  (\aw_{ \uparrow d}^{\dag\ a_3} 
  \aw_{ \downarrow d}^{\dag\ a_4}
  \beta_{ \uparrow d}^{\dag\ a_5} 
  \beta_{ \downarrow d}^{\dag\ a_6}
  \cdots
  \beta_{ \uparrow d}^{\dag\ a_9} 
  \beta_{ \downarrow d}^{\dag\ a_{10}})
   \, ,
\\
\OS_2 \equiv  
  \epsilon_{a_1\cdots a_{10}} 
  (\aw_{ \uparrow u}^{\dag\ a_1} 
  \aw_{ \downarrow u}^{\dag\ a_2}
  \beta_{ \uparrow u}^{\dag\ a_3} 
  \beta_{ \downarrow u}^{\dag\ a_4}
  \cdots
  \beta_{ \uparrow u}^{\dag\ a_7} 
  \beta_{ \downarrow u}^{\dag\ a_8})
  (\aw_{ \uparrow d}^{\dag\ a_9} 
  \aw_{ \downarrow d}^{\dag\ a_{10}})
   \, .
\end{align}
The state of the doubly magic nucleus is given by 
\begin{align}
&
 (\OS_1)^3 |0\rangle \ ,
&&
 (\OS_1)^2\OS_2 |0\rangle \ ,
&&
 \OS_1(\OS_2)^2 |0\rangle \ ,
&&
 (\OS_2)^3 |0\rangle \ . 
\end{align}
In the states above, $ (\OS_1)^3 |0\rangle$ and $(\OS_2)^3 |0\rangle$ 
cannot be constructed from protons and neutrons since they have too much down (up) quarks.%
\footnote{%
These states correspond to nuclei which contain $\Delta$, 
as they contain too many up or down quarks for nuclei only with protons and neutrons. 
They will have larger energy than those only with protons and neutrons 
if we take into account the first order perturbation \eqref{V}. 
A similar argument can be applied for other cases of magic numbers $2,8,20$. 
For larger magic numbers, more precise analysis on the isospin representation 
and perturbative corrections is needed to exclude similar unnecessary states, which is left for future studies. 
} 
If the nucleus has more neutrons than protons, 
the state is given by 
\begin{align}
|\psi_0\rangle
= (\OS_1)^2\OS_2 |0\rangle \ , 
\label{28}
\end{align}
which reproduces the quark numbers for
the pair of 2 protons and 8 neutrons. 
Each of $\OS_1$ and $\OS_2$ is made of 10 creation operators and is singlet of $SU(A)$ with $A=10$,
and consists of 2 up (down) quarks and 8 down (up) quarks, each of which is the magic number.
So, if one adds one more baryon, it needs $\gamma^\dagger$ which costs more energy than $\beta^\dagger$.
The stability of the state \eqref{28} of the nuclear matrix model is expected to show the doubly magic number $(Z,N)=(2,8)$.
However, there remains a subtlety. The state with 2 protons and 8 neutrons
is not really equivalent to \eqref{28}, because the state should have the isospin 3 and so 
is in the irreducible representation of the isospin. 
Therefore \eqref{28}, which is not an
eigenstate of the total isospin operator, cannot be the expected proton-neutron state.
The state \eqref{28} should be mixed with some other states and will become the
proton-neutron state with the doubly magic numbers $(Z,N)=(2,8)$.%

Since we have already constructed the nucleon states in Sec.~\ref{sssec:nucleon}, 
we can compose the nuclear states from the nucleon operators $\mathcal N$ which is defined as 
\begin{align}
 \mathcal N^{abc}_{\dot \alpha i,(N)} 
 = \mathcal F_{\dot \alpha i}^{\dot \beta_1 j_1 \dot \beta_2 j_2 \dot \beta_3 j_3}
 \left((\AX^{\dag})^N \aw^{\dag}\right)_{\dot \beta_1 j_1}^a 
 \left((\AX^{\dag})^N \aw^{\dag}\right)_{\dot \beta_2 j_2}^b
 \left((\AX^{\dag})^N \aw^{\dag}\right)_{\dot \beta_3 j_3}^c \ , 
 \label{ON}
\end{align}
where $\mathcal F$ gives $I=J=\frac{1}{2}$ states from 
three fundamental representations of spin and isospin, 
and indices $a$, $b$ and $c$ are totally symmetric under the permutation. 
The state with $(Z,N) = (2,8)$, for example, is given by 
\begin{align}
 |\psi_0\rangle 
 &= 
 \epsilon_{a_1\cdots a_{10}}
 \epsilon_{b_1\cdots b_{10}}
 \epsilon_{c_1\cdots c_{10}}
 \mathcal N^{a_1 b_1 c_1}_{ \uparrow p, (0)}
 \mathcal N^{a_2 b_2 c_2}_{ \downarrow p, (0)}
 \notag\\
 &\quad\times
 \mathcal N^{a_3 b_3 c_3}_{ \uparrow n, (0)} 
 \mathcal N^{a_4 b_4 c_4}_{ \downarrow n, (0)}
 \mathcal N^{a_5 b_5 c_5}_{ \uparrow n, (1)} 
 \mathcal N^{a_6 b_6 c_6}_{ \downarrow n, (1)}
 \cdots
 \mathcal N^{a_9 b_9 c_9}_{ \uparrow n, (1)} 
 \mathcal N^{a_{10} b_{10} c_{10}}_{ \downarrow n, (1)} |0\rangle
   \, . 
\end{align}
This state actually contains \eqref{28}, and the relation to the nucleon state is clear, 
though the relation to the magic number is not obvious. 

In order to see the relation to the magic number, 
we try to construct the nuclear state of 3 protons without $X^I$ excitations, 
and show that it is impossible. 
Each proton consist of 2 up quarks and 1 down quark, 
and the state consists of 6 up quarks and 3 down quarks takes the following form;  
\begin{align}
 |\psi_0\rangle 
 = 
 \epsilon_{a_1 a_2 a_3}
 \epsilon_{b_1 b_2 b_3}
 \epsilon_{c_1 c_2 c_3}
 (\aw_{\dot \alpha_1,u}^{\dag\ a_1} 
 \aw_{\dot \alpha_2,u}^{\dag\ a_2} 
 \aw_{\dot \alpha_3,d}^{\dag\ a_3}) 
 (\aw_{\dot \beta_1,u}^{\dag\ b_1} 
 \aw_{\dot \beta_2,u}^{\dag\ b_2} 
 \aw_{\dot \beta_3,d}^{\dag\ b_3}) 
 (\aw_{\dot \gamma_1,u}^{\dag\ c_1} 
 \aw_{\dot \gamma_2,u}^{\dag\ c_2} 
 \aw_{\dot \gamma_3,d}^{\dag\ c_3}) \ , 
 \label{3pq}
\end{align}
since 3 up quarks, $\aw_{\dot \alpha,u}^{\dag\ a}$, cannot form the antisymmetric combination. 
This state cannot be constructed from the nucleon (proton) operator \eqref{ON}, 
since 3 baryon indices of $\mathcal N^{abc}_{\dot \alpha ,p}$ must be contracted 
with different $\epsilon$ symbols, but one of 3 ``nucleon'' in \eqref{3pq} consists of 3 up quarks. 
If we try to construct the state from 3 proton operators without $X^I$, $\mathcal N^{abc}_{\dot \alpha ,p,(0)}$, 
the state is expressed as 
\begin{equation}
 |\psi_0\rangle 
 = 
 \epsilon_{a_1 a_2 a_3}
 \epsilon_{b_1 b_2 b_3}
 \epsilon_{c_1 c_2 c_3}
 \mathcal N^{a_1b_1c_1}_{\dot \alpha_1 ,p,(0)}
 \mathcal N^{a_2b_2c_2}_{\dot \alpha_2 ,p,(0)}
 \mathcal N^{a_3b_3c_3}_{\dot \alpha_3 ,p,(0)} \ ,  
\end{equation}
it vanishes identically because it contains at least one antisymmetric combination of 3 up quarks, 
\begin{equation}
 \epsilon_{a_1 a_2 a_3}
 \epsilon_{b_1 b_2 b_3}
 \epsilon_{c_1 c_2 c_3}
 (\aw_{\dot \alpha_1,u}^{\dag\ a_1} 
 \aw_{\dot \alpha_2,u}^{\dag\ a_2} 
 \aw_{\dot \alpha_3,u}^{\dag\ a_3}) = 0 \ . 
\end{equation}
Thus the nuclear state with 3 protons cannot be formed without $X^I$, 
implying that the first magic number for proton is 2. 
It should be noted that the state in terms of $\mathcal N$ would not be the eigenstate of the Hamiltonian, 
but should be mixed with other state to form an irreducible representation. 
The state may have contribution from $\Delta$ as nuclei may contain it with small probability. 
The precise decomposition for the irreducible representation of the isospin will be left for
our future study.

We have shown 
the magic numbers $2,8,20,\cdots$ for doubly magic nuclei with $Z=N$, from the nuclear matrix model
with $N_c=3$ and $N_f=2$.
The derivation of the more general magic numbers, either for proton or neutron, 
has difficulties such as the decomposition of the irreducible representation of the isospin, 
and is more involved. 
A complete derivation will be reported in future communications.

%%%%%%%%%%%%%%%%%

\subsubsection*{Relation to the nuclear shell model}

At this stage, we are ready to explain a similarity to the nuclear shell model.
In the shell model, nucleons with isospin $=1/2$ and spin $=1/2$ 
are put in a trapping potential, and those nucleons occupy discretized 
energy levels given by the potential. 
Since the potential is assumed to be rotationally symmetric,
the energy levels are specified by the angular momentum $L$ of the nucleon,
as $L(L+1)$, as well as the radial excitation number. 
The magic numbers of the shell model originate in the difference of the largest 
angular momentum of the constituent nucleon, the shells. 
The wave function of the nucleon for the angular part is given explicitly by the wave function 
\begin{align}
\psi_L\equiv x^{I_1} \cdots x^{I_L} \ , 
\label{anguwave}
\end{align} 
which is the eigen function of the angular momentum operator $L^2$,
where $x^I$ is the spatial coordinate $(I=1,2,3)$. 

The wave function \eqref{anguwave}
has the same form as our $\OX$ used for the construction of the ground state 
of the nuclear matrix model, thus the coincidence with the nuclear shell model
is evident, although the principles about
how the wave function itself is constructed is quite different from
the nuclear shell model.% 
\footnote{%
For spin and flavor, our model is described by those of quarks, not of nuclei, 
and hence, has a problem when only one of the number of the up or down quark is in the magic number, 
while the nuclear shell model is described only in terms of nuclei. 
}
It should be emphasized that
in our model the fundamental structure of the ground state is provided not only 
by the constraint \eqref{NPhys} for the operator $w$ (which has a similarity to the
nuclear shell model), but also by the $SU(A)$ gauge invariance constraint (which is
not present in any other nuclear physics models and is
a direct consequence of the D-brane picture).%
\footnote{%
These two constraints together form the universal constraint \eqref{Const}.
}

Note that the nuclear states we constructed are not bound states of excited nucleons.
Although the insertion of $\OX_n(a^\dag)$, \eqref{OX}, in the nuclear states may look exciting a quark
momentum inside a single baryon so that it forms an excited baryon, it is not the case,
because the $(X^I)_a{}^b$ operator always acts on two different baryons at the same time.
The latter can be understood from the fact that $X^I$ becomes trivial for the single baryon case
$A=1$. Therefore, it is natural to think of the nuclear states we constructed
as bound states of ordinary nucleons, rather than those of excited nucleons.

The magic numbers in the shell model are based on the two ingredients: the shape of the 
trapping potential, and the spin-orbit coupling. Basically these two notions can emerge 
from the nuclear matrix model, but it is beyond the scope of this paper. We make a brief comment
on the spin-orbit coupling here. 
In our model, the orbital angular momentum operator $\hat L^I$ 
and spin angular momentum operator $\hat S^I$ are given by 
\begin{align}
 \hat L^I_\text{total} 
 &= 
 \,:\!\tr X^J \PX^K \!: \epsilon^{IJK} 
 = \AX^{\dag\,J}_A \AX^K_A \epsilon^{IJK} \ , 
\\
 \hat S^I_\text{total} 
 &= 
 \frac{1}{2}\!:\!w \tau^I \pi \!:\! + \frac{1}{2}\!:\!\bar w \tau^I \bar \pi \!:\, 
 = 
 \frac{1}{2} \aw^{\dag\,a}_{\dot \alpha i} (\tau^I)^{\dot \alpha}{}_{\dot\beta} \aw_a^{\dot \beta i}
 + \frac{1}{2} \bar \aw^{\dag\,\dot \alpha i}_a 
    (\tau^I)_{\dot \alpha}{}^{\dot\beta} \bar\aw^a_{\dot \beta i} \ . 
\end{align}
The operators above are those for whole of the system, 
and are given in the form of the trace of $A\times A$ matrices of the baryon $U(A)$, 
\begin{align}
 \hat L^{I\,b}_a
 &= 
 \frac{1}{2} \AX^{\dag\,Jc}_a \AX^{Kb}_c \epsilon^{IJK} 
 + \frac{1}{2} \AX^{\dag\,Jb}_c \AX^{Kc}_a \epsilon^{IJK} \ , 
\label{Lmatrix}
\\
 \hat S^{I\,b}_a 
 &= 
 \frac{1}{2} \aw^{\dag\,b}_{\dot \alpha i} (\tau^I)^{\dot \alpha}{}_{\dot\beta} \aw_a^{\dot \beta i}
 + \frac{1}{2} \bar \aw^{\dag\,\dot \alpha i}_a 
    (\tau^I)_{\dot \alpha}{}^{\dot\beta} \bar\aw^b_{\dot \beta i} \ . 
\label{Smatrix}
\end{align}
The operators for each baryon (nucleon) would be defined as the diagonal components of the matrices above. 

If we look at the interaction term \eqref{V} of the nuclear matrix model,
one of the most important parts has the form
\begin{align}
V_{XXww} \sim
\cc \epsilon^{IJK} X^J X^K  \bar w \tau^I w \ , 
\label{ADHMV}
\end{align} 
which appears to be similar to the spin-orbit coupling. The form $\bar w \tau^I w$ measures the
spin of the nucleons.  Furthermore, the angular orbital operator has the structure of 
being quadratic in $X$ and made by $\epsilon^{IJK}$, which is shared by the above. 
In terms of the creation and annihilation operators, 
the interaction term above is expressed as 
\begin{align}
 V_{XXww} 
 &\sim 
 \frac{\cc}{\mw^2}
 \epsilon^{IJK} \AX_A^{\dag\,J} \AX_B^{K} f^{AB}{}_C\, 
 \aw^{\dag\,a}_{\dot \alpha i} (\tau^I)^{\dot \alpha}{}_{\dot\beta} (t^C)_a{}^b \aw^{\dot \beta i}_b 
\notag\\
 &\sim 
 \frac{\cc}{\mw^2}
 \epsilon^{IJK} \left(\AX_a^{\dag\,J\,c} \AX_c^{K\,b} - \AX_c^{\dag\,J\,b} \AX_a^{K\,c}\right)\, 
 \aw^{\dag\,a}_{\dot \alpha i} (\tau^I)^{\dot \alpha}{}_{\dot\beta} \aw^{\dot \beta i}_b \ . 
\label{XXww1st}
\end{align}
Here, we have picked up only the terms with the same numbers of the creation and annihilation operators, 
which are relevant to the first order perturbation.  
Nevertheless, the complete equivalence between the ADHM interaction \eqref{ADHMV} and
the spin-orbit interaction is not seen at this stage. 
The orbital part, which is related to $X^I$, in the interaction \eqref{XXww1st}, 
is the antisymmetric part of the baryon $SU(A)$, 
while the orbital angular momentum \eqref{Lmatrix} is the symmetric part of $SU(A)$. 
It should be noted that the spin-orbit coupling in the nuclear shell model 
is not that of the total spin and total orbital angular momentum, but that in each nucleon, 
and possibly have a different form than $\tr \hat L^I \hat S^I$. 

Another candidate of the spin-orbit coupling in our model 
would be from the second order perturbation. 
We consider the correction term \eqref{XXww2}. 
Now we take the contraction of a pair of $w$ and a pair of $X^I$, 
\begin{align}
   \epsilon^{IJM} \AX_A^J f^{AE}{}_C\,
   \epsilon^{KLM} \AX_B^{\dag\,K} f^{BE}{}_D\, 
   \aw^{\dag\,a}_{\dot \alpha i} (\tau^I\tau^L)^{\dot \alpha}{}_{\dot\beta} 
    (t^C t^D)_a{}^b \aw^{\dot \beta i}_b \ . 
\label{XXww2nd}
\end{align}
Products of the two generators of $SU(N)$ are expressed as 
\begin{align}
 t^A t^B = \frac{1}{2N} \delta^{AB} + \frac{1}{2} \left(i f^{ABC} + d^{ABC}\right) T^C \ , 
\end{align}
where $f$ and $d$ are totally antisymmetric and symmetric in all indices, respectively. 
By picking up the symmetric part of the baryon $SU(A)$ 
and the antisymmetric part of the spin $SU(2)$ in \eqref{XXww2nd}, 
we obtain 
\begin{align}
 \epsilon^{IJK} \AX_A^{\dag\,J} \AX_A^{K} \, 
 \aw^{\dag\,a}_{\dot \alpha i} (\tau^I)^{\dot \alpha}{}_{\dot\beta} \aw^{\dot \beta i}_a 
 + \epsilon^{IJK} \AX_A^{\dag\,J} \AX_B^{K} d^{ABC} \, 
 \aw^{\dag\,a}_{\dot \alpha i} (\tau^I)^{\dot \alpha}{}_{\dot\beta} (t^C)_a{}^b \aw^{\dot \beta i}_b  \ , 
\end{align}
where the first term is nothing but $\tr \hat L^I \hat S^I$. 
The second term would be relevant to the spin-orbit coupling 
since it will pick up some components of $\tr \hat L^I \hat S^I$ and 
also symmetric under the exchange of the $SU(A)$ indices $A$ and $B$ of $X^I$. 

It should be noted that many other terms appear when \eqref{XXww2} is rewritten into 
the normal-ordered form, which may give non-negligible contributions. 
We here just point out that (a candidate of) the spin-orbit coupling appears 
in the second order perturbation.

%%%%%%%%%%%%%%%%%%%%%%%%%%%%%%%%%%%%%%%%%%%%%%%%%%%%%%%%%%%%%%%%%%%%%%%%%%%%%%%%
%%%%%%%%%%%%%%%%%%%%%%%%%%%%%%%%%%%%%%%%%%%%%%%%%%%%%%%%%%%%%%%%%%%%%%%%%%%%%%%%
%%%%%%%%%%%%%%%%%%%%%%%%%%%%%%%%%%%%%%%%%%%%%%%%%%%%%%%%%%%%%%%%%%%%%%%%%%%%%%%%

\section{Summary and discussions}

In this paper, we have proposed a new quantization procedure to study 
the nuclear spectra from the nuclear matrix model in holographic QCD. 
We have shown that our procedure gives quantum states which 
agree with nuclear states for small baryon numbers. 
We have calculated the spectra of hyperons. 
Our method gives a mass formula which is very similar to the Gell-Mann--Okubo formula, 
and provides the hyperon spectra in remarkable agreement with experimental data. 
We have also considered the dibaryons and baryon resonance. 
Our procedure provides decent predictions for H-dibaryon and di-Omega. 
The results for baryon resonance also show good agreement with the observation. 
Furthermore, we have shown that the model partially reproduces the magic numbers
for doubly magic nuclei with $Z=N$. 

In our new procedure, the spectra can be calculated as the eigenvalues of the Hamiltonian. 
In this model, the equation of motion of the gauge field gives a constraint. 
We have imposed this constraint on the eigenstates. 
Then, the eigenstates which satisfy the constraint 
have turned out to have appropriate properties as states of nuclei. 
We have introduced an effective potential for nucleons and 
consider the perturbative expansion. 
The states at the leading order have an appropriate statistics 
and charges and thus are identified as nuclear states. 

In this paper, we have focused on the nuclei with small baryon numbers. 
The spectra for small baryon numbers can be analyzed mostly only from the physics of $w$, 
and this $w$ field is related to the quarks in the nuclei. 
We first have explored the allowed states with small baryon numbers in our model, 
and shown the agreement with the known nuclear states. 
We further studied the three-flavor case and looked at the hyperon spectrum. 
We have assumed that the $SU(3)$ flavor symmetry is broken only by the $w$-mass difference, 
and that the interaction terms have no additional source of the flavor symmetry breaking. 
Then, our model provides a mass formula for hyperons which 
is very similar to the Gell-Mann--Okubo formula. 
Our result for hyperons shows significant agreement with the experimental data. 

At the leading order of the perturbation of the model, 
our quantization gives the result that two dibaryon states share the same smallest mass. 
One of them is identified to deuteron, but the other has spin $J=3$. 
The dibaryon with $I=0$, $J=3$ is known as $D_{03}$, 
which has a larger mass than the deuteron. 
In order to solve this problem, we have considered 
the second order perturbation and found a correction term 
which gives spin dependence to the mass. 
Taking this correction term into account, 
we further have made predictions for the masses of H-dibaryon and di-Omega.  
The predictions are at the reasonable energy scale. 

We have also investigated the baryon resonance. 
The model naturally includes the excitations of pairs of $w$ and $\bar w$, and they  
are identified with the excitations inside the baryons. 
It should be noted that $w$ and $\bar w$ 
are naively interpreted as the quark and anti-quark, respectively, 
but their pairs may also represents something different in our model, 
for example, the internal motion of quarks. 
Here, we considered the excitations of a pair, 
and calculated the spectra of the baryon resonance. 
Although the differences to the experimental data are larger than those for hyperons, 
our results correspond to some of the resonance states in the experiments. 

In this paper, we have not studied details on the physics of $X^I$
which describes the motion of baryons in the nuclei. 
However, even from the basic part of our procedure, 
it can be seen that our model reproduces, at least, small magic numbers for nuclei with $Z=N$. 
Our model also has a similarity to the nuclear shell model. 
In order to show more complete agreement with the shell model, 
our model should have an effect similar to the spin-orbit coupling. 
In this paper, we have just shown a few possible origins of the spin-orbit coupling in our model. 
More detailed analyses on effects of $X^I$ will be reported in future communications. 

We assumed that the baryons are sufficiently close to each other 
for the matrix model Hamiltonian to be valid. 
The potential of \eqref{action} yields attractive force even for a large separation,
but in reality the interactions between the baryons should disappear, so this means that 
the action \eqref{action} is not suitable for describing baryons at a large separation. 
Actually, when the baryon vertices are sufficiently separated, we need to consider the contributions of all the fields including the ones which we have neglected in \eqref{action}.
See \cite{Banks:1996vh} for a derivation of interactions between separated D-branes.
Unfortunately the computation of this potential is difficult in our model and 
we have not studied it in this paper. 
Calculation of more detailed potential and studies on the stability of 
the bound state is left for future studies.

One of the interesting results of this paper is that the nuclear states naturally appear 
as a result of the D-brane interaction and the quantum charge constraint. 
The latter in fact puts our analysis different from the
previous analyses based on the soliton picture of baryons in holographic QCD 
(Ref.~\cite{Hata:2007mb} and the subsequent study).
In addition, this paper is focused on the sector with small baryon numbers,
while in nuclear physics the sectors with large atomic numbers are of equal importance.
In \cite{Hashimoto:2011nm} the large atomic number limit was considered in the same model
with ignoring the effect of $w$ fields. Here in this paper we treated $w$ as the fundamental 
excitations, so the unified treatment of \cite{Hashimoto:2011nm} and the present quantization
method is necessary for further investigation of heavy nuclei.

In principle, our quantization method can be applied also to the infinite system
with a finite chemical potential for the baryon number. In the soliton approach of the
holographic QCD, the quark matter is made of a solitonic lattice. Our approach may 
be an alternative, but based on the same D-brane picture. It would be interesting to
use our quantization method for the quark/baryon matter and see how the phase diagram 
can be derived, while checking any consistency with the soliton approach.
We leave these issues to the future work.

\section*{Acknowledgments}

The authors would like to thank Atushi Hosaka, Kaori Kaki and Akihiko Matsuyama 
for valuable discussions.
The work of K.H. is supported 
in part by JSPS KAKENHI Grants No.~JP15H03658, 
No.~JP15K13483, and No.~JP17H06462.
The work of Y.M. is supported in part by
the Ministry of Science and Technology, R.O.C.\ 
(project no. 107-2119-M-002-031-MY3)
and by National Taiwan University (project no.\ 105R8700-2).
The work of Y.M. is also supported 
in part by JSPS KAKENHI Grants No.~JP17H06462.
The work of T.~M. is supported in part by Grant-in-Aid for Young Scientists B (No. 15K17643) from JSPS.

%%%%%%%%%%%%%%%%%%%%%%%%%%%%%%%%%%%%%%%%%%%%%%%%%%%%%%%%%%%%%%%%%%%%%%%%%%%%%%%%
%%%%%%%%%%%%%%%%%%%%%%%%%%%%%%%%%%%%%%%%%%%%%%%%%%%%%%%%%%%%%%%%%%%%%%%%%%%%%%%%

\appendix

\section*{Appendix}

%%%%%%%%%%%%%%%%%%%%%%%%%%%%%%%%%%%%%%%%%%%%%%%%%%%%%%%%%%%%%%%%%%%%%%%%%%%%%%%%
%%%%%%%%%%%%%%%%%%%%%%%%%%%%%%%%%%%%%%%%%%%%%%%%%%%%%%%%%%%%%%%%%%%%%%%%%%%%%%%%
%%%%%%%%%%%%%%%%%%%%%%%%%%%%%%%%%%%%%%%%%%%%%%%%%%%%%%%%%%%%%%%%%%%%%%%%%%%%%%%%

\section{Generalization of the model for dibaryons}
\label{app:gen}

Here, we consider a modification of our model. 
In Sec.~\ref{sec:Dibaryon}, the second order perturbation is studied 
as an candidate of correction terms. 
Here, we consider the most general 4-point interactions of $w$, 
and show that the possible interaction terms 
would be quadratic Casimir operators of the symmetries and 
square of the number of the excitations of $w$. 
Then, we will study the dibaryon spectrum by using a simple ansatz | 
the interaction terms are given by \eqref{VFCasimir} and \eqref{VS}, 
and the coupling constants are independent of $A$. 

%%%%%%%%%%%%%%%%%%%%%%%%%%%%%%%%%%%%%%%%%%%%%%%%%%%%%%%%%%%%%%%%%%%%%%%%%%%%%%%%
%%%%%%%%%%%%%%%%%%%%%%%%%%%%%%%%%%%%%%%%%%%%%%%%%%%%%%%%%%%%%%%%%%%%%%%%%%%%%%%%

\subsection{Most general 4-point interactions of $w$}

In this section, we consider the most general 4-point interactions of $w$ and $\bar w$ 
which are invariant under the baryon $U(A)$, flavor $SU(N_f)$, 
and rotation in 3-dimensional space. 
As $w$ and $\bar w$ are complex scalars 
which are conjugate to each other, 
the 4-point interactions, in general, should take the form of 
\begin{equation}
 w w \bar w \bar w \ . 
\end{equation}
The fields $w$ and $\bar w$ have indices of 
the fundamental representations of the
baryon $U(A)$, flavor $SU(N_f)$ and spin $SU(2)$ symmetries. 
Since the generators in the fundamental representations of $U(N)$ (or $SU(N)$)
can be replaced by the exchange of the indices 
by using the formula \eqref{U(N)} (or \eqref{SU(N)}), 
all 4-point interactions which satisfy the symmetries 
can be obtained simply by contracting the indices with each other. 
Possible combinations for the contraction are 
to contract all 3 indices in the same pairs of $w$ and $\bar w$; 
\begin{equation}
 (w^a_{\alpha i} \bar w^{\alpha i}_a) 
 (w^b_{\beta j} \bar w^{\beta j}_b) \ , 
\end{equation}
or 1 of 3 indices is contracted in the different pair 
than the other 2 indices; 
\begin{align}
& 
 (w^a_{\alpha i} \bar w^{\alpha i}_b) 
 (w^b_{\beta j} \bar w^{\beta j}_a) \ , 
&%\\
& 
 (w^a_{\alpha i} \bar w^{\beta i}_a) 
 (w^b_{\beta j} \bar w^{\alpha j}_b) \ , 
&%\\
& 
 (w^a_{\alpha i} \bar w^{\alpha j}_a) 
 (w^b_{\beta j} \bar w^{\beta i}_b) \ . 
\end{align}

In a similar fashion to the arguments in Sec.~\ref{ssec:4pt-w}, 
the 4-point interaction terms above can be rewritten in terms of 
the generators of the spin $SU(2)$, flavor $SU(N_f)$ and baryon $SU(A)$ symmetries. 
Then, the most general interaction terms are given 
by an linear combination of the following terms; 
\begin{align}
 &\left(w \bar w\right)^2 \ , 
&
 &\left(w \tau^I \bar w\right)^2 \ , 
&
 &\left(w t_f^M \bar w\right)^2 \ , 
&
 &\left(w t^A \bar w\right)^2 \ , 
\end{align}
where $\tau^I$, $t_f^M$ and $t^A$ are the generators 
of the spin $SU(2)$, flavor $SU(N_f)$ and baryon $SU(A)$ symmetries, 
and all indices are contracted in each parenthesis. 
For the physical ground states, 
only the creation and annihilation operators for $w$ are relevant 
at the linear order perturbation of the interaction term above. 
If $w$ has the same mass $\mw$ for all flavors, 
$w$ and $\bar w$ can simply be replaced by $\aw^\dag$ and $\aw$, 
with a factor of $\mw^{-1/2}$. 
Here, the constraint \eqref{Const} implies that 
the physical state must be singlet of the baryon $SU(A)$ symmetry, 
and the quadratic Casimir of baryon $U(A)$ 
gives the square of the overall $U(1)$ charge. 

Therefore, the following 3 interaction terms are the
most general 4-point interactions of $w$ and $\bar w$: 
the square of the number of the excitations of $w$, 
\begin{equation}
 V_N = (\aw^{\dag\, a}_{{\dot \alpha} i} \aw^{{\dot \alpha} i}_a) 
 (\aw^{\dag\, b}_{{\dot \beta} j} \aw^{{\dot \beta} j}_b) \ , 
\label{VN0}
\end{equation}
which simply gives the constant $N_c^2A^2$, 
the quadratic Casimir of spin $SU(2)$, 
\begin{equation}
 V_S = \frac{1}{4} (\aw^{\dag\, a}_{{\dot \alpha} i} (\tau^I)^{\dot \alpha}{}_{\dot \beta} \aw^{{\dot \beta} i}_a) 
 (\aw^{\dag\, b}_{\dot \gamma j} (\tau^I)^{\dot \gamma}{}_{\dot \delta} \aw^{\dot \delta j}_b)  \ , 
\label{VS0}
\end{equation}
and the quadratic Casimir of the flavor $SU(N_f)$ ,
\begin{equation}
 V_F = (\aw^{\dag\, a}_{{\dot \alpha} i} (t_f^M)^i{}_j \aw^{{\dot \alpha} j}_a) 
 (\aw^{\dag\, b}_{{\dot \beta} k} (t_f^M)^k{}_l \aw^{{\dot \beta} l}_b) \ . 
\label{VF0}
\end{equation}

%%%%%%%%%%%%%%%%%%%%%%%%%%%%%%%%%%%%%%%%%%%%%%%%%%%%%%%%%%%%%%%%%%%%%%%%%%%%%%%%
%%%%%%%%%%%%%%%%%%%%%%%%%%%%%%%%%%%%%%%%%%%%%%%%%%%%%%%%%%%%%%%%%%%%%%%%%%%%%%%%

\subsection{Dibaryon spectrum from a simple ansatz}

Here, we consider the dibaryon spectrum by using an alternative ansatz. 
Since the most general 4-point interactions of $w$ give 
the quadratic Casimir operators of the symmetries, 
the simplest ansatz for the mass formula would be 
a linear combination of the Casimir operators. 
The physical states are invariant under the baryon $SU(A)$ symmetry, 
and the Casimir operator of the baryon $SU(A)$ would not have important contribution. 
Hence, we consider a linear combination of the Casimir operators of the flavor and spin symmetries, 
which become $V_F$ and $V_S$ after taking the effect of mass difference $\ms\neq\mw$, 
with arbitrary coupling constants $\rc$ and $\rc'$. 
We assume that the coupling constants are independent of $A$. 
We also introduce an additional constant, which comes from the mass of the baryon D4-branes, 
and assume that it is simply proportional to the number of the baryon D4-branes, $A$. 
Then, the mass formula is given by 
\begin{align}
 M_\text{nucl} 
 &= 
 \tilde{M}_{\rm D4} A + 4\tilde{\lambda} (1-\delta)C_f
 -\left( M_S \, \delta - 2 \tilde{\lambda}\,\delta\left(4A - 3 + \delta - 2\delta A\right)\right) Y
\notag \\
&\quad
+4\tilde{\lambda}\, \delta
\left(I(I+1) -\frac14 Y^2 \right) 
+ \tilde{\lambda}\, \delta^2 \left(2-\frac{1}{A}\right) Y^2
\notag\\&\quad
+  4\tilde \lambda' 
 \bigl[ (1- \delta) J(J+1) 
  + \delta s_I(s_I+1) - \delta \left(1- \delta\right) s_S(s_S + 1) 
 \bigr] 
 \ . 
\label{M_nucl}
\end{align}
It should be noted that H-dibaryon \eqref{H-dib} always has 
a smaller mass than the total mass of 2 hyperons in the $SU(3)$ invariant case. 
As H-dibaryon state is singlet of all the symmetries, 
$V_F$ and $V_S$ vanish, and this H-dibaryon has mass of $2 \tilde M_\text{D4}$ 
in the $SU(3)$ invariant case, namely, for $\ms = \mw$. 
Since hyperons have non-zero $V_F$, or equivalently $V_S$, 
the mass of this H-dibaryon is less than the total mass of 2 hyperons, for $\ms = \mw$.

The result is shown in Table~\ref{table:rhyp-di1}. 
We fit our formula to 8 hyperons, deuteron and $D_{03}$. 
The data of $D_{03}$ is necessary to determine $\rc'$ 
as we discussed in Sec.~\ref{ssec:Dibaryon-Result}. 
The parameters are determined as 
\begin{align}
&\tilde{M}_{\rm D4} = 977\, \mbox{[MeV]}, \quad
M_S = 595\, \mbox{[MeV]}, \quad
\notag\\\
&\tilde{\lambda}= 14.5 \, \mbox{[MeV]}, \quad
\tilde{\lambda}' = 11.8 \, \mbox{[MeV]}, \quad
\delta = 0.374 \, .
\end{align}

\begin{table}[tb]
\begin{center}
  \begin{tabular}{|c|cccc|}
  \hline 
  Octet &  N(939) & $\Lambda$(1116)  & $\Sigma$(1193)  & $\Xi$(1318)   \\ \hline \hline
     GMO \eqref{GMO} &939&1117&1183&1328 \\ \hline
     Our \eqref{M_nucl} & 919 & 1100 & 1178 & 1328 \\ \hline
  \end{tabular}
  
  \vspace{5mm}  

 \begin{tabular}{|c|cccc|}
  \hline 
  Decuplet    &
     $\Delta$(1232) & 
     $\Sigma^*$(1385) & 
     $\Xi^*$(1533) & 
     $\Omega$(1672)  \\ \hline \hline
     GMO \eqref{GMO} &1238&1383&1528&1673 \\ \hline
     Our \eqref{M_nucl} & 1234 & 1377 & 1526 & 1682 \\ \hline
  \end{tabular}
  \vspace{5mm}  

 \begin{tabular}{|c|ccc|ccc|cc|}
  \hline 
  Dibaryon    &
     $D$(1876) & 
     $D_{12}$(2160?) & 
     $D_{03}$(2370) & 
     $D_{10}$ & 
     $D_{21}$ & 
     $D_{30}$ & 
     $H$ & 
     $\Omega \Omega$ \\ \hline \hline
     Our \eqref{M_nucl} 
& 
 1896 
&
     2202 
& 
 2370 
& 
     1918
& 
     2245 
& 
     2499
&
 1954 
&
 3292 \\ \hline
  \end{tabular}

  \end{center}
   \caption{A numerical fit of the hyperon and dibaryon spectrum by our formula \eqref{M_nucl}.}
  \label{table:rhyp-di1}
\end{table}

%%%%%%%%%%%%%%%%%%%%%%%%%%%%%%%%%%%%%%%%%%%%%%%%%%%%%%%%%%%%%%%%%%%%%%%%%%%%%%%%
%%%%%%%%%%%%%%%%%%%%%%%%%%%%%%%%%%%%%%%%%%%%%%%%%%%%%%%%%%%%%%%%%%%%%%%%%%%%%%%%
%%%%%%%%%%%%%%%%%%%%%%%%%%%%%%%%%%%%%%%%%%%%%%%%%%%%%%%%%%%%%%%%%%%%%%%%%%%%%%%%

\section{Mass of $X$}\label{app:Mass}

Though the potential $[X^I,X^J]^2$ has flat directions classically, 
it provides a bound state at quantum level. 
In this paper, we approximate the potential by the harmonic potential with mass $\MX$.
When we focus on the $X^I$ part of the model ignoring 
interaction with $w$ and $\bar w$, 
\begin{align}
 H_X &= \frac{1}{2} \tr (\PX^I)^2 
  - 2 \cc \tr \left[X^I , X^J \right]^2 \ , 
\label{H-appB}
\end{align}
then what we mean by the approximation is the following relation
\begin{equation}
- 2 \cc \tr [X^I,X^J]^2  \; \sim \;   \frac{1}{2} \MX^2 \tr (X^I)^2 \ . 
\label{crit0}
\end{equation}

Indeed, it has been studied and discussed in many references how the
commutator squared potential is studied by perturbative expansion around a harmonic potential in diverse Yang-Mills type matrix models 
\cite{Mandal:2009vz,Hotta:1998en, Morita:2010vi, Mandal:2011hb, Kabat:1999hp,Kabat:2000zv,Kabat:2001ve,Iizuka:2001cw,Nishimura:2001sx,Kawai:2002jk,Nishimura:2002va}.%
\footnote{%
In supersymmetric theories, the flat direction of the $[X^I,X^J]^2$ potential is protected  for supersymmetric states and the approximation \eqref{crit0} does not work.
Since supersymmetry is broken in our model, we expect that the approximation \eqref{crit0} is suitable.
} 

There are various possible methods to approximate the commutator  squared potential.
One method would be suitable to calculate a physical quantity, 
but another method would be more appropriate for another quantity. 
These different quantities in the different methods for the same physical observables would converge, if these methods work and we can compute sufficiently higher order of the perturbative expansion.

We do not specify which method we use in our model \eqref{action-o}, and we just (temporary) rewrite the model  as \eqref{H0} by assuming that at least one of the approximation methods works and it provides a suitable mass term.%
\footnote{%
Of course, this is not satisfactory. In order to judge whether one method is reliable, we need to compute various quantities at least several order of the perturbative expansion and see if they converges asymptotically.
This question is beyond our scope of the present paper. 	
}

In order to strengthen our treatment about the mass term, we list and review some of the approximation methods by considering the model \eqref{H-appB} (without the $w$ field)  in Sec.~\ref{B:list}.
You will see that $X^I$ fields indeed obtain a mass term dynamically in these methods.%
\footnote{
One potential confusion is that the obtained values of the mass $m$ of $X^I$ depend on the approximation schemes.
Actually, it does not immediately mean that these schemes are inconsistent, since the masses introduced in these methods are not physical observable.
In Sec.~\ref{B:comment}, we comment on a possible argument on how different mass $\MX$ could give the same answer for physical observables.
}

\vspace{15mm}

\subsection{Dynamical mass of $X^I$ in various approximation methods \label{B:list}}

\subsubsection*{Auxiliary field method}
We can introduce an auxiliary field in our model \eqref{H-appB} and obtain the effective mass 
as the solution of the effective mass in the classical limit. 
The Hamiltonian \eqref{H-appB}  is now expressed as 
\begin{equation}
 H_X = \frac{1}{2} (\PX^I_A)^2 + \frac{1}{2} B^{IJ}_{AB} X^I_A X^J_B 
 - \frac{1}{32\cc} (K_{AB,CD}^{IJ,KL})^{-1} B^{IJ}_{AB} B^{KL}_{CD} \ ,  
\end{equation}
where 
\begin{equation}
 K^{IJ,KL}_{AB,CD} = \epsilon^{IKM} \epsilon^{JLM} f^{ACE} f^{BDE}\ . 
\end{equation}
The equation of motion for the auxiliary field $B$ is given by 
\begin{equation}
 B^{IJ}_{AB} = 8 \cc K^{IJ,KL}_{AB,CD} X^K_C X^L_D \ . 
\end{equation}
We separate $B$ to its expectation value and quantum fluctuation, 
and then, the fluctuation would correspond to the perturbation $V_X$. 
The expectation value is calculated as 
\begin{equation}
 \langle B^{IJ}_{AB}\rangle = 8 \cc K^{IJ,KL}_{AB,CD} \langle X^K_C X^L_D \rangle 
 = \frac{16 A \cc}{3(A^2 -1)} \langle X^K_C X^K_C \rangle \delta^{IJ} \delta^{AB} \ . 
\end{equation}
Here, we have assumed that the state is invariant under rotation and baryon $U(A)$ symmetry 
and only invariant part of $\langle X^I_A X^J_B \rangle$ survives. 
Then, the effective mass is identified as 
\begin{equation}
 \MX^2 = \frac{16 A \cc}{3(A^2 -1)} \langle X^K_C X^K_C \rangle \ . 
\end{equation}
If $X^I$ can be approximated by the harmonic oscillator with the effective mass $\MX$, 
The 2-point function $\langle X^I_A X^J_B \rangle$ is estimated as 
\begin{equation}
 \langle X^I_A X^J_B \rangle = \frac{1}{2m} \delta^{IJ} \delta_{AB} \ , 
\label{<XX>}
\end{equation}
and then, the consistency condition for the effective mass $\MX$ is obtained as 
\begin{equation}
 \MX^3 = 8 A \cc \ . 
\end{equation}

It should be noted that the result depends on the definition of the auxiliary field $B$. 
Alternative definition of the auxiliary field $B$ would be 
\begin{equation}
 H_X = \frac{1}{2} (\PX^I_A)^2 + \frac{1}{2} B_{AB} X^I_A X^I_B 
 - \frac{1}{64\cc} (K_{AB,CD})^{-1} B_{AB} B_{CD} \ ,  
\end{equation}
where 
\begin{equation}
 K_{AB,CD} = f^{ACE} f^{BDE}\ . 
\end{equation}
The expectation value of $B$ is calculated as 
\begin{equation}
 \langle B_{AB}\rangle = 16 \cc K_{AB,CD} \langle X^I_C X^I_D \rangle 
 = \frac{16 A \cc}{(A^2 -1)} \langle X^I_C X^I_C \rangle \delta^{AB} \ . 
\end{equation}
The effective mass is now estimated as 
\begin{equation}
 \MX^3 = 48 A \cc \ . 
\end{equation}

Here, we calculate the expectation value $\langle X^I_A X^J_B \rangle$ 
from the 0-th order Hamiltonian of effective mass $\MX$. 
Since $X^I$ now has interaction with fluctuation of auxiliary field $B$, 
which should be taken into account perturbatively.

\subsubsection*{Large-$D$ and Large-$A$ approximation}

It is known that the potential $[X^I,X^J]^2$ provides 
the mass gap of $X^I$ in a large-$D$ and large-$A$ limit, $A,D \gg 1$, 
where the index $I$ of $X^I$ runs from 1 to $D$ by generalizing the space dimension \cite{ Mandal:2009vz, Hotta:1998en, Morita:2010vi, Mandal:2011hb, Azuma:2014cfa}. 
The procedure is roughly the same as the auxiliary field method.
If we take the large-$D$ and $A$ limit such that $D,A \to \infty$ with a fixed $DA\lambda$, the fluctuation of the auxiliary field introduced in \cite{Hotta:1998en, Mandal:2009vz}  around the expectation value is suppressed by powers of $1/D$ and $1/A^2$. 
The mass gap $\MX$ is calculated as 
\begin{align}
	m^3=&8 D A \lambda \left( 1 +\frac{3}{D}\left( \frac{7 \sqrt{5}}{30}-\frac{9}{32} \right)  +O(1/D^2,1/A^2) \right) \nonumber \\
	=&29.8 \times A \lambda \left(1+O(1/D^2,1/A^2)\right) \ , 
	\label{1/D-NO}
\end{align}
where the number of the spatial dimension is $D=3$ in our case.
See \cite{Mandal:2009vz} for more details.

The model \eqref{H-appB} has been computed numerically in \cite{Aharony:2004ig, Aharony:2005ew,  Kawahara:2007fn, Azuma:2014cfa}, and
we can see good agreements with the large-$D$ expansion.
Although our model does not have so large $D$ but $D=3$, the errors would be $1/D^2 =1/3^2 \sim 10\%$ and the numerical results indeed appear within this error \cite{Azuma:2014cfa} when $A$ is large.

Note that we cannot evaluate the value of mass $m$ directly in the numerical computations. 
	However, many observables calculated by the numerical methods and the large-$D$ expansion agree, and this agreement supports the existence of the mass gap predicted by the large-$D$ expansion.

Although the large-$D$ expansion works well in the model \eqref{H-appB}, what we are really interested in is the model including $w$ and $\bar{w}$ fields and the $ N_c \int \tr A_t$ term \eqref{action-o}.
As we have seen, they strongly constrain the allowed states, and the mass $m$ should be modified for these states.
For evaluating such state dependent modifications, the large-$D$ expansion may not be so convenient.
Besides, we are also interested in the small $A$ cases, 
and the large-$A$ approximation in \eqref{1/D-NO} may not be so reliable.
For these reasons, we explore other methods to approximate the model \eqref{action-o}.

\subsubsection*{Variational principle for test wavefunction}

A good approximation for the wave function 
can be found by using the variational principle. 
Since the ground state has minimal energy, 
the expectation value of the energy $E = \langle H_X \rangle$ 
for any other state than the exact ground state is 
larger than the exact energy of the ground state. 
This implies that a wave function with smaller 
energy expectation value would give better approximation. 
We consider the wave function of the ground state 
for the harmonic oscillator with a frequency $\MX$  as an approximation. 
The frequency $\MX$ should be chosen such that $E$ becomes minimum, 
or equivalently, 
\begin{equation}
	\partial_\MX E = 0 \ . 
\end{equation}
The energy $E$ is calculated as 
\begin{align}
	E 
	&\sim
	\frac{1}{2}\langle \PX^I_A\PX^I_A \rangle 
	- 2 \cc \langle [X^I,X^J]^2 \rangle 
	\notag\\
	&= 
	\frac{3}{4} \left(A^2 - 1\right)\MX  
	+ \frac{6 A (A^2-1)\cc}{\MX^2} \ . 
\end{align}
Then, the frequency $\MX$ is obtained as 
\begin{equation}
	\MX^3 = 16 A \cc \ . 
	\label{m-VP}
\end{equation}

Now, we have found that the wave function for 
the harmonic oscillator with \eqref{m-VP} gives a good approximation 
for the ground state of the Hamiltonian \eqref{H-appB}. 
This wave function will be appropriate for the leading order 
wave function of the perturbative expansion. 
Then, the Hamiltonian should also be expanded accordingly 
with the effective mass \eqref{m-VP}.

\subsubsection*{Mean field approximation}

Here, we treat $X^I_A X^J_B$ as an operator. 
The interaction term $[X^I,X^J]^2$ is the 2-point interaction of the operator $X^I_A X^J_B$. 
We approximate it by interaction to the mean field, 
or equivalently the expectation value $\langle X^I_A X^J_B \rangle$. 
The interaction term is expressed as 
\begin{equation}
- 2 \cc [X^I,X^J]^2 
= 2 \cc K^{IJ,KL}_{AB,CD} X^I_A X^J_B X^K_C X^L_D   \ , 
\end{equation}
and can be expanded as 
\begin{align}
2\cc K^{IJ,KL}_{AB,CD} X^I_A X^J_B X^K_C X^L_D  
&= 
2\cc K^{IJ,KL}_{AB,CD} \langle X^I_A X^J_B \rangle \langle X^K_C X^L_D \rangle  
\notag\\&\quad
+ 4\cc K^{IJ,KL}_{AB,CD} \langle X^I_A X^J_B \rangle \delta (X^K_C X^L_D)  
\notag\\&\quad
+ 2\cc K^{IJ,KL}_{AB,CD} \delta (X^I_A X^J_B) \delta (X^K_C X^L_D)  \ , 
\end{align}
where $\delta (X^I_A X^J_B)$ is the fluctuation around the expectation value; 
\begin{equation}
\delta (X^I_A X^J_B) = X^I_A X^J_B - \langle X^I_A X^J_B \rangle \ . 
\end{equation}
We approximate the interaction term to the linear order of the fluctuation; 
\begin{align}
& 2\cc K^{IJ,KL}_{AB,CD} \langle X^I_A X^J_B \rangle \langle X^K_C X^L_D \rangle  
+ 4\cc K^{IJ,KL}_{AB,CD} \langle X^I_A X^J_B \rangle \delta (X^K_C X^L_D)  
\notag\\
&\qquad= 
4\cc K^{IJ,KL}_{AB,CD} \langle X^I_A X^J_B \rangle X^K_C X^L_D  
- 2\cc K^{IJ,KL}_{AB,CD} \langle X^I_A X^J_B \rangle \langle X^K_C X^L_D \rangle \ . 
\end{align}
Now the interaction to the mean field is nothing but the effective mass $\MX$; 
\begin{equation}
\frac{1}{2} \MX^2 (X^I_A)^2 = 
4\cc K^{IJ,KL}_{AB,CD} \langle X^I_A X^J_B \rangle X^K_C X^L_D \ . 
\end{equation}
Since the expectation value $\langle X^I_A X^J_B \rangle$ 
for the harmonic oscillator is given by \eqref{<XX>}, 
the effective mass $\MX$ is given by 
\begin{equation}
\MX^3 = 16 A \cc \ . 
\end{equation}

This procedure is essentially the same as the auxiliary field method. 
Here, we just expand $X^I_A X^J_B$ around the expectation value, 
but the mean field can also be treated as an auxiliary field.

\subsubsection*{Equivalence for expectation values of potentials}

A simple method to determine the mass $m$ is to identify the equivalence relation \eqref{crit0} 
to an equality for the expectation value 
\begin{equation}
- 2 \cc \left\langle\tr [X^I,X^J]^2\right\rangle = \frac{1}{2} \MX^2 \left\langle\tr (X^I)^2\right\rangle \ , 
\end{equation}
where the expectation values are calculated by using $|\psi^{(0)}\rangle$, 
an eigenstate of the 0-th order Hamiltonian \eqref{HX0}. 
It should be noted that the effective mass $\MX$ depends 
on the state $|\psi^{(0)}\rangle$. 
The state $|\psi^{(0)}\rangle$ is not an unique reference state, 
but should be the physical state under the consideration. 
Therefore, the effective mass $\MX$ depends on the physical state. 

For example, we consider the state $|\psi_0\rangle$ 
which has no excitations of $X^I$. 
Then, the expectation value is calculated as 
\begin{align}
 0 = - \frac{1}{4} \MX + \frac{2 A \cc}{\MX^2}  \ , 
\end{align}
and then, $\MX$ is calculated as 
\begin{equation}
 \MX^3 = 8 A \cc \ . 
\end{equation}

The condition above is defined at the first order of the perturbation. 
As was described in Sec.2, we rewrite the total Hamiltonian for $X$ as
\begin{align}
 H_X &= H_{X0} + V_X \ , 
\label{HX}
\\
 H_{X0} 
&= 
 \frac{1}{2} \tr (\PX^I)^2 + \frac{1}{2} \MX^2 \tr (X^I)^2  \ , 
\label{HX0}
\\
 V_X 
&= 
  - \frac{1}{2} \MX^2 \tr (X^I)^2 
  - 2 \cc \tr \left[X^I , X^J \right]^2 \ . 
\label{VX}
\end{align}
By the condition above, the first order correction to the energy vanish, 
\begin{equation}
 \langle V_X \rangle 
 = 
 - 2 \cc \left\langle\tr [X^I,X^J]^2\right\rangle 
 - \frac{1}{2} \MX^2 \left\langle\tr (X^I)^2\right\rangle = 0 \ . 
\end{equation}
Although it is sufficient just to fix the effective mass at the leading order, 
further tuning of $\MX$ will give better convergence of the perturbation.  
A best criterion at higher order may be to cancel correction terms at each order. 
We separate the ``counter term'' in $V_X$ as 
\begin{equation}
 V_X 
 = 
  - \frac{1}{2} \sum_n \MX^2_n \tr (X^I)^2 
  - 2 \cc \tr \left[X^I , X^J \right]^2 \ , 
\end{equation}
where $\sum_n \MX_n^2 = \MX^2$, and take $\MX_n$ to cancel 
the $n$-th order correction of the perturbative expantion of the energy, $E = \langle H \rangle$. 
Then, the leading order energy is formally fine-tuned to agree with the exact energy 
to arbitrary order.

\subsubsection*{Effective mass from contraction terms}

Another method to determine $m$ in the expression \eqref{HX} 
would be to identify the equivalence relation \eqref{crit0} 
to the relation for normal ordered operators. 
By taking the normal ordering of $X^I$, 
the interaction term is expressed as 
\begin{align}
 - 2 \cc \tr [X^I,X^J]^2 
 &= 2 \cc \left(\epsilon^{IJK} X^I_A X^J_B f^{AB}{}_{C}\right)^2 
\notag\\
 &= 
 2 \cc : \left(\epsilon^{IJK} X^I_A X^J_B f^{AB}{}_{C}\right)^2 : 
 + \frac{8 A \cc}{\MX} : X^I_A X^I_A : 
 + \frac{2 A \cc}{\MX^2} \delta^{II} \delta_{AA}   \ . 
\label{X^4NO}
\end{align}
Then, the second term will be identified with the effective mass term. 
Therefore, the effective mass $\MX$ is given by 
\begin{equation}
 m^3 = 16 A \cc \ . 
\end{equation}

Generalization to higher order is straightforward. 
We take the normal ordering in the expression of the higher order correction, 
and then, the terms at the quadratic order of (normal ordered) $X^I$ 
are identified to the mass term. 
This method is an analogue of the renormalization in field theory. 
The contractions in the normal ordering appear as loops in Feynman diagram, 
and divergence from the loops is canceled by the counter term. 
Here, the contraction term itself is canceled by finite ``counter term'' | the first term in \eqref{VX},  
and the only effective mass term $\MX$ remains. 

Although this criterion is very clear for vacuum state, which has no excitation of $X^I$, 
Generalization to excited states would not be very clear. 
We need to take the contraction to the excitation in the state, 
and quartic terms also give contribution to the effective mass term.

\subsubsection*{Principle of minimal sensitivity}

Another method to determine $\MX$ in \eqref{HX} 
is ``Principle of Minimal Sensitivity'' \cite{Stevenson:1981vj}.%
\footnote{%
An application of this method to the bosonic matrix quantum mechanics was studied in \cite{Kabat:1999hp}. 
}
Since the exact solution does not depend on the effective mass $\MX$, 
the physical quantities can have only very small $\MX$-dependence if the approximation is good. 
Here, we simply fix $\MX$ by the condition 
\begin{equation}
 \partial_\MX E = 0 \ , 
\label{PMS}
\end{equation}
where $E$ is the eigenvalue of the Hamiltonian. 
At the leading order for the ground state 
this condition is equivalent to the variational principle, and 
gives 
\begin{equation}
 \MX^3 = 16 A \cc \ . 
\end{equation}

For higher order $E$ includes the higher order correction of the perturbative expansion.

\subsubsection*{Effective action}

In this criterion, the effective mass is identified with 
that in the effective action. 
The kinetic term in the effective action  in \eqref{HX0} is given by 
the inverse of the 2-point function, whose perturbative expansion 
is schematically expressed as 
\begin{align}
 G 
 &= 
 G_0 + G_0 \Sigma G_0 + G_0 \Sigma G_0 \Sigma G_0 + \cdots 
\notag\\
 &= 
 \left(G_0^{-1} - \Sigma\right)^{-1} \ , 
\end{align}
where $G_0$ is the unperturbed 2-point function. 
The original kinetic term at 0-th order of the perturbative expansion 
is inverse of the unperturbed 2-point function $G_0^{-1}$. 
If the effective mass $\MX$ is chosen to be identical to 
that in the effective action, 
the 2-point function $G$ equals to the unperturbed 2-point function $G_0$. 
This condition implies that the perturbative correction 
of the 2-point function $\Sigma$ vanishes identically. 
For the vacuum state $|\psi_0\rangle$ which has no excitations of $X^I$, 
the first order correction is calculated as 
\begin{align}
 \Sigma 
 &= 
 8 \MX^2 \langle\psi_0|X^I_A X^I_A|\psi^{(1)}\rangle  \ , 
\end{align} 
where the first order correction to the wave function $|\psi^{(1)}\rangle$ is expressed as 
\begin{align}
 |\psi^{(1)}\rangle = \sum_{n\neq 0}|\psi_n\rangle 
 \frac{1}{E_0 - E_n}\langle\psi_n|V_X|\psi_0\rangle \ , 
\end{align}
where $E_n$ is the energy of the states $|\psi_n\rangle$, 
and then, we obtain 
\begin{align}
 \Sigma 
 &=  4 \MX \langle\psi_0|(\AX^I_A)^2  
 \left[- \frac{1}{4} \MX (\AX^{\dag\,J}_B)^2 
 + \frac{4 A \cc}{\MX^2} (\AX^{\dag\,J}_B)^2 \right]|\psi_0\rangle \ , 
\end{align} 
where we have used \eqref{X^4NO}. 
Then the condition $\Sigma=0$ gives the effective mass 
\begin{equation}
 \MX^3 = 16 A \cc \ . 
\end{equation}

%%%%%%%%%%%%%
\subsection{Comments on the perturbative expansion}\label{B:comment}

In the previous subsection, we listed and reviewed various criteria for calculating the mass $\MX$
which give good approximations of the potential $[X^I,X^J]^2$. 
They provide us with different values for $\MX$, but it would not be inconsistent because the mass itself is not a physical observable.
Each of the approximation schemes gives better approximations 
in different situations, under different assumptions or for different physical observables. 

It may be instructive to introduce an alternative viewpoint here. 
In Sec.~\ref{sec:Model}, we split the original Hamiltonian \eqref{H} 
into the free Hamiltonian with the mass \eqref{H0} and the perturbation \eqref{V}. 
Although we are considering the perturbation around $H_0$, which contains $\MX$, 
we are calculating the expectation values for the total Hamiltonian $H$, which is independent of $\MX$. 
Thus, in principle, the physical quantities would not depend on $\MX$ itself, 
if the terms of all order could be taken into account. 
If the perturbation works well and converges, then even with starting
with different values of $\MX$ it is expected that the final value of the physical observable coincide, 
when higher order corrections are taken into account. 
This expectation was verified in Yang-Mills-type matrix models \cite{Kabat:1999hp,Kabat:2000zv,Kabat:2001ve,Iizuka:2001cw,Nishimura:2001sx,Kawai:2002jk,Nishimura:2002va}, in fact.
Note that,  if the expansion is truncated at some order, 
the values of the physical observables depend on $\MX$. 
It is expected that, at a sufficiently high order,  there would be some range of the value of $\MX$ in which the 
$\MX$ dependence of the quantities is suppressed, and they would approximate the quantities 
calculated at all order.

We have just considered the leading order perturbation in this appendix. 
Due to the lack of performing higher order calculations at the moment, 
in this paper, we have not studied concrete values of excitation energy due to the insertion of
the massive $X$ operator to the states. Instead, in this paper we only studied the 
physical consequence of our assumption that the $X$ commutator potential is approximated by
a harmonic potential, for example for the discussion of the magic numbers.
It would be very interesting to study the higher order corrections, 
to see which perturbation has a better convergence for our nuclear matrix model, 
and whether the perturbation actually allows a region of $\MX$ for the independency of $\MX$ on
physical observables.

%%%%%%%%%%%%%%%%%%%%%%%%%%%%%%%%%%%%%%%%%%%%%%%%%%%%%%%%%%%%%%%%%%%%%%%%%%%%%%%%
%%%%%%%%%%%%%%%%%%%%%%%%%%%%%%%%%%%%%%%%%%%%%%%%%%%%%%%%%%%%%%%%%%%%%%%%%%%%%%%%
%%%%%%%%%%%%%%%%%%%%%%%%%%%%%%%%%%%%%%%%%%%%%%%%%%%%%%%%%%%%%%%%%%%%%%%%%%%%%%%%

\section{More on dibaryon spectra}
\label{app:more}

In Sec.~\ref{sec:Dibaryon}, we calculated the dibaryon spectra. 
The coupling constant $\rc'$ was treated as a free parameter and determined 
by fitting the mass formula \eqref{M_nucl3} to the experimental data. 
We also assumed that the constant part of the mass formula is proportional to $A$. 

Here, we directly use the result from the second order perturbation \eqref{V-2nd}, 
and determine the mass $\MX$ by using one of the criteria in App.~\ref{app:Mass}. 
The mass formula is now given by 
\begin{align}
 M_\text{nucl} 
 &= 
 {M}_{0}(A) - M_S \left[(-3 + 2 \delta) A + \delta Y\right] 
\notag \\
&\quad
 + 4\tilde{\lambda} \mw^2 \tilde V_F 
+  16 c(A) \frac{\rc^2 \mw^4}{\MX^3} \left(V_S - A \tilde V_F \right)
 \ , 
\label{M_nucl4}
\end{align}
where $M_{0}(A)$ comes from the mass of the baryon D4-branes. 
If $M_{0}(A)$ is simply the mass of free D4-branes, 
it satisfies $M_0(A) = A M_\text{D4}$, as we assumed in Sec.~\ref{sec:Dibaryon}. 
However, $M_{0}(A)$ contains non-trivial contributions from 
the binding energy of the D4-branes. 
Here, we take this effect into account and treat $M_0(A)$ 
as an arbitrary function of $A$. 
As we have discussed in Sec.~\ref{sec:Dibaryon}, 
$c(A)$ is given by $c(1) = 0$ and $c(2) = 1$. 

The parameters $\ms$, $\rc$, $\delta$ and $M_0(1) = M_\text{D4}$ 
are determined by fitting \eqref{M_nucl4} to the experimental data 
as we have done in Sec.~\ref{sec:Hyperon}, 
and obtained as \eqref{parameters}. 
We fix $M_0(2)$ such that \eqref{M_nucl4} 
reproduces the mass of deuteron, and calculate 
the other dibaryon masses from \eqref{M_nucl4}. 
The result of dibaryon masses for $\MX^3 = 16 A \cc$, $\MX^3 = 8 A \cc$ and $\MX^3 = 6 A \cc$ 
are shown in Table~\ref{table:diB-1/2}. 
Note that the deuteron mass trivially agrees with observation, 
it is the only input to fit the dibaryons mass formula to experimental data. 

The effective mass $\MX^3 = 6 A \cc$ is not obtained in App.~\ref{app:Mass}, 
but is a special case in which the spin and flavor symmetries would be enhanced 
because the coefficient of their Casimir from the interaction terms will be the same in this case. 
Besides, the result from the global fit \eqref{para-diB} and \eqref{para-diB-cont} in Sec.~\ref{sec:Dibaryon} 
corresponds to the effective mass close to this point, $\MX^3 \simeq 11 \cc$. 

For $\MX^3 > 6 A \cc$, the isospin dependence of the mass dominates 
as in the result of the linear order perturbation in Sec.~\ref{sec:Linear}. 
For $\MX^3 < 6 A \cc$, however, the spin dependence becomes more important than that of the isospin, 
and dibaryons with the isospin larger than the spin have smaller mass. 
This is different from results of experiments. 
Also, it might cause a problem in the perturbation 
that the spin dependence $V_S$, which comes from the second order perturbation, 
provides larger contribution than the isospin dependence $V_F$, 
which appears from the linear order perturbation. 
Another important point would be $\MX^3 = 4 A \cc$, 
where $V_F$ from the second order perturbation cancels 
that in the linear order perturbation. 
If $\MX^3 < 4 A \cc$, the dibaryons with larger isospin have smaller mass, 
which is quite different from experiment. 
The isospin dependence $V_F$ from the second order perturbation 
becomes larger than that from the linear order. 
Therefore, the perturbative expansion would be invalid for $\MX^3 < 4 A \cc$. 
Most of the criteria in App.~\ref{app:Mass} give $\MX^3 = 16 A \cc$ or larger, 
and hence the perturbation would be valid even though 
the contribution from the second order is comparable to that from the first order | 
the coefficient of $V_F$ from the second order is 
$-1/4$ times of that at the first order for $\MX^3 = 16 A \cc$.

\begin{table}[tb]
\begin{center}

 \begin{tabular}{|c|ccc|ccc|cc|}
  \hline 
  Dibaryon    &
     $D$ & 
     $D_{12}$ & 
     $D_{03}$ & 
     $D_{10}$ & 
     $D_{21}$ & 
     $D_{30}$ &
     $H$ & 
     $\Omega \Omega$ 
\\ \hline \hline
Experiment
&
1876
& 
2160?
&
2370
&
--
&
--
&
--
&
--
&
--
\\\hline
     Our ($\MX^3 = 16 A \cc$) 
& 
     (1876)
&
     2075
& 
     2001
& 
     2001
& 
     2325
& 
     2748
& 
     1862
& 
     3330
\\ \hline
     Our ($\MX^3 = 8 A \cc$) 
& 
     (1876)
&
     2075
& 
     2125
& 
     1926
& 
     2175
& 
     2424
& 
     1969 
& 
     3221
\\ \hline
     Our ($\MX^3 = 6 A \cc$) 
& 
     (1876)
&
     2075
& 
     2208
& 
     1876
& 
     2075
& 
     2208
& 
     2047
& 
     3148
\\ \hline\hline
Threshold 
&
 N$+$N
&
 N$+ \Delta$
&
 $\Delta$$+\Delta$
&
 N$+$N
&
 N$+ \Delta$
&
 $\Delta$$+\Delta$
&
 $\Lambda$$+\Lambda$
&
 $\Omega$$+\Omega$
\\\hline
Thres.(Exp.) 
& 
 1878
&
 2171
&
 2464
& 
 1878
&
 2171
&
 2464
&
 2232
&
 3344
\\\hline
Thres.(Ours)
& 
 1882
&
 2181
&
 2480
& 
 1882
&
 2181
&
 2480
&
 2230
&
 3352
\\\hline
\end{tabular}

\caption{%
A numerical fit of the hyperon and dibaryon spectrum by our formula \eqref{M_nucl3}. 
Since the deuteron mass is the only input for dibaryons, 
it trivially agrees with observation. 
The mass of $D_{03}$ in the recent experiment \cite{Bashkanov:2008ih, Adlarson:2011bh} is 2370~MeV. 
Signal of $D_{12}$ is also observed around 2160~MeV though it is not very clear if it is a bound state. 
}
\label{table:diB-1/2}
  \end{center}
\end{table}

\subsubsection*{Without mass renormalization}

Here, we take all terms from the contraction between 
the creation and annihilation operators into account. 
We calculate the hyperon and dibaryon spectra by using \eqref{VFcont} and \eqref{VScont}. 
We first make the fitting for the hyperons and the result is the same as 
those in Table.~\ref{table:rhyp}. 
The mass $\mw$ and $\ms$ are different from those in \eqref{parameters}, 
since they do not include the effects of the contraction terms, 
and are obtained as 
\begin{align}
\ms = 761\, \mbox{[MeV]}, \quad
\tilde{\lambda}= 24.9 \, \mbox{[MeV]}, \quad
\delta = 0.339 \, .
\label{para-cont}
\end{align}
The masses $\mw$ and $\ms$ are different from those in \eqref{parameters}, 
since they are not renormalized in this calculation. 
As the difference is only whether the masses are renormalized or not, 
the rescaled coupling constant $\rc$ is the same as \eqref{parameters}. 
The ratio of $\ms$ to $\mw$ is also the same as \eqref{parameters}. 
This would be because the mass renormalization comes from the contractions 
between the creation and annihilation operators and independent of the hypercharge $Y$. 

The mass formula for dibaryons is now given 
by \eqref{M_nucl4} with \eqref{VFcont} and \eqref{VScont}. 
We determine $M_0(2)$ with the deuteron mass. 
Since the contraction terms are independent of the spin $J$ and isospin $I$, 
the results for dibaryons without strangeness are the same as those without contraction terms, 
Table~\ref{table:diB-1/2}. 
Only the spectra of H-dibaryon and di-Omega are affected by the mass renormalization, 
which are shown in Table~\ref{table:diB-1/2-cont}.

\begin{table}[tb]
\begin{center}
 \begin{tabular}{|c|ccc|}
  \hline 
  $m^3$ 
&
  $6 A \cc$
&
  $8 A \cc$
& 
  $16 A \cc$
\\ \hline\hline
   $H$ 
& 
   1933
&  
   1788 
& 
   1570
\\ \hline
     $\Omega \Omega$ 
&
     3038
&
     2960
&
     2842
\\ \hline
  \end{tabular}
\caption{
Masses of H-dibaryon and di-Omega by using \eqref{M_nucl3} with \eqref{VFcont} and \eqref{VScont}.}
\label{table:diB-1/2-cont}
  \end{center}
\end{table}

\end{document}